\documentclass[11pt,letterpaper]{cslreport}
\pdfoutput=1
\usepackage{graphicx}
\usepackage{amsmath}
\usepackage{rotating}
\usepackage{url}
\usepackage{alltt}
\usepackage{color}
\usepackage{xcolor}
\usepackage{boldline,multirow}
\usepackage{colortbl}
\usepackage{fancyhdr}
\usepackage{tabularx}
\usepackage{listings}
\usepackage{amssymb}
\usepackage{fancybox}
\usepackage{gensymb}
\newlength{\hsbw}
 \newenvironment{smallsession}{\begin{flushleft}
  \setlength{\hsbw}{\linewidth}
  \addtolength{\hsbw}{-\arrayrulewidth}
  \addtolength{\hsbw}{-\tabcolsep}
  \begin{tabular}{@{}|c@{}|@{}}\hline 
  \begin{minipage}[b]{\hsbw}
  \begingroup\small\mbox{ }\\[-1.8\baselineskip]\begin{alltt}}{\end{alltt}\endgroup\end{minipage}\\\hline 
  \end{tabular}
  \end{flushleft}}

\usepackage[inline]{enumitem}
\newlist{inlinelist}{enumerate*}{1}
\setlist*[inlinelist,1]{%
  label=(\arabic*),
  itemsep=5pt,
}

\usepackage{soul}
\usepackage{hyperref}
\hypersetup{
    colorlinks=true,
    linkcolor=blue,
    filecolor=magenta,
    urlcolor=cyan,
    pdftitle={DesCert Phase 1 Report for DARPA ARCOS},
    pdfpagemode=FullScreen,
    }

\usepackage{wrapfig}
\definecolor{light-gray}{gray}{0.95}
\newcommand{\code}[1]{\colorbox{light-gray}{\texttt{#1}}}
\usepackage{verbatimbox}
\usepackage{fancyvrb}
\ttfamily

\title{DesCert: Design for Certification}
\acknowledge{Supported by DARPA under agreement number FA8750-20-C-0226.  The views and conclusions contained herein are those of the authors and should not be interpreted as necessarily representing the official policies or endorsements, either expressed or implied, of  DARPA or the U.S. Government.  }
\author{\begin{tabular}[h]{lll}
  Natarajan Shankar & Devesh Bhatt & Michael Ernst \\
  Minyoung Kim & Srivatsan Varadarajan & Suzanne Millstein \\
  Jorge Navas & Jason Biatek & \\
  Huascar Sanchez & Anitha Murugesan & \\
          & Hao Ren & \\
  SRI International  & Honeywell Research & University of Washington \\
\end{tabular}
}
\cslreportnumber{SRI-CSL-2022-01}
\begin{document}

\maketitle

\begin{abstract}
  The goal of the DARPA Automated Rapid Certification Of Software (ARCOS) program is to
  ``automate the
evaluation of software assurance evidence to enable certifiers to determine rapidly that system risk
is acceptable.''  As part of this program, the DesCert project focuses on the assurance-driven development of
new software.  The DesCert team consists of SRI International, Honeywell Research, and the University of Washington.
We have adopted a formal, tool-based approach to the construction of software artifacts that are supported by
rigorous evidence.  The DesCert workflow integrates evidence generation into a design process that goes from
requirements capture and analysis to the decomposition of the high-level software requirements into
architecture properties and software components with assertional contracts, and on to software that
can be analyzed both dynamically and statically.  The generated evidence is organized by means of an
assurance ontology and integrated into the RACK knowledge base.
\end{abstract}

\newpage
\pagenumbering{roman}
\tableofcontents
\listoffigures

\chapter{Introduction}\label{chap:introduction} %shankar
\setcounter{page}{0}  %%% NOT NUMBEREDxgotx
\pagenumbering{arabic}
\pagestyle{plain}
Software has become a core component of many safety-critical systems.
Software is slowly taking over functions that were previously handled by electromechanical devices.
In cars, antilock braking systems use software to delicately and rapidly ``pumping the brakes''
in order to avoid skidding due to the wheels locking up.  Many of the timing, control,  safety, climate control, and
infotainment subsystems
of the vehicle are now handled largely by software.
The use of software supports
greater flexibility, resilience, and versatility in the design and maintenance of a system.
While software is not subject to wear-and-tear,
it can be a significant source of system failure due to bugs and security vulnerabilities since
even a small design or coding error can have big consequences.

Software assurance is a way of ensuring the safety of a system containing software components before
it is deployed.  In the context of system safety,  a \emph{hazard} is a
potentially dangerous event caused by a failure.  A \emph{failure} is a deviation from the intended
behavior.  A failure can be caused by an \emph{error} in the functioning of a component.  This error
itself might result from a \emph{fault} such as a missing check in the software.
An \emph{assurance case} captures the rationale behind the safety of the software
in the context of the system design, its concept of operations, and the mission needs.
The assurance case ensures that all of the hazards, the consequential ways in which the software might
cause the system to fail, have been mitigated in the design.
An assurance case for software consists of claims supported by arguments and evidence.
The claims must cover the software requirements,  design, architecture, code, and platform (including hardware, communication network, systems software, and other libraries)
to demonstrate that software behavior is fit for its intended purpose and does not compromise system safety.

At the highest level, the desired qualities of the software are
captured in the requirements spanning both the functional requirements
and the performance characteristics.  It is well-known that many
software projects fail because of erroneous or ill-defined
requirements.  The requirements process starts with an enumeration of the hazards
and an argument that this enumeration is exhaustive.  The software requirements
are then constructed to specify the intended behavior while mitigating these hazards.
The requirements capture properties
of the behavior of the software including assumptions on the environment, actions that
the software must take in response to inputs, and states that the software must avoid.
Some properties are \emph{generic} in that they are satisfied by a general class of systems.
For example, the software must free from deadlock, type violations, and runtime errors.
Other properties are specific to the intent of the system such as ensuring the accuracy of
the navigational data.  The requirements must be shown to be consistent, complete, and verifiable.
The software design correctly delivering these requirements decomposes into the \emph{architecture}, namely the generic
way in which software functions are decomposed into interacting modules, and the \emph{low-level requirements} (LLRs)
which are the contracts imposed on the software modules.  Models play a critical role in the design and
assurance of a software design capturing environmental assumptions, platform assumptions, operator behavior,
as well as the behavior of the physical components including the plant (the physical system being controlled),
sensors, and actuators.  These models must be sufficiently accurate approximations of the real behavior
of these components and subsystems given the architecture that mediates the interaction between
the software and the physical world.  The hardware platform together with
the operating system, libraries, and glue code  must be shown to correctly implement the architecture.
The source code for the software components must satisfy the contracts and be correctly compiled into executable code
that is properly composed within the architecture.  Integrating all of these claims and evidence into an watertight
assurance case for the system is therefore a significant challenge.

What is the purpose of an assurance case? The claims, argument, and evidence presented in the assurance case can be
examined, probed, and attacked in isolation to ensure that they do indeed make a compelling pre-deployment
case for system safety.  These steps in the argument can fail for a variety of reasons: missing, incorrect,
or ambiguous requirements or assumptions; untrustworthy tools; imprecise or mismatched semantics;
invalid justifications; incorrect designs; and buggy code.
The strength or the \emph{efficiency} of an assurance case is the ease with which
a skeptical reviewer can identify the flaws in the case by examining its individual steps without
having to test or analyze the fully integrated system.  Efficiency is achieved by using languages and
notations with precise semantics, trustworthy tools that are compatible with the semantics, and reusable
artifacts such as testing theories, code generators, proof assistants,  typecheckers, static analyzers, and compilers.
If,  in spite of the efficiency of the assurance case, no flaws are found,
then the software along with its assurance case can be confidently certified for deployment.

If our goal is to reduce the cost of software certification while
increasing its rigor, then the steps in the assurance argument
sketched above must be efficient in the above sense.  It must be easy
for evaluators to identify any gaps or flaws.  If the argument fails,
the source of the failure must be egregious.  This means
that we have to rely on precise claims established by reusable tools
and design concepts like a sound testing theory, requirements analyzers,
model checkers, typecheckers, architecture frameworks, code
generators, and compilers.  The deployment of such reusable tools and concepts
amplifies the falsification space, since the generic claims associated
with these tools, e.g., the soundness of a typechecker, can be
falsified by any instance where the tool fails even outside of its
use within the system under assurance.  Clearly, it can more costly to
establish the high-level claims such as the soundness of model
checkers, typecheckers, and static analyzers, but this cost can be
amortized over the multiple uses of the tool.  Efficiency is not merely a
characteristic of an assurance argument, but a design heuristic that
favors design choices such as rigorous architectural frameworks and
semantically precise notations that amplify the salience of any design flaws.

There are numerous standards for evaluating assurance cases for certification.
IEC 62304 is a standard for the certification of medical device software that
prescribes the evidence artifacts needed for the device to be declared safe for
devices that operate at hazard levels where either no injury is possible (Class A),
no serious injuries are possible (Class B), and death or serious injuries are possible (Class C).
The automotive industry  standard ISO 26262 for functional safety spanning the system
design lifecycle covering requirements, software, and hardware~\cite{26262}.  MIL-STD-882E prescribes
the processes and procedures for certifying system safety for military aerospace systems~\cite{882E}\@.
In the civil aviation setting, SAE ARP4754 describes the guidelines for developing Civil Aircraft
and Systems~\cite{ARP4754}, and SAE ARP 4761 outlines the techniques to be used in system safety assessment~\cite{ARP4761}.
These are used in conjunction with RTCA standards such as DO-254 for hardware design assurance~\cite{DO254}
and DO-178C for software safety standards~\cite{DO178C}\@.  DO-178-C spells out around 71 assurance objectives that must be
fully or partially met in order to certify software for hazard levels ranging from catastrophic (A) to
minor (D).

Many safety standards are prescriptive in terms of development
processes, assurance objectives, and assessment procedures.  These activities
only indirectly impact system safety.  Overarching Properties (OPs) takes a
more direct view of safety assessment in terms of three large
objectives~\cite{UnderstandingOAP}\@.  One, that the \emph{intent} of the system has been accurately
captured.  Two, the system as designed is \emph{correct} with respect to the
intent.  Three, that any additional functionality in the system that
is not traceable to intent must be \emph{innocuous} or \emph{acceptable},
  i.e., must be shown to not
impact system safety.  The OPs approach can be used in conjunction
with other standards as long as the arguments can be made to align
with the intent, correctness, and innocuity objectives.

\begin{figure}
  \centering
\includegraphics[width = .6\linewidth]{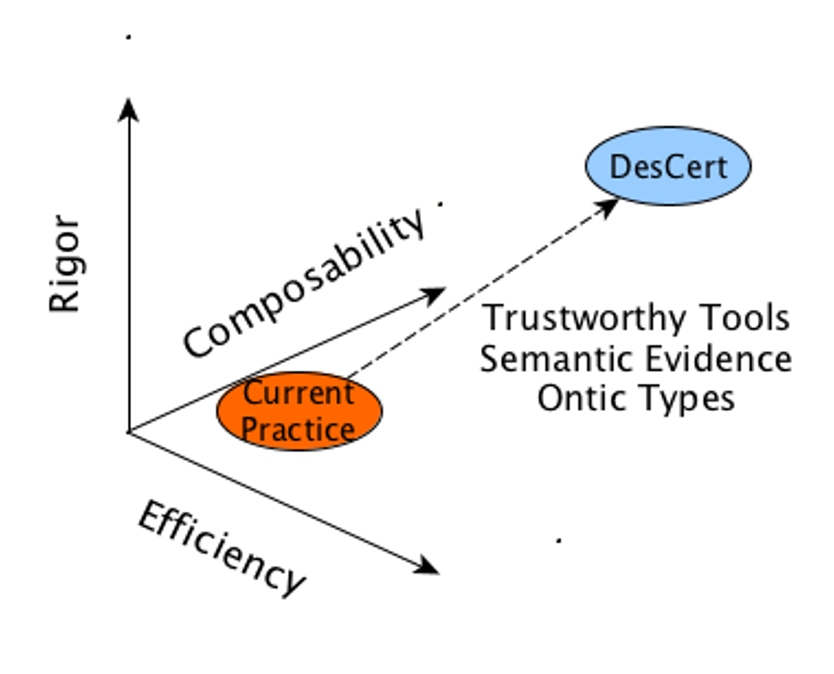}
  \caption{DesCert Project Goals}
  \label{fig:goals}
\end{figure}
Software certification complying with the above standards is an arduous labor-intensive process of collecting
evidence and constructing an assurance case that passes muster with the evaluators.  In safety-critical
settings, certification is the most expensive aspect of software development in terms of time and money.
The ARCOS program aims to reduce the cost and increase the efficiency of developing and certifying
safety-critical software.  The program decomposes assurance case construction into three phases: evidence
generation, evidence curation, and assurance case development.  The DesCert project addresses the
evidence generation phase with a focus on new software development.
As shown in Figure~\ref{fig:goals},  the  project is directed at achieving dramatic improvements
in all phases of high-assurance software development
in three specific dimensions: the formal rigor underlying the claims and evidence associated with the
software artifacts, the composability of evidence generated at different levels of abstraction,
and the efficiency with which the evidence can be integrated into an assurance argument and assessed
by an evaluator.  Toward  this goal, the DesCert approach emphasizes
\begin{enumerate}
\item Automated tools that generate rigorous, semantically coherent evidence
\item Languages and type systems, such as ontic typing, that capture the intent of the design
\item Compositional certification of complex systems
\item Efficient, easily verifiable arguments for capturing software and system safety
\item Reusable and trustworthy automated tools and techniques  with a low amortized certification cost
\item Continuous evidence generation and validation throughout the design lifecycle.
\end{enumerate}

The DesCert team consists of SRI International, Honeywell Research, and the University of Washington.
The DesCert continuous certification workflow integrates a number of requirements modeling, design, and code
analysis tools, including
the CLEAR requirements description language~\cite{bhatt2018clear,hall2018clear}, the Text2Test for generating test cases and models from requirements,
the PVS interactive proof assistant, the  Sally model checker, the Radler Architecture Definition Language,
the Checker Framework for pluggable typechecking, the Randoop system for generating unit tests,
the Daikon system for learning putative assertional properties of software, and the SeaHorn static analyzer.
For continuous assurance, we have developed the Baseline DesCert workflow tool for monitoring and maintaining
the status of the assurance artifacts.  This tool also interacts with the RACK tool for curating
and ingesting assurance artifacts in accordance with an assurance ontology.

The DesCert approach to evidence generation for new software loosely follows the DO-178C objectives.
We employ the CLEAR language for capturing software High-Level Requirements (HLRs) in terms of the
input/output state-machine behavior of software components.  The HLRs are used to generate
test inputs that drive the input space of the operations used in the requirements and the
possible state changes, along with test oracles for judging whether the software component
exhibits the right behavior on the test inputs.  The design of the software complying with the
HLRs consists of the software architecture with its model of computation and interaction
as well as the component low-level requirements.  We use the Sally model checker~\cite{conf/nfm/DutertreJN18}
to analyze the
CLEAR HLRs both at the component level and the integrated behaviors in the context of
the architectural assumptions.  The software architecture is captured within the multi-rate,
quasi-periodic model of computation used by Radler~\cite{conf/memocode/LarrieuS14,conf/memocode/LiGS15}.  The use of Radler allows HLRs to be
established from architecture properties and component code contracts.  We employ static and
dynamic analysis tools for analyzing the code for generic properties such as well-typedness and
the absence of certain classes of runtime errors, as well as for specific properties expressed by
precondition/post-condition contracts.  Ontic type systems are used to connect data values with the
phenomenal quantities that they represent, e.g., ground speed vs. air speed, terrain altitude vs.
barometric altitude, private vs. publishable data, encrypted vs. unencrypted messages,
and filtered vs. unfiltered sensor inputs.  These ontic labels
are used in a consistent way at the requirements, design, and code levels to detect a broad
class of data misuse errors that are not detected by ordinary type systems.   The assurance workflow
is implemented and monitored in a framework called Baseline DesCert that maintains the claims and
evidence as the design evolves.

We present an outline of the DesCert approach to evidence generation within an
assurance-driven methodology for the development of  safety-critical software
in Chapter~\ref{chap:approach}\@.  The details of our approach are fleshed out in subsequent chapters.
In Chapter~\ref{chap:challenge}, we describe the case study of our system under assurance, namely, an Advanced Fail Safe (AFS) module of the ArduCopter platform that is used to detect and recover from specific classes of failures.
Chapter~\ref{chap:ontology} outlines our evidence ontology.  The details of the assurance tools and their
integration are given in Chapter~\ref{chap:tools}\@.  The Baseline DesCert continuous assurance flow manager is
described in Chapter~\ref{chap:baseline}\@.  The details of evidence generation using the DesCert methodology
on the AFS case study are laid out in Chapter~\ref{chap:evidence}\@.  Concluding observations are given in
Chapter~\ref{chap:conclusions}\@.

\chapter{The DesCert Approach}\label{chap:approach} %shankar
In the DesCert project, we have adopted an Assurance-Driven Development (ADD) approach to software where
the design objective is to create software that is supported by semantically rigorous evidence that
can be incorporated into an efficient argument.  The evidence artifacts we create are aligned with the
DO-178C objectives but are also compatible with the goals of intent, correctness, and innocuity
associated with Overarching Properties.  We illustrate our approach to rigorous evidence generation with
an Advanced Fail Safe module for the Arducopter rotorcraft platform.  

\section{Assurance-Driven Development (ADD)}\label{sec:ADD}

The products of engineering have been engines, bridges, buildings, factories, planes, and automotives.
These products are designed based on abstractions of chemical, structural, mechanical, and
electrical laws and processes that are observable with reasonable precision and
reproducible with uncanny accuracy.  Software is a different kind of beast.  Each piece of software
is \emph{sui generis}\@.  Software-intensive systems have to meet a stringent range of
demands spanning functionality, performance, reliability, persistence, security, and maintainability.
Software designs can be exceedingly complex.  Such systems can be composed of many, deeply nested layers of abstraction.  There are few
general laws that help in modeling and understanding software behavior.  The operation of
software is typically only indirectly observable.  Software-related failures can range from annoying issues like memory
leaks and poor search results to design flaws, security holes, and bugs.  Even minor errors
like numerical overflows can have catastrophic consequences, as was illustrated in the
failure of the Ariane-5 launch.   Software-intensive systems exhibit high internal
structural complexity, as well as high external complexity in meeting a broad range of requirements
and use cases.  Since it is not possible for testing to span the internal and external complexity, it is
important to independently certify the behavior of the software for its intended use.
Certifying software-intensive systems is hugely expensive due to the combination of
internal and external complexity.

Performing a \emph{post facto} certification of a completed software project suffers
from confirmation bias.  Any gaps uncovered late in the software lifecycle
during the certification process are likely to be too costly to fix.
The DesCert project approaches certification as an integral part of the software
lifecycle.  We achieve software certification through an assurance-driven development (ADD) methodology
where
\begin{itemize}
\item The primary objective of the development is an assurance argument that can be maintained along with the software, and
\item The claims, arguments, and supporting evidence are developed and refined through the software lifecycle
\end{itemize}

\begin{figure}[!htb]
\centering
\includegraphics[scale=.55]{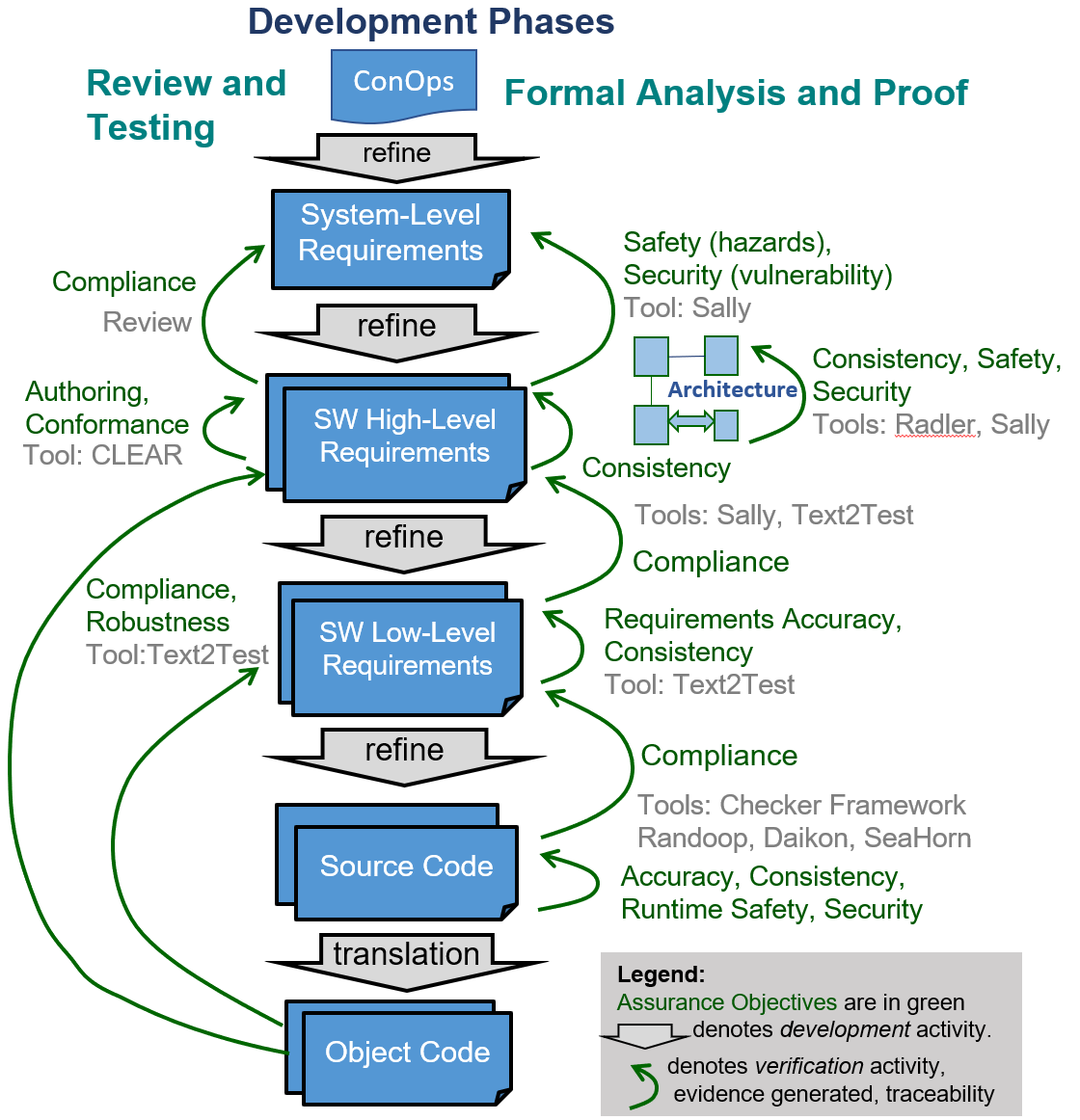}
\caption{DesCert Assurance Driven Development Methodology}
\label{fig:assurance-driven-development}
\end{figure}

The ADD methodology tracks the software development artifacts (high-level requirements, architecture,
low-level requirements, source code, object code) through each stage of the lifecycle.
Tools are employed at each stage to generate evidence to show that
\begin{enumerate}
\item The refinement of the development artifacts from one stage to the next is correct.
\item The development artifacts produced in each stage are accurate and consistent,
      and exhibit runtime safety and security properties.
    \end{enumerate}

    In Phase 1 of the ARCOS program, we have focused on the ADD infrastructure in the form of requirements
    definition and analysis, architecture and build systems, trusted tools for static and dynamic code analysis,
    and continuous assurance and evidence integration workflows.

\section{Designing for Efficient Assurance Arguments}\label{sec:efficient}

On 2 September 2006, an RAF Nimrod  XV230 ``suffered a catastrophic mid-air fire'' while flying
in Helmand province, Afghanistan.  All fourteen people aboard the plane died.
The fire happened 90 seconds following air-to-air refuelling (AAR).
The Nimrod, developed from the de Havilland Comet,
has been flying since 1969
but the AAR had been added by BAE first in 1982 and upgraded in 1989, and
certified on the basis of a \emph{safety case} developed
by BAE in consultation with QinetiQ during 2001-2004.
The cause of the fire was a fuel leak around the AAR
that was ignited by contact with an exposed (due to frayed/inadequate insulation)
element of the cross-feed (CF) duct (1969-75) and Supplementary Conditioning Pack (SCP) duct
(1979-84) that transported hot (470\degree C) air.  The cross-feed duct was
placed dangerously close to a fuel tank.  The accident was the subject of an investigation by
Charles Haddon-Cave~\cite{Nimrod09}\@.  The report from this investigation pointed out a number of things
that were wrong with the design of the AAR system as well as the ``safety case''.
A key point noted by the report is
\begin{quote}
\emph{As a matter of good engineering
  practice, it would be extremely unusual (to put it no higher) to
  co-locate an  exposed source of ignition with a potential source of
  fuel, unless it was designated a fire zone and provided with
  commensurate protection. Nevertheless, this is what occurred within
  the Nimrod.   }
\end{quote}
    
The report also observes that:
\begin{quote}
\emph{
     A  Safety  Case  itself  is  defined  as ``a  structured  argument,
     supported  by  a  body  of  evidence,  that  provides a
     compelling, comprehensible and valid case that a system is safe
     for a given application in a given environment''.}

\emph{     The basic aims, purpose and underlying philosophy of Safety Cases
     were clearly defined, but there was limited practical guidance as
     to how, in fact, to go about constructing a Safety Case.
     \ldots
     If the Nimrod Safety Case had been properly carried out, the loss of XV230 would have been avoided.
   }
 \end{quote}

 The Nimrod XV230 accident and a number of other accidents demonstrate that
 failures can arise from a combination of many sources: poor regulation, 
 inept management, bad design, defective engineering, inadequate maintenance,
 and improper operation.  If a safety case or assurance argument covered 
 each of these sources of failure in their full complexity, it would fail 
 to be convincing.  No evaluator would have the resources to draw out all of
 the flaws in such a complex assurance case.  With the Nimrod, the safety
 case as presented drew attention away from the simple requirement that
 \emph{fuel and ignition should not interact outside the combustion chamber.}
 This would mean that any heat sinks would need to be physically and
 thermally isolated from fuel both as part of the design and maintenance.
 Any deviation from this restriction would be easily detected.  More importantly,
 the need to develop such an efficient argument for safety serves as a design
 heuristic: \emph{avoid design features that lack efficient
   arguments.  }

 An efficient argument requires designs with proven background theories,
 large safety margins, trusted tools and
 processes, secure platforms, and architectures that offer strong guarantees.  The trusted processes
 must include the use of powerful modeling and analysis tools for analyzing requirements.
 For example, in the DesCert project, we use the CLEAR modeling language to capture requirements
 in a precise and cogent notation.  The requirements are analyzed through model checking and
 testing.  Ontic types are used to connect a data representation with the quantity it represents.
 Ontic type analysis can rule out a number of potential design flaws such as sending unencrypted
 data over a public channel, allowing sensitive operations to operate on tainted inputs,
 giving unauthenticated users access to sensitive data, or performing calculations with
 raw, unfiltered sensor input.  The software architecture must also support strong claims for
 isolating functionality so that they can
 only interact through the ``official channels'' supported by the architecture.  The architecture
 can thus guarantee correct timing and functional behavior while protecting against Denial of Service (DoS)
 and side-channel attacks.  Similarly, a certified build system and a secure platform can guarantee
 the provenance and fidelity of the software that is executed.  In Descert, we rely on Radler to offer
 guarantees on the architecture and the build process.  We also employ static and dynamic analysis tools to
 deliver strong guarantees for the safe execution of C, C++, and Java code used in the implementation.
 We also gain efficiency by employing  a safety monitor to observe the execution of the system and ensure that
 the safety policies are not being violated.  

 We can illustrate the contrast between an efficient and inefficient argument with a few simple examples.
 The use of  a separation kernel to guarantee isolation between processes yields an efficient argument
 in contrast to a fine-grained argument for separation bases on analyzing the memory accesses of the
 individual processes.  The former argument rests on the correctness of the separation kernel, which might 
 be a substantive assurance exercise, but one with a much larger falsification space.  Claims about the separation kernel can be evaluated once and reused multiple times even within the same assurance argument, whereas any analysis of the memory accesses
 would have to be repeated for each instance.  Similarly, one could use a programming language with a strong type
 system such that a type-safe program cannot crash.  This argument depends on assurance regarding
 the soundness of the typechecker, where the cost of this assurance case can be amortized over multiple uses
 of the programming language.  The claim for the soundness of the typechecker also fails easily if there is any bug,
 including those that are irrelevant to the programs in the current project.  The key design strategies for
 efficient arguments are 
\begin{itemize}
\item Precise claims
\item Validatable models and assumptions
\item Amortized cost through trusted and reusable design tools/artifacts
\item Architectural separation of concerns
\item Rigorous chain of reasoning and evidence
\end{itemize}

Eventually, when we have enough data on certification costs, we can quantify
the efficiency of the argument by comparing the amortized cost of an efficiently argued claim
against the cost of an inefficient, one-time argument.  
 In summary, DesCert evidence generation supports strong, reusable claims through the use of the CLEAR requirements
 notation, Sally model checking, Radler architecture definition, Ontic type analysis, and powerful static and
 dynamic code analysis.  

\section{Motivation: The Eight Variables Model}\label{sec:8var}

The ADD methodology shown in Figure~\ref{fig:assurance-driven-development} illustrates the above aspects of assurance with corresponding development and verification activities. The development artifacts in successive lifecycle stages are shown in the center column of this figure, with verification activities denoted by green arcs on the left and right sides. The arcs going from a development artifact to a higher level one show compliance --- i.e., the correctness of refinement (Goal 1). The self arcs analyze a particular development artifact to show accuracy, consistency, runtime safety, and security (Goal 2).

Tools are used to automate the verification activities at each stage of the lifecycle, thus enabling incremental, continuous assurance throughout the lifecycle. The use of automated tools also enables a systematic process of iterative refinement and defect removal in early lifecycle stages. For example, requirements consistency and completeness defects, reported by Text2Test tool, can be iteratively removed before proceeding to the next stage.
In this respect, the DesCert approach is similar to recent agile and test-driven-development methods as opposed to the traditional waterfall method. Chapter~\ref{chap:baseline} describes the continuous assurance flow automation in DesCert.

The ADD methodology employs both review and testing based methods (shown on the left side of Figure~\ref{fig:assurance-driven-development}) and formal methods (shown on the right), both supported by tools. The two types of methods can support complementary assurance objectives or can be used for the same objectives to lower the testing burden/cost and to increase confidence. The concept of \emph{properties}, essential to this approach, is described in Section~\ref{sec:property-desc}. The tools and their usage to generate evidence is described in Chapter~\ref{chap:tools}.

\begin{figure}
  \centering
  \includegraphics[width=.6\linewidth]{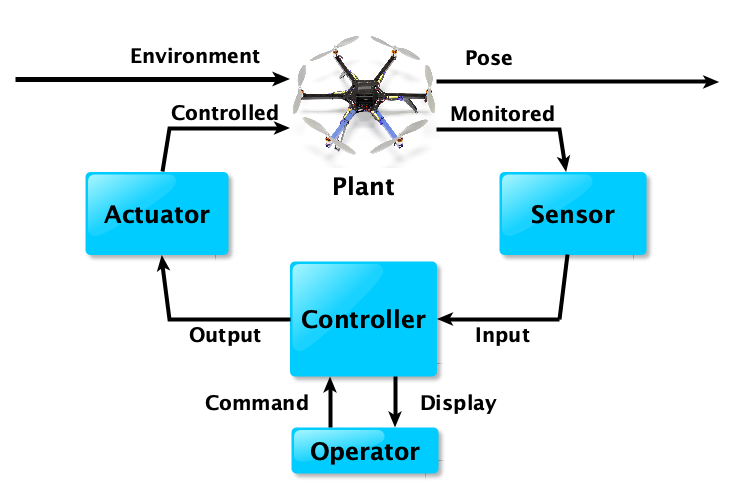}
  \caption{The Eight Variables Model}
  \label{fig:8var}
\end{figure}

Our specific instantiation of the assurance-driven development (ADD) %DO-178C guidelines
follows our Eight Variables Model (8VM) of
cyber-physical systems.  The model as shown in Figure~\ref{fig:8var} categorizes the
classes of components, agents, or actors in the design and the variables that capture the observable behavior
of these components and their interactions.\footnote{The variables here are just labels for the
  observable behaviors of the actors and their interactions, and should not be confused with
  program variables. }
In a typical cyber-physical system, there is a physical
plant, such as a vehicle or a building.  The Pose of the plant is a class of variables that
includes the position, orientation, temperature, etc. of the plant.  The plant is also interacting
with an external physical World covering the terrain, wind, friction, and other factors.
We can measure some of the physical variables, namely, the Monitored ones,
in the Plant and Environment through Sensors.  The Sensor observes these physical values of the
Monitored variables and writes these observations as digital values of the class of Input variables.
The Input variables
are processed by the Controller, i.e., the software component, along with any Operator Commands,
to produce the Output to the actuator and updates to the operator Display.  The Output drives
the Actuator to produce the Controlled input to the Plant.  Note that the Environment,
Pose, Command, and Display are the externally observable variables, and the Controlled, Monitored, Input,
and Output variables are internal.  The Environment, Pose, Controlled, and Monitored variables
are physical variables.  The Input, Command, Display, and Output variables are digital variables.
Also, the Controlled, Monitored, Input, and Output variables are Parnas's original Four Variables.

We typically work with models of the
World, Plant, Sensor, and Actuator, and even the Operator.  The top-level
claim for the whole System has the form below
\begin{alltt}
     WorldModel(Environment) AND
     PlantModel(Environment, Control, Pose, Monitored) AND
     SensorAccuracy(Monitored, Input) AND
     ActuatorResponse(Output, Control) AND
     ControllerOutput(Input, Command, Output, Display) AND
     OperatorModel(Display, Command)
   IMPLIES
   Requirement(Command, Environment, Pose, Display)
\end{alltt}

In DesCert Phase 1, we are primarily focused on the Software Requirements which are
captured by the ControllerOutput predicate.  The physical variables
can be continuous or switched continuous.  In the latter case, the
variable is characterized by a series of epochs defined by the
switching times $\mathit{switch}(i)$.  For each time $t$, there is a
function $\mathit{epoch}(t)$ that indicates the epoch to which the
time $t$ belongs.  For example, the torque delivered by the engine might be a
switched variable that is switched by means of gear shifts.
The digital variables are switched but, unlike physical variables, the values
are latched within each epoch.  For example, the variable \texttt{input} 
is a sampled sequence of values of the physical variable \texttt{monitor}.
The \texttt{SensorAccuracy} predicate constrains the discrepancy between the
value of the physical variable and its sampled counterpart.

\begin{figure}
  \centering
  \includegraphics[width=.6\linewidth]{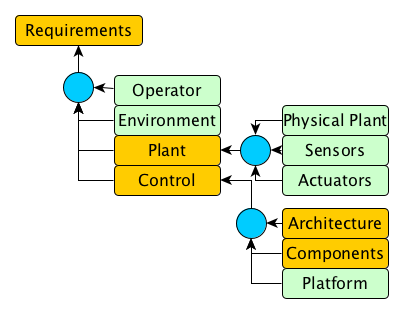}
  \caption{Assurance Argument Structure}
  \label{fig:argument}
\end{figure}

The software requirements in the \texttt{ControllerOutput} specification are
implemented by the Architecture and the Component Contracts (the Low Level Requirements or LLRs).  The Architecture consists
of the Logical Architecture and the Physical Architecture.  The Logical Architecture
specifies the nodes (with their periods, step functions, their list of published and subscribed topics)
and the topics.  The Physical Architecture maps nodes to virtual machines on actual physical
hardware platforms, and topic channels to specific communication mechanisms.
The assurance argument is factored so that the ControllerOutput specification is entailed by the
Component Contracts and the Logical Architecture.  The Physical Architecture can be shown to
imply the assumptions made in the Logical Architecture regarding scheduling jitter, worst-case execution time (WCET) and communication latencies and throughput.  The semantic structure of the assurance argument is
described in Figure~\ref{fig:argument}\@.  In supporting this argument, we need to ensure
that we also have empirical or analytical evidence to support the argument nodes (the circles) as well.

The validity of the argument rests on the strengths of the inference nodes.  These inference nodes
can be \emph{inductive}, as in supporting a claim with test evidence,
\emph{deductive}, as in deductively decomposing a claim into subclaims,
or \emph{probabilistic}, as when the probability of failure is computed.
The inference nodes represent argumentation patterns that are used repeatedly and
hence compatible with the efficient argument goal.

   \section{The Radler Architecture Framework}\label{sec:Radler}

The Radler framework employs a distributed quasi-periodic model of
computation where individual computation nodes interact within a
publish/subscribe architecture~\cite{conf/memocode/LarrieuS14,conf/memocode/LiGS15}.
The logical architecture consists of
nodes and topics.  Each node publishes on a collection of topics,
where each topic has at most one publisher node, and subscribe to
another collection of topics.  The nodes execute periodically with
minimum and maximum bounds on the period between two successive
executions.  In each execution, a node reads its subscription mailbox
to extract the inputs to which it applies a step function.  The
outputs of the step function are then published to the corresponding
subscriber mailboxes.  A number of useful properties can be derived
directly from Radler logical architecture.  The physical architecture
maps the nodes to virtual machines which are themselves mapped to
physical machines, and the mailbox semantics is implemented using
physical communication channels and buffers.  Within the Radler
architecture, behavioral properties of the state machines can be
established using the step function precondition/postcondition
contracts and the logical architecture.  The step function code can be
independently analyzed for compliance with the contract, the
worst-case execution time bounds, and for generic properties such as
the absence of runtime errors or ontic type violations.  Traceability
information can be maintained tracking requirements to state machines
which are decomposed into Radler nodes with step functions implemented
by code.  Radler also provides a certified build system so that the
executables are created with the glue code needed for execution and
interaction.

   A Radler software architecture is specified in the Radler Architecture Definition Language (RADL)
   as a publish/subscribe system with periodically executing nodes publishing on bounded latency channels.
   An example \texttt{.radl} file with the architecture definition is shown in Section~\ref{sec:RADL}\@.
   A RADL architecture definition consists of a \emph{logical architecture} and a \emph{physical architecture}\@.
   The logical architecture specifies \emph{nodes} and \emph{topics}.  Each topic has a unique publisher node
   and a message type.  Each node specifies a set of topics to which it subscribes along with the buffer sizes,
   and a set of topics on which
   it publishes.  In addition, the node descriptions captures the minimum/maximum period, the expected latency
   on each subscribed channel, and the step function that maps the subscription message buffers to published
   messages.  The channel latency is measured from when the publisher starts executing its step function
   so it includes the worst-case execution time.
   A node might also be attached to and exchanging data with zero or more devices.  The physical architecture
   maps nodes to processors and virtual machines and maps each topic channel between a publisher/subscriber pair
   for a topic to a physical channel on a communication bus connecting the two endpoints.

   From the logical architecture, we can derive a number of useful theorems that have been formally verified using PVS.
   These theorems can be used to derive ArchitectureProperties that capture the end-to-end latencies and other timing properties
   in the architecture.  By factoring out these properties, we can modify the logical architecture while preserving
   the Architecture properties so as to maintain the structure and validity of the assurance argument.
   \begin{alltt}
LogicalArchitecture(Nodes, Topics)
IMPLIES
ArchitectureProperties(Input, Command, Output, Display, Nodes, Topics)
   \end{alltt}
   Let $\mathit{min}(n)$ and $\mathit{max}(n)$ be the minimum and maximum periods of node $n$, and
   let $D_{mn}$ be the message latency between publisher $m$ and subscriber $n$.  We then have the following claims:
   \begin{enumerate}
   \item If $\mathit{min}(m) > D_{mn}$, then $n$ received messages from $m$ in the same order in which they were sent. (No Overtaking)
   \item Subscriber $n$ can conservatively detect the failure of $m$ by observing when $k$ consecutive periods have transpired
     with no new messages from $m$ for $k . \mathit{min}(n) > D_{mn} + \mathit{max}(m)$. (Failure Detection)
   \item Messages can be lost because the buffer is over-written by newer messages from the publisher.
     Under the assumption of No Overtaking and a buffer size of $L$ messages, no more than
     $M-L$ consecutive messages can be lost for smallest $M$ such that $M.\mathit{min}(m) > D_{mn} + \mathit{max}(n)$.
     This is a crucial property that ensures that in each step, the subscriber $n$ sees at least one of every $M - L + 1$
     consecutive messages sent by $m$, and that $n$'s buffer contains the last $L$ messages received. (Bounded Message Loss)
   \item The age of a message from $m$ to $n$
     is the time elapsed between when it is published by $m$ and and the time it
     is processed by the subscriber $n$.  The maximum age of a message is bounded by $D_{mn} + \mathit{max}(m)$\@. (Bounded Age)
 \end{enumerate}

 The upshot of these architecture properties is that if a publisher is signalling a condition, then it has to be aware
 that due to Bounded Message Loss, the condition has to be signalled in at least  $M - L + 1$ consecutive messages
 in case $M - L$ of these messages are over-written in the buffer.  Conversely, a subscriber's published values are
 based on inputs with published age constrained by the Bounded Age assumption.  In the case of the room temperature controller,
 the latter assumptions bounds the delay between sensing the temperature and the actuation of the heater.  When combined
 with bounds on the leakage rate and the heating rate, and the component contracts, the architecture properties
 can be used to  demonstrate that the room temperature eventually stabilizes to a range between \emph{Min} and \emph{Max},
 and remains stably in this interval.  In general, temporal properties like this can be derived from the combination
 of the step function contracts and the architecture properties.

   The \texttt{ArchitectureProperties} follow from the logical
   architecture description, which in turn is satisfied by the mapping
   to the physical architecture and the properties we assume of the
   physical platform.

   \begin{alltt}
PlatformProperties(VirtualMachines, TransportMedium) AND
PhysicalArchitectureMapping(Nodes, VirtualMachines,
                            Topics, TransportMedium) 
IMPLIES LogicalArchitecture(Nodes, Topics)
\end{alltt}

The Radler architecture framework also includes a software build system that takes as input the logical and
physical architecture definition and the associated source files for the step functions and creates a collection
of executable binary images that can be launched on the physical platform.  The software running on the
platform implement the architecture in terms of the nodes executing periodically and communicating on the
topic channels.  The Radler build system also adds monitors to check that the specified latency bounds are
not breached and adds a flag to the message to indicate if the contents are based on stale inputs (which can
occur even with normal behavior) or missing inputs (which is abnormal).  We extended Radler to integrate
nodes defined using Java code running on a Java Virtual Machine (JVM).  We specifically added a Java-defined
node executing the BeepBeep3 safety monitoring framework.  Unlike the AFS node which executes recovery
actions, the safety monitor is a passive component that ensures that any failure event or combination of
events does trigger the appropriate recovery action.  

\section{A Motivating Example: Room Temperature Regulation}\label{sec:thermostat}

\begin{figure}[htb]
  \centering
  \includegraphics[width=.8\linewidth]{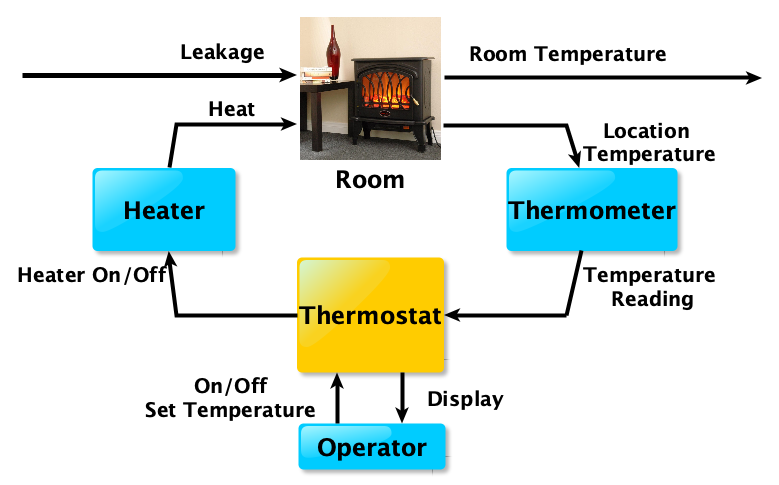}
  \caption{Eight-Variable Model for Thermostat Room Temperature Controller}
  \label{fig:8var-thermo}
\end{figure}

We can illustrate the argument template using a simple example of a thermostat-based room temperature
controller which captures the structure of the argument and the forms of evidence.  The temperature
controller can be mapped on to the eight-variables model as shown in Figure~\ref{fig:8var-thermo}.
The thermostat turns the heater on or off, and the thermometer senses the room temperature at a specific
location.  The operator can switch the thermostat on or off and set the desired temperature.  

If we consider a simple system like a thermostat, the main requirement is that it
maintains the room temperature around the set temperature between 
\texttt{Low} and \texttt{High} by switching the thermostat on when the sensed temperature falls below
$\mathtt{Low} + \Delta$, and off when the sensed temperature exceeds $\mathtt{High}  - \Delta$\@.
There might be additional requirements, for example, that there is some hysteresis built into the
switch for the heater so that it is not damaged by being switched on and off too frequently.  
We assume (\texttt{SensorAccuracy}) that there is an error of $\epsilon$ in sensing the temperature, and
 the room temperature
 can rise (\texttt{ActuatorResponse}) or fall (\texttt{PlantModel}) at no more than a rate of $rho$
 degrees per second.   We also assume that when the heater is on, the room temperature rises
 at a rate of at least $\rho^-$ degrees per second, for $0 < \rho^- < \rho$\@. 

 The thermostat example is not as trivial as it might seem.   Assumptions about the sensor,
 actuator, plant, and environment are needed to achieve the desired behavior.  Additionally,
 the architectural model contributes timing latencies that need to be factored into the argument.  
 The desired property is  that the 
room temperature is maintained between $\mathtt{Low}$ and $\mathtt{High}$ when the thermostat is on.  
However, this might not hold at the initial point when the thermostat is switched on.
If the initial temperature is already above $\mathtt{High}$, then there is no way to force the
temperature to within the acceptable range since the system only heats and does not cool.  
When the initial temperature $\theta_0$ is below $\mathtt{Low}$, 
it will take some time, at least $\frac{\mathtt{Low} - \theta_0}{\rho^-}$, from when the heater is switched on 
before the temperature converges to within this bound.  Since there is a latency of at most $\tau$, in sensing the
temperature and switching on the heater, the temperature could drop to $\theta_0 - \rho\tau$ within this time,
we need to allow at least $\frac{\mathtt{Low} - \theta_0 - \rho\tau }{ \rho^-} $ seconds following the switching on of
the thermostat,  for convergence to have occurred.  Once the thermostat has been switch on and enough
time has elapsed for the room temperature to converge to the acceptable range, it can be shown that
it remains within this range.  This is because, when the temperature is below $\mathtt{Low} + \Delta$, then
either the heater is already switched on, and the temperature is rising, or it is off.  The latter condition
can only arise when temperature was above $\mathtt{Low} + \Delta - \epsilon$ at least $\tau$ seconds ago.
As long as $\frac{\mathtt{Low} + \Delta - \epsilon}{ \rho^+} < \tau$, we can ensure that the temperature does not
drop below $\mathtt{Low}$ before the heater is switched on.  Symmetrically, we see that as long as
$\frac{\mathtt{High} - \Delta + \epsilon}{\rho^+} < \tau$, the temperature does not exceed $\mathtt{High}$\@.
If we ensure a $\mathtt{High} - \Delta$ exceeds $\mathtt{Low} + \Delta$ by at least a positive quantity
$\gamma > 2\epsilon$, then we can fulfil the hysteresis requirement that the heater not be switch on or off
too frequently.  This is because there will be a gap of at least $\frac{\gamma - 2\epsilon}{\rho^+}$ between
the heater being switched on and the off, or vice-versa.

\begin{figure}[htb]
  \centering
  \includegraphics[width=.8\linewidth]{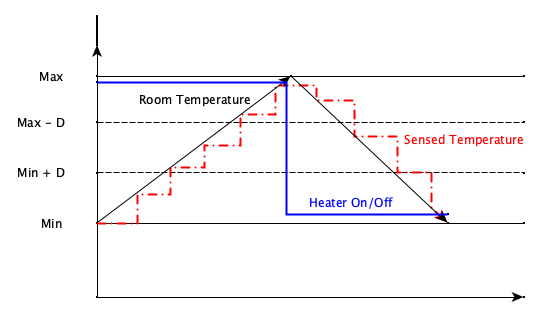}
  \caption{Thermostat Behavior in Radler}
  \label{fig:thermostat}
\end{figure}
The correct behavior of the temperature controller depends on some crucial architecture properties of the Radler model.
The interaction between the physical room temperature, the digital sampled sensed temperature, and the
activation of the heater is shown in Figure~\ref{fig:thermostat}\@.
The thermometer, the thermostat controller can be viewed as independent nodes in the architecture.
The thermometer samples the room temperature at a rate, say, of 10Hz.  The thermostat switches the
heater on or off depending on whether the detected temperature falls below $\mathtt{Low} + \Delta$ or
exceeds $\mathtt{High} - \Delta$\@.  
A temperature reading of $\hat{\theta}$ might
represent an actual temperature $\theta$ in the range $\hat{\theta} \pm \epsilon$ and drift by at most $\rho/10$ between readings.  
 When the sensed
 temperature falls below $Low + \Delta$, it will be sampled within a tenth of a second.
 If we assume that the message latency between the temperature sensor that the heater controller is
 at most a tenth of a second.  If we assume that the thermostat is operating at 2Hz, then
 the end-to-end delay between the temperature falling below $\mathtt{Low} + \Delta$ and the
 heater being switched off could exceed $.5 + .1 + .1 = .7$ seconds.  
 In order for
the thermostat to regulate the room temperature within the $[\mathtt{Low}, \mathtt{High}]$ interval,
we have assume that rate of change of temperature, both during heating and cooling, is bounded.
We need to ensure that the delays introduced by the sampling rates and message communication is
sufficiently small that the temperature remains within the safe interval in the time between the
temperature triggers $\mathtt{Low} + \Delta$ and $\mathtt{High} - \Delta$ are detected, and the heater
is turned on or off.

In the above argument, we are relying on the Radler architecture properties as well as component contracts regarding
the step functions associated with the individual nodes.  These component contracts are just precondition/post-condition
pairs associated with these step functions relative to the mailbox inputs on their subscribed channels
and the outputs on their published channels.  For example, the thermostat contract is that it must direct
the heater to be switched on (respectively, off) when the sensed temperature falls below $\mathtt{Low} + \Delta$
(respectively, exceeds $\mathtt{High} - \Delta$).  

The argument supporting the \texttt{ControllerOutput} claim can be stated as
\begin{alltt}
  ArchitectureProperties(Input, Command, Output,
                         Display, Nodes, Topics) AND
  ComponentContracts(Input, Command, Output, Display, Nodes)
  IMPLIES ControllerOutput(Input, Command, Output, Display)
   \end{alltt}

   As we saw in the case of the thermostat, assumptions SensorAccuracy, ActuatorResponse, WorldModel, PlantModel,
   and OperatorModel might all factor into the design of the Controller since certain couplings between
   variables might hold only because of these \emph{external} constraints.  As an example,  an interlock in the
   operator console might mean that certain commands are impossible in specific modes.
   The key takeaways are that
   \begin{enumerate}
   \item The Eight Variables model decomposes the argument for a system into precisely stated modeling assumptions
     about the  environment, physical plant, sensors, actuators, and operator under which the software requirements
     must be met.
     \item Writing software requirements even for simple systems such as a temperature controller can be quite subtle. 
     \item The model of computation used in the architecture allows independent software components to be
       be independently developed with their individual component contracts.
     \item The argument that the software requirements have been implemented follows from the architecture
       properties, the logical architecture, and the component contracts.
     \item The assurance case must also demonstrate that the physical architecture satisfies the assumptions
       regarding scheduling, worst-case execution time, and channel latencies specified in the logical architecture. 
   \end{enumerate}

      The ArduCopter system is more complicated than
   a thermostat, but the structure of the decomposition into claims
   and subclaims remains the same, and much of the underlying reasoning follows the same pattern.

\section{Aligning the Assurance Argument Structure with DO-178C Guidance}

\begin{figure}[htb]
  \centering
  \includegraphics[width=.6\linewidth]{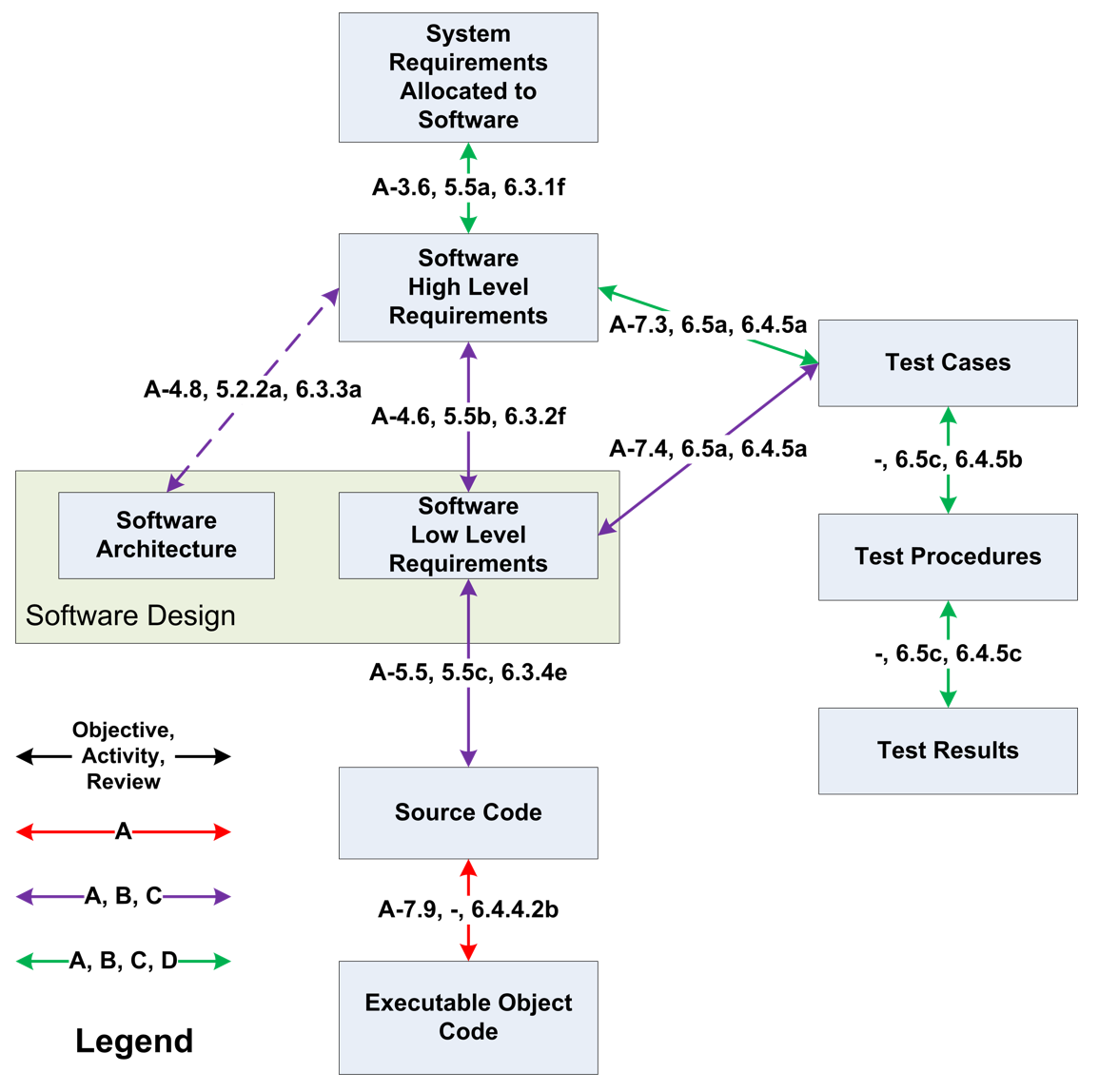}
  \caption{The DO-178C objectives and argument structure}
  \label{fig:do178c}
\end{figure}

The assurance driven workflow and argumentation structure shown in Figure~\ref{fig:assurance-driven-development} aligns with the DO-178C scaffold as shown in Figure~\ref{fig:do178c}\@.
The RTCA DO-178C guidelines specify certification objectives based on five design assurance
levels (DAL) that correlate with the impact of anomalous behavior: Catastrophic (Level A), Hazardous (Level B),
Major (Level C), Minor (Level D), No Effect (Level E). Figure~\ref{fig:do178c} (taken from \url{https://en.wikipedia.org/wiki/DO-178C}) shows the objectives and activity for each DAL, with required traceability between artifacts.

The assurance argument in our case is for the AFS component, and it consists of claims for
\begin{enumerate}
\item Tool validity: These will support the validity of the claims and counterexamples generated by the individual tools as supported by tool qualification%regression suites.
\item Validity of high-level requirements: Consistency, completeness, verifiability, and compliance with system-level requirements, including traceability.
\item Validity of mapping from high-level requirements to low-level requirements (LLR), including traceability: Radler architecture + Sally models.
\item Validity of source code: absence of runtime errors, and compliance with LLR, including traceability.
\item Validity of object code: generation of tests from high- and low-level requirements and test execution on object code, including traceability.
\end{enumerate}

\begin{figure}[htb]
  \centering
  \includegraphics[width = .8\linewidth]{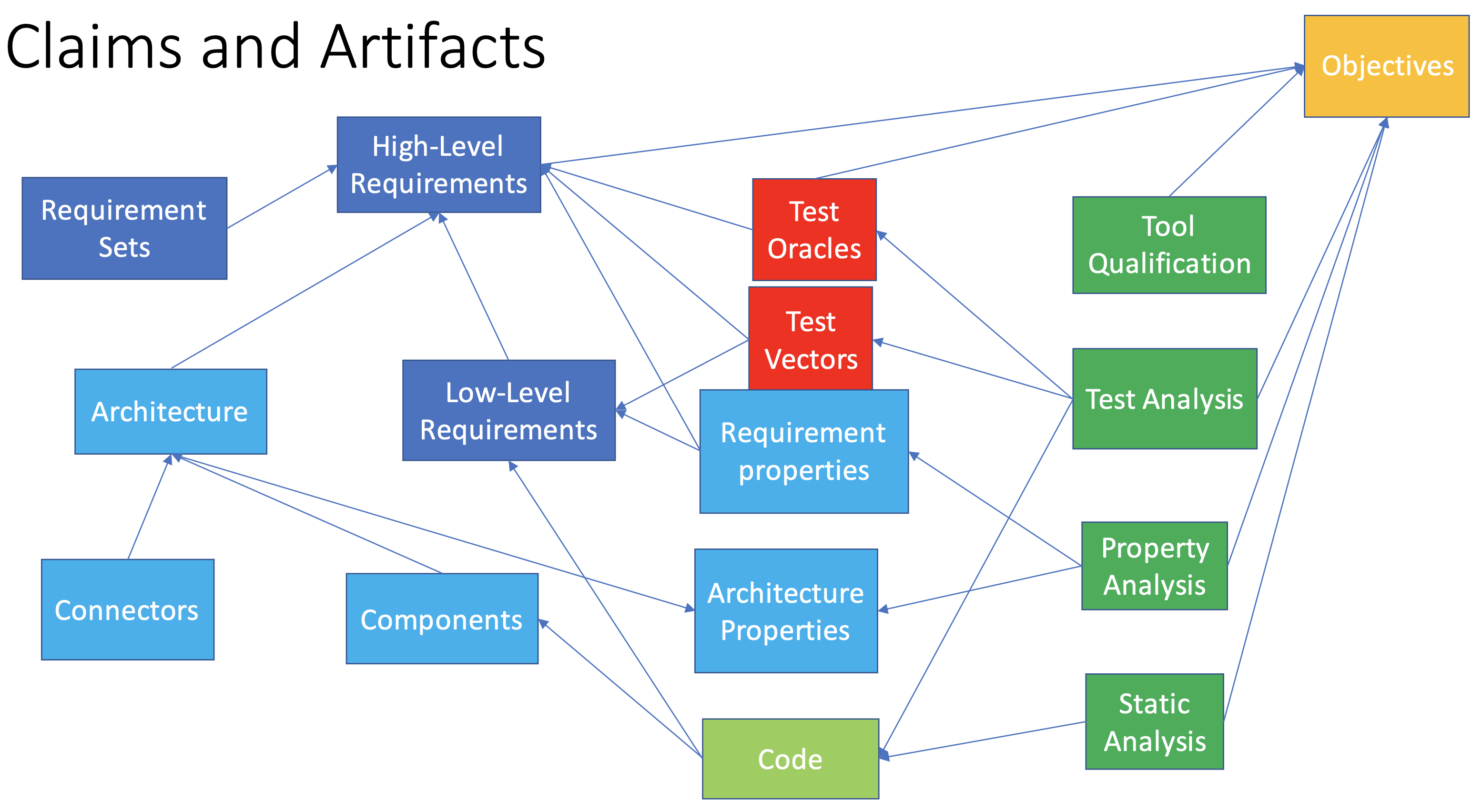}
  \caption{DesCert Assurance Claims and Artifacts}
  \label{fig:claimsArtifacts}
\end{figure}

The claims and artifacts are captured in Figure~\ref{fig:claimsArtifacts} (claims are in green colored boxes on the right, other boxes denote evidence artifacts).
The evidence generated to support the above claims include
\begin{enumerate}
\item Tool Qualification evidence for CLEAR, Text2Test, Clear2Sally, Sally, Radler, Seahorn, Randoop, Daikon, Toradocu, and the Checker Framework.
\item High-level requirements in CLEAR partitioned module-wise into Requirement sets.
\item Test oracles
\item Test suites
\item Radler architecture properties supported by test traces
\item Sally model-checking claims and counterexamples.
\item Code analysis: static and dynamic analysis evidence.
\end{enumerate}

\section{Assurance Ontology}\label{sec:Ontology}

The assurance artifacts created and maintained in the DesCert project are ingested into RACK.
These artifacts are either data (papers, requirements, test cases, analysis results, architecture definitions, proofs,
and code, or metadata (requirement labels, traceability, tool configuration, file handles).
The data is provided in the form of files in a separate directory.  The metadata is ingested into RACK.
The DesCert ontology is defined in the SADL language.
In representing the evidence data in the TA2 Rack-in-a-Box framework, we employ the Provenance
ontology schema (see Figure~\ref{fig:prov_core}) consisting of  entities \textbf{Agents}, \textbf{Activities}, and \textbf{Artifacts},
and relations:
\begin{enumerate}
\item ActedOnBehalfOf(Agent, Agent).
\item WasAssociatedWith(Activity, Agent)
\item WasAttributedTo(Entity, Agent)
\item WasDerivedFrom(Entity, Entity)
\item Used(Activity, Entity)
\item WasGeneratedBy(Entity, Activity)
\item WasInformedBy(Activity, Activity)
\end{enumerate}

In the TA2 ontology, the activities are System development, Requirements development, Hazard Identification,
Code development, Test development, Test execution.  For DesCert, we add the Software Architecture and
Low-Level Requirements activities.   The evidence schemas we employ consist of
\begin{enumerate}
\item High-Level Requirements and Test Development using CLEAR and Text2Test to develop Requirements sets mitigating hazards. 
\item   Property Checks using Sally Tool covering both specific and generic properties associated with requirement sets 
\item   High-Level Requirement Analysis by Text2Test Tool generating Requirements Analysis results
\item   Software Architecture and Code Contract (Low-Level Requirements) Development using the Radler Architecture Definition Language (RADL) and build system as well as specific software libraries
\item   Property Analysis of Source Code by SeaHorn connecting code to Low-Level Requirements on code components
\item   Property Analysis of Source Code by Randoop and Daikon connecting code to Low-Level Requirements on code components
\item   Property (Type) Analysis of Source Code by Checker Framework connecting code to Low-Level Requirements on code components
\end{enumerate}

We have also extended the ontology to connect properties with the corresponding DO-178C objectives.  
The DesCert ontology is described in detail in Chapter~\ref{chap:ontology}.

\section{ArduCopter Challenge Problem}\label{sec:ardu}

\begin{figure}[htb]
  \centering
  \includegraphics[width = .8\linewidth]{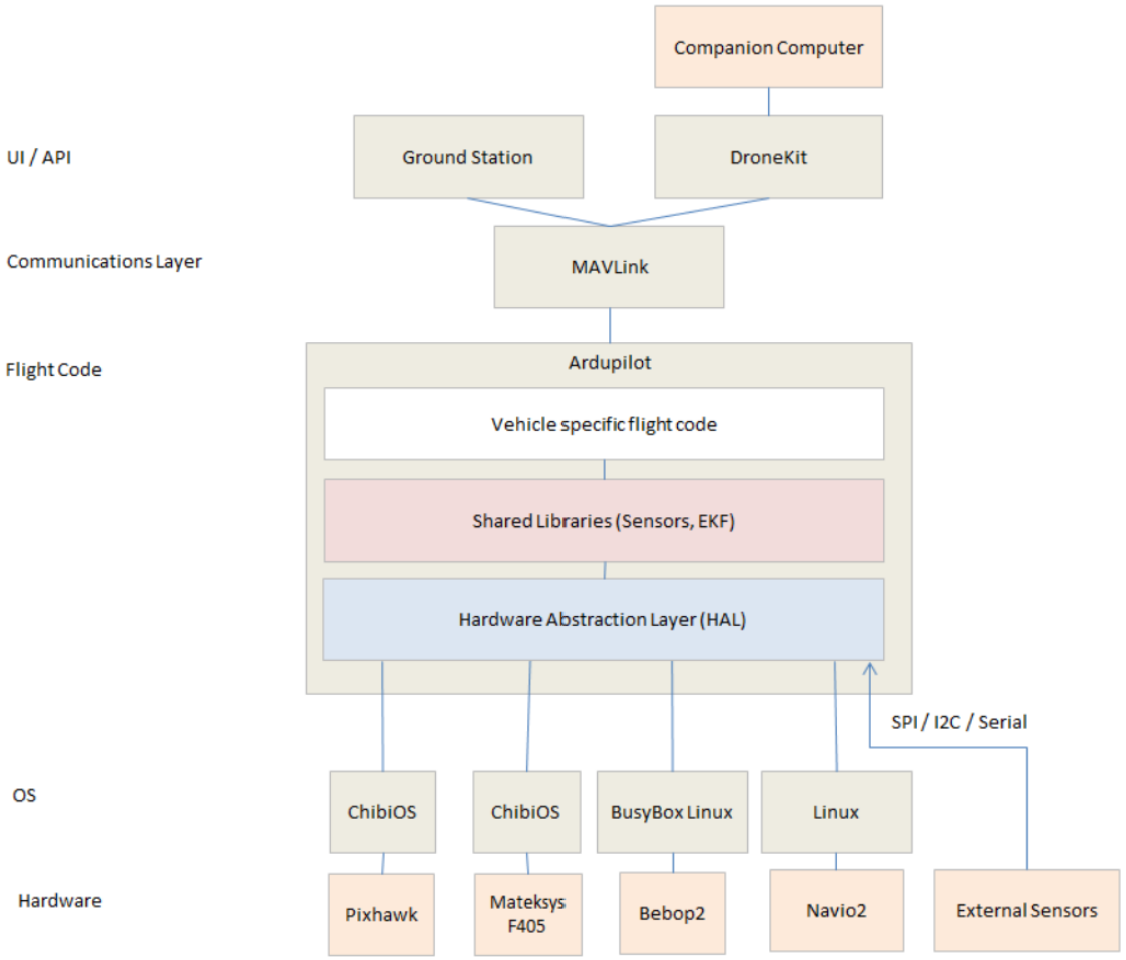}
  \caption{ArduPilot Platform Architecture (From \url{https://ardupilot.org/dev/docs/learning-ardupilot-introduction.html
})}
  \label{fig:ArduPilot}
\end{figure}

The DesCert approach to assurance-driven development of new software is prototyped using the ArduPilot platform.
The ArduPilot is an open source platform for controllers for a range of vehicles including rovers, fixed-wing aircraft,
and rotorcraft.  In DesCert, we employ the ArduCopter instantiation of the ArduPilot.  The platform architecture
for the ArduPilot is shown in Figure~\ref{fig:ArduPilot}\@.  The architecture has a Hardware Abstraction Layer (HAL)
that supports a number of hardware/OS platforms, a shared library for control-related computations, and vehicle-specific
code which in our case is the ArduCopter rotorcraft.  The platform supports a number of simulation engines, and in our
project we use the Software-In-The-Loop (SITL) simulator.  

Though the ArduCopter has a basic Advanced Fail Safe (AFS) functionality for recovering from glitches, it is embedded
into the main control loop.  We decided to define an independent AFS functionality that uses data from the primary
control software.  Our AFS system is located on the companion computer which is connected to the primary computer
through MAVLink.  We deployed our Radler architecture running on the Robot Operating System  (ROS)
by introducing a gateweay Radler node representing the interface with the primary ArduPilot platform. 
The communication between the gateway node and the primary computer employs the MAVROS transport channel.

The assurance case study for  the Advanced FailSafe (AFS) component of the
ArduCopter focused on a concept of operations (ConOps)  where the
ArduCopter autonomously executes  a mission plan by flying through a sequence
of way points at specified altitudes.
The AFS monitor detects events such as range violations, geofence breaches, GPS loss, communication loss, and battery depletion to trigger appropriate recovery actions.   
  When a potential failure event is detected, the AFS monitor executes
a recovery action to keep the vehicle safe either by returning to the launch site, hovering in place, or
landing.  The Radler architecture for the challenge problem also integrated BeepBeep3,
a Java application for safety monitoring.  
The details of the AFS Challenge Problem are spelled out in Chapter~\ref{chap:challenge}.

\section{Assurance Tools}\label{sec:tools}

As already noted, a design workflow supporting efficient arguments requires trusted tools
with semantically coherent interfaces that can be composed for evidence generation.
The DesCert workflow employs CLEAR as a notation for capturing high-level behavioral
software requirements (HLRs).  CLEAR requirements capture temporal properties specifying the
reactive behavior of state machines such as the AFS module.  They also specify certain
safety and timing properties that must be satisfied by the state machines.
From the CLEAR requirements, we use the Text2Test tool to generate test inputs and
test oracles for the state machines in the form of controllable inputs and observable outputs.
Test generation is based on a testing theory for exploring and monitoring
the implementation of each operator used in the requirements definition.
The Text2Test tool also generates transition system models from the CLEAR requirements
in the Sally language.  These transition system models can be individually or jointly
analyzed for temporal properties, specifically invariants, using the Sally model checker.
The software HLRs are refined to a design given by the Radler architecture which
implements each state machine component as a Radler node.  We have used 
SRI's Prototype Verification System (PVS)~\cite{Owre95:prolegomena}, an interactive proof assistant,
to verify certain key architectural propertes of the Radler architecture.
As we showed in Section~\ref{sec:thermostat}, these architecture properties can be used to
refine the HLRs in terms of precondition/post-condition contracts on the step functions
employed by the nodes.  These contracts as well as generic properties of the source code
such as type correctness and the absence of certain classes of runtime errors are established
using dynamic and static analysis.  The static analysis tools include the Checker Framework for
annotated Java code, and SeaHorn for LLVM bit code.  The dynamic analysis tools include the
Randoop unit test generator and the Daikon analyzer for likely program assertions, as well as the
Text2Test tool.   We also integrate BeepBeep3 as an runtime safety monitor.  

While we generated some modest tool
qualification evidence, we did not take a serious stab at tool qualification.  A rigorous tool qualification
following the guidelines in RTCA DO-330 would be an extremely costly exercise that would distract us from
the research goal of developing a proof-of-concept evidence generating design workflow.
The DesCert assurance tools are summarized in the table in Figure~\ref{fig:toolsuite}\@. 

\begin{figure}[htb]
  \centering
{\small\begin{tabular}{{|l|l|l|}}\hline
  Phase & Tools & Artifacts\\\hline
Tool Qualification & Self Analysis   & Test+Analysis \\\hline
  System Requirements & Operational Scenarios &
                                                \begin{tabular}{{l}}
                                                  Consistency\\ Completeness
                                                \end{tabular}
\\\hline
  Hazard Analysis  & Sally & Model Checking \\\hline
  Software High Level Requirements &
                                     \begin{tabular}{{l}}
                                       CLEAR\\ Text2Test\\ PVS\\ Sally
                                     \end{tabular}
 & \begin{tabular}{{l}}
                                                  Consistency\\ Completeness\\Validation
   \end{tabular}
         \\\hline
  Software Low Level Requirements & \begin{tabular}{{l}}
                                       SeaHorn\\ Randoop\\ Daikon\\ Checker Framework
                                     \end{tabular}  &  Static/Dynamic Analysis\\\hline
  Executable Object Code & \begin{tabular}{{l}}
                                                  BeepBeep3\\Text2Test
                                                \end{tabular} & Safety Monitoring  \\\hline
\end{tabular}}
\caption{DesCert Tool Suite}\label{fig:toolsuite}
\end{figure}

\section{Current Limitations of DesCert Assurance Methodology}\label{sec:limitations}

The DesCert assurance-driven development workflow is aimed at creating a paradigm for the automated certification
of safety-critical systems.  The Phase 1 effort was largely exploratory.  We centered our workflow on the
creation of designs that supported efficient arguments.  The Radler model of computation plays a key role
in facilitating an efficient argument structure.  Our assurance-driven development follows
the structure of an assurance case complying with the guidance in DO-178C.  We track several of the
Level C and D objectives and traceability relations suggested by the DO-178C standard.  However, our
approach generates evidence during the design lifecycle where the goal of the design is the
creation of a software system supported with the design and assurance artifacts.  This is in contrast
to a \emph{post facto} approach to certification where the evidence chain is constructed to comply with the
DO-178C objectives as a postscript to the design lifecycle.   We also focus on constructing evidence that
targets the behavior of the software and not the process by which it is constructed and analyzed.
Since we generate evidence from different phases of the design, we target evidence that is semantically
coherent so that the behavioral models and claims at the different levels are consistent with each other.
In particular, any software failure can be connected to a flaw in the assurance argument constructed
from the evidence.

Broadly, the DesCert approach to assurance-driven development starts
with the formalization of the intent of the system in the form of
precise requirements defined in the CLEAR language.  The analysis of
the requirements captured in CLEAR cover both generic properties that
must hold of any requirements as well as specific properties that
constrain the software application under certification.  We use ontic
type annotations to capture the representational intent of the data
objects consistently throughout the design.  The software design is
centered around a choice of an architectural model of computation,
which in our case is the Radler framework.  The architecture, defined
in the Radler Architecture Definition Language (RADL), captures the
logical architecture in the form of nodes operating
quasi-periodically, and communicating over publish-subscribe channels
with specified latency bounds.  The RADL physical architecture maps
the nodes to processes within a virtual machine and the topic channels
are implemented through mailboxes connected to their publishers
through a transport protocol.  Radler architectures are flexible about
how the physical architecture is actually realized as long as the
period, communication, and communication latency assumptions are
satisfied.  The software services provided by each node are
implemented as step functions with their own
precondition/post-condition contracts.

While the above outline of a high-assurance design process can be made fully rigorous,
there are some limitations with the Phase 1 work that need to be addressed in future work.
\begin{enumerate}
\item CLEAR has only been applied to a limited set of case studies.  Both the behavioral
  language and the background libraries of useful operations need to be expanded.
\item We have not yet defined an Ontic type framework that spans the design stages from requirements to source code.
\item The experience with translating CLEAR state machine requirements to Sally is limited to a few examples.  This translation will need to be expanded to handle complex requirements.
\item Sally itself only implements model checking for invariants and cannot handle more complex temporal properties.
\item The soundness of the translation from CLEAR to Sally needs to be certified.
\item Though we have a broad and mature suite of tools, only the HiLite tool has gone through a tool qualification process.
\item While we have the source code analysis tools and did apply them to isolated examples, we did not make a systematic
  effort into generate and integrate evidence from the analysis of source code components since the High-Level
  Requirements and Design levels took up a fair amount of effort.  
\item The Baseline DesCert continuous integration workflow only has a small number of plug-ins, mainly Randoop and Daikon,
  and we will be working to expand the number of plug-ins in future work. 
\end{enumerate}

% We are creating a evidence-generating software design pipeline for new software spanning
% system requirements, high-level requirements, architecture design, low-level requirements,
% source code development, and object code creation.  The overall structure of the workflow, which is loosely
% modeled on the DO-178C standard, captured in the table below.

% {\tiny\begin{tabular}{{|l|l|l|}}\hline
%   Phase & Tools & Artifacts\\\hline
% Tool Qualification & Self Analysis   & Test+Analysis \\\hline
%   System Requirements & CLEAR &  Consistency + Completeness\\\hline
%   Software High Level Requirements & Text2Test, PVS, Sally & Consistency+Completeness+Validation\\\hline
%   Software Low Level Requirements & Sally & Model Checking Analysis\\\hline
%   Hazards  & Sally & Model Checking Analysis \\\hline
%   Software Source Code Components & Static Analysis + Testing & Test + Analysis\\\hline
%   Executable Object Code & Dynamic Analysis/Monitoring & Test \\\hline
%       \end{tabular}}

\chapter{Phase 1 Challenge Problem: Advanced Fail Safe (AFS) Case Study}\label{chap:challenge}%Vatsan (AFS, Architecture, ConOps, High Level Description of Requirements)
We are using the open source ArduPilot (\cite{ardupilot})
platform for the the Phase 1 case study/challenge problem. The ArduPilot Project provides an advanced, full-featured and reliable open source autopilot software system. The first ArduPilot open code repository was created in 2009 - since then it has been developed by a team of diverse professional engineers, academics, computer scientists, and other members of our global community. It is capable of controlling almost any vehicle system imaginable: conventional and VTOL airplanes, gliders, multirotors, helicopters, sailboats, powered boats, submarines, ground vehicles and even Balance-Bots.

ArduPilot is a deeply tested and trusted autopilot system and the open-source code base means that it is rapidly evolving, always at the cutting edge of technology development, whilst sound release processes provide confidence to the end user. With many peripheral suppliers creating interfaces, users benefit from a broad ecosystem of sensors, companion computers and communication systems. Since the source code is open, it can be audited to ensure compliance with security and secrecy requirements. The software suite is installed in vehicles from many manufacturers, such as many from our Partners, and more broadly throughout the global autonomous systems industry. It is also used for testing and development by large institutions and corporations such as NASA, Intel and Insitu/Boeing, as well as countless colleges and universities around the world.

ArduPilot works with a wide variety of hardware platforms\footnote[1]{\url{https://ardupilot.org/copter/docs/common-autopilots.html}},  as well as a range of simulators\footnote[2]{\url{https://ardupilot.org/dev/docs/simulation-2.html}}. We selected the rotorcraft \emph{ArduCopter} platform in the context of the Software-In-The-Loop (SITL) simulator since it has navigation software functionality that is similar (in principle) to the NAV system in legacy Boeing AH-64 Apache platform provided by TA4 in ARCOS, and our results in ArduPilot can then be easily reproduced/repeated. Further we can faithfully simulate the actual software execution on an actual platform in a SITL simulation thereby alleviating the need to deal with time-consuming hardware/platform integration issues. We focus on the \emph{Advance Fail Safe}  (AFS) runtime monitor software component that checks where the ArduCopter is flying within the established limits while executing the mission plan and initiates planned contingency recovery actions when any violations are detected.

To summarize, we chose this ArduPilot platform for the following reasons: (i) allows for easy access, modification and distribution of source code associated with different software pieces (ii) has representative complexity of the typical functionality on the aircraft (iii) easy to to demonstrate enabling certification technologies developed within ARCOS for the evaluation teams using releases of pre-installed necessary software in Virtual Machines (iv) easy to integrate and packages solution of certification tool  technologies along with the systems and software that needs to certified.

\section{Concept of Operations}
\label{sec:conops}

\begin{figure}[!hbtp]
\centering
\includegraphics[scale=.40]{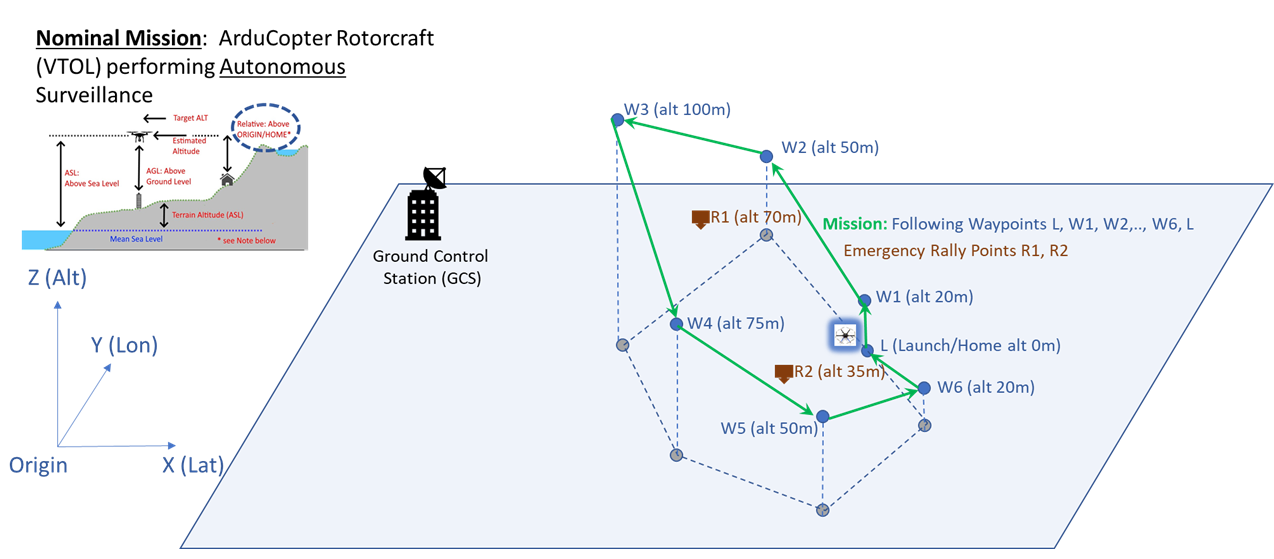}
\caption{Phase 1 Mission Operation}
\label{fig:conops_mission}
\end{figure}

As shown in Figure \ref{fig:conops_mission}, we envision an ArduCopter Rotorcraft (VTOL) performing Autonomous Surveillance mission during a \emph{nominal operation} where \emph{no contingency situations} are encountered during operation and the operation is to be performed in a location with communication proximity to a Ground Control Station (GCS) that expected to oversee the ArduCopter operation. All coordinates shown are in triplet \emph{(latitude, longitude, altitude)} and in \emph{meters} as units. Also altitude is specified relative height above origin, which is also assumed to be the \emph{home} position or site of \emph{launch} $L$ i.e. where mission begins as well as site of return once the mission is completed.  The mission involved following a sequence of waypoints $L, W1, W2, W3, ..., W6, L$.  After reaching the launch position, the craft will land. 

As part of the mission configuration, \emph{Rally Points} are also specified e.g. $R1, R2$. These are pre-specified points where the ArduCopter can proceed to, as alternative to Home point  or waypoint, during emergencies or contingencies. For example,  ArduCopter can proceed to the closest Rally Point, rather than proceeding all the way back to the Home position, and can loiter at that location, and perform an automated landing there.

\begin{figure}[!hbtp]
\centering
\includegraphics[scale=.40]{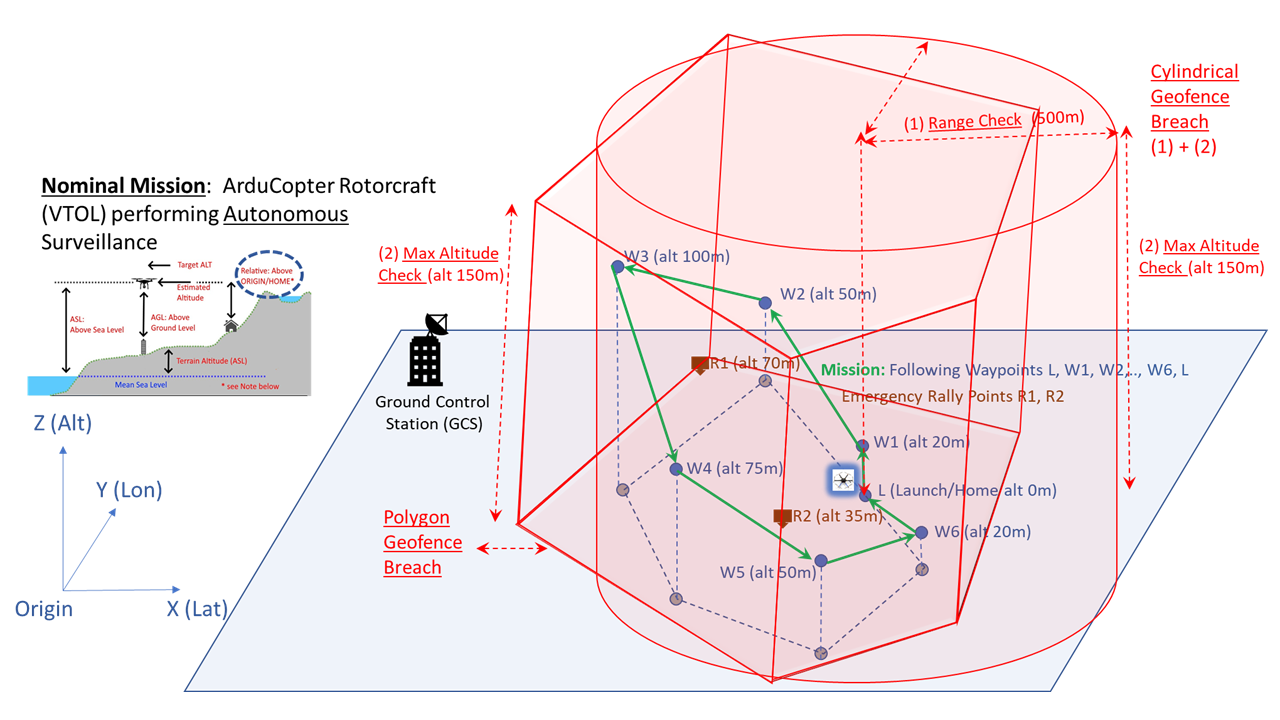}
\caption{Geofences during Mission Operation}
\label{fig:conops_mission_with_fences}
\end{figure}

ArduPilot supports alarms generated by several types of Fences (boundaries described by latitude/longitude and/or altitude) to prevent the vehicle from traveling higher or further than desired, or into unwanted areas. Types of Fences supported varies by vehicle. Upon Fence breach, selectable actions are taken. As part of mission configuration and illustrated in Figure \ref{fig:conops_mission_with_fences}, the ArduCopter has Geo-Fences pre-specified. The Cylindrical Geo-Fence is a simple “tin-can” centered around home. Cylindrical Geo-Fence restriction has two checks associated with it (i) max altitude check (height of cylinder) and (ii) a range check  (radius of cylinder). Additionally, there is an arbitrary shaped Polygonal Geo-Fence restriction to ensure ArduCopter flies  specific locations within the polygonal boundary. Cylindrical and  Polygon Geo Fence Breach Checks are both \emph{Inclusion} fences to keep vehicle from flying “out-of the fence”.  Note, \emph{Exclusion} fence available but not utilized in the current mission configuration setup to keep vehicle from flying “into the fence”. 

\section{Mission Scenario and Assumptions}
\label{sec:mission_scenario}

The primary objectives of the mission are for the ArduCopter to :

\begin{itemize}

\item Successfully complete a surveillance mission that is pre-configured before start and includes takeoff from home/launch, visit all waypoints in sequence and finally return to launch and land \textit{(Objective 1)}

\item Complete the mission with full autonomous control of the ArduCopter :
\begin{itemize}
\item Without any remote pilot assistance from Ground Control System (GCS) \textit{(Objective 2)}
\item Without any loss of control of the ArduCopter in both nominal and off-nominal/contingency/emergency situations enumerated apriori \textit{(Objective 3)}
\item Without physically losing track of the ArduCopter whereby GCS is notified of the location of the ArduCopter during the whole mission duration \textit{(Objective 4)}
\end{itemize}

\item Complete the mission operation safely:
\begin{itemize}
\item By limiting potential hazards (e.g. collision) risks exposed to humans, properties and other airborne assets during the operation of the mission by restricting the Arducopter to flying within a pre-configured Geofence i.e. Operational Safety Zone \textit{(Objective 5)}
\item Without crashing (destroying) the ArduCopter during takeoff, cruising through the waypoints or during landing \textit{(Objective 6)}
\end{itemize}

\end{itemize}

\textit{We would ideally like to satisfy all missions objectives, if possible, and if there are conflicts/trade-offs then we require mission objectives prioritization (from high to low) be in the order: (5), (6), (3), (4), (1), (2)}. As an example of prioritization, landing immediately at some point when insufficient battery emergencies occurs rather than going towards at Home/Launch shows prioritization of objectives (5) \& (6) over (1).

There are a variety of \emph{assumptions} related to the validity of the mission related configuration:
\begin{itemize}
\item Mission Waypoints trajectory are correctly specified in a loop: $L \rightarrow W1 \rightarrow W2 \rightarrow \cdots \rightarrow Wn \rightarrow L$
\item Geofence configuration  $(\mathit{max\_altitude, range, polygon})$ is correctly specified:
\begin{itemize}
	\item 	Cylinder $(2.\pi.\mathit{range} \times \mathit{max\_altitude})$, Polytope $(2D\ \mathit{Polygon \times max\_altitude})$ defined
	\item $\mathit{Operational\ Safety\ Zone} = \mathit{Cylinder} \cap \mathit{Polytope}$ i.e. Common 3D space intersecting Cylinder and Polytope must be non-zero 
	\item \textit{Minimizes potential hazards (e.g. collision) risks} exposed to humans, properties and other airborne assets as long as the operation of the mission is restricted within this operational safety zone
\end{itemize}
\item Mission Waypoints trajectory is feasible if it completely within the Geofence:
\begin{itemize}
	\item All waypoints and points en-route between waypoints are completely within $\mathit{Operational\ Safety\ Zone} = \mathit{Cylinder} \cap \mathit{Polytope}$
	\item All waypoints and points between waypoints along $L \rightarrow W1 \rightarrow W2 \rightarrow \cdots \rightarrow Wn \rightarrow L$ satisfy (1) range check , (2) max\_altitude check and (3) polygon check
\end{itemize}
\item Emergency Rallypoints $R1, R2, \cdots Rn$ are suitably chosen:
\begin{itemize}
	\item All rallypoints are within Geofence and satisfy (1) range Check , (2) max\_altitude check and (3) polygon check
	\item In case of reacting to a specific emergency/off-nominal/contingency situation, rather than having to only proceed to the Launch/Home position always (which might be far off depending on where in the trajectory ArduCopter is currently flying), Rallypoints are suitable “alternate” location choices for the ArduCopter to proceed to e.g. within Line of Sight (LOS) of Ground Control Station (GCS)
	\item Apriori configure within the mission “closest” Rallypoints associated different waypoint and paths between waypoints. E.g. $R1$ Rallypoint for $L \rightarrow W1 \rightarrow W2 \rightarrow W3 \rightarrow W4$  and $R2$ for $W4 \rightarrow W5 \rightarrow W6 \rightarrow W7$
\end{itemize}

\item Mission Waypoints and trajectory en-route between waypoints as well as Rallypoints are within communication range of the Ground Control Station (GCS) in normal/nominal situations. Note: Off-nominal complete loss of communication as well as discontinuity in service (intermittent service) is expected

\item Mission Waypoints and trajectory en-route between waypoints as well as Rallypoints typically have access to GPS satellites and GPS Fix signals for navigation in normal/nominal situations. Note: Off-nominal complete loss of GPS signal as well as discontinuity in service (intermittent service) is expected. In these situation the on-board navigation (e.g. EKF filter) is expected to supply location information continuously to the ArduCopter to coast for a while using only primary inertial sensor IMU (accelerometer, gyros) measurements until aiding sensor GPS fix signals may be obtained later (no guarantee for that GPS Fix signal to happen).

\item Remaining Battery Energy Level is sufficiently provisioned/budgeted and configurations appropriately setup for $[100\% > T_{NOM}  > T_{RTL} > T_{LAND} > 0\%]$ as shown in Figure \ref{fig:battery_levels}
\begin{figure}[!hbtp]
\centering
\includegraphics[scale=.90]{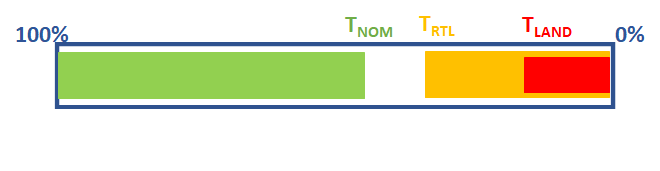}
\caption{Battery Energy Levels}
\label{fig:battery_levels}
\end{figure}
\begin{itemize}
	\item Nominally the full mission can be completed with $(100\%- T_{NOM})$ battery energy and some spare battery level remaining at $T_{NOM}$.  No guarantees in off-nominal situations requiring tough maneuvers, battery drainage etc.
	\item $T_{RTL}$ remaining battery threshold level is so chosen such $T_{RTL}$ battery energy is good enough with spare to return to launch/home and do the necessary vehicle maneuvers to go from anywhere within mission trajectory in a normal/nominal condition. No guarantee in off-nominal situations
	\item $T_{LAND}$ battery threshold level is so chosen such $T_{LAND}$ battery energy is good enough with spare to vertically land on the ground immediately and safely from any altitude within max\_altitude in normal/nominal conditions
\end{itemize}
\end{itemize}

\section{System and Software Architectures}
\label{sec:SySoArch}

\begin{sidewaysfigure}[!hbtp]
\centering
\includegraphics[scale=.55]{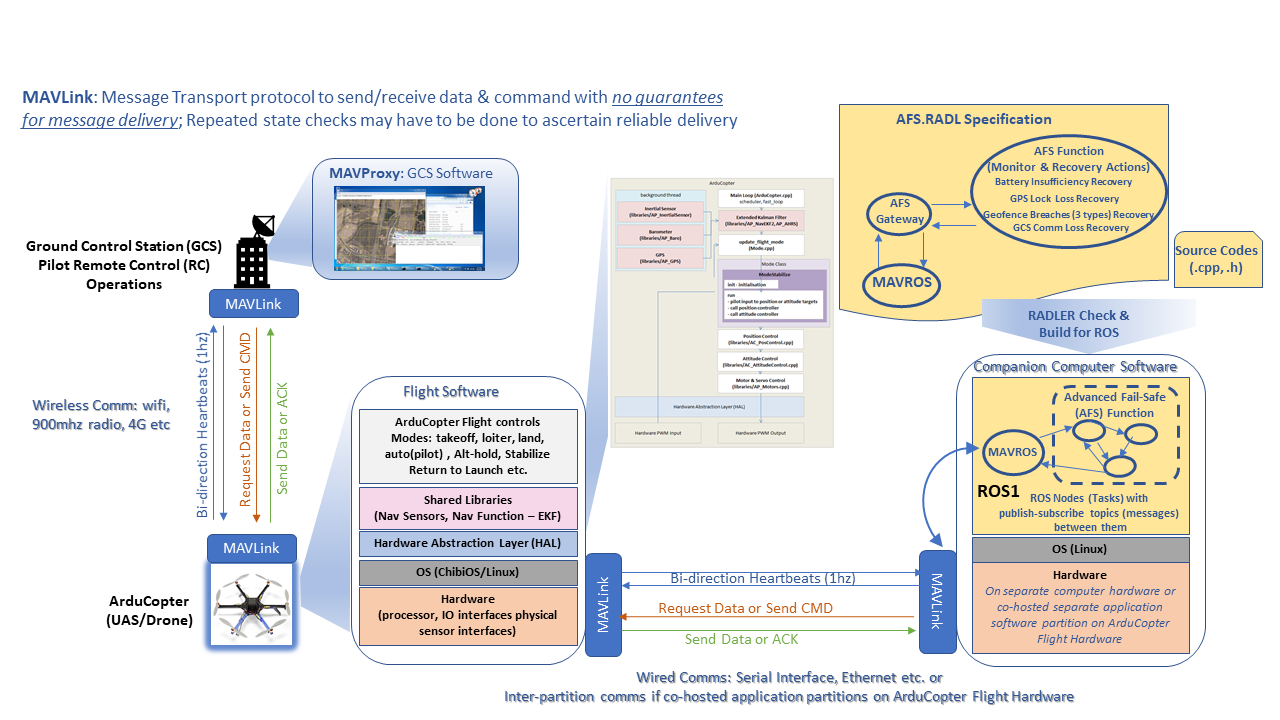}
\caption{System Architecture with Advanced Fail-Safe (AFS)  Software Component}
\label{fig:sys_sw_arch}
\end{sidewaysfigure}

For the phase 1 challenge problem, we envision a system architecture built with associated software components depicted in Figure \ref{fig:sys_sw_arch}.  In particular, phase 1 focus will be limited to generation of evidences for compliance and certification a single software component associated with \emph{Advanced Fail-Safe (AFS)} function, shown to the top right of Figure \ref{fig:sys_sw_arch}.  The vision is that the Ground Control Station (GCS) performs remote pilot operations (monitors and potentially does control interventions) of the ArduCopter using some wireless channel of communication (5G/4G, Radio on WiFi, SATCOM etc). We use MAVLink as the message transport protocol for the communication between GCS and ArduCopter including a bi-directional periodic heartbeats, status updates, sending/receiving command, data and acknowledgments (ACKs). Note that MAVLink offers no guarantees for message delivery and repeated state checks may have to be done by underlying software to ascertain reliable delivery. 

The GCS Software used MAVProxy and Flight Software (on-board software on ArduCopter) has the vehicle specific flight code and in phase 1, we \emph{do not} make any changes to these code and use the code as is. As shown in the bottom left of Figure \ref{fig:sys_sw_arch}, the flight software has multiple layers/parts: (1) Implements the ArduCopter flight control modes including: Takeoff, Loiter, Land, Auto (autopilot), Altitude-Hold, Stabilize, Return-to-Launch (RTL) etc. (ii) Shared Libraries to support multiple drivers for various navigation sensors e.g. GPS, IMU and other inertial sensors, barometer, camera etc (iii) Navigation function implemented using Extended Kalman Filter (EKF), Position/Attitude and Motor/Servo control etc for teh different mode specifications (iv)  hardware abstraction layer (HAL) to support portability  to lots of different platforms. and (v) OS support for Linux etc and various hardware support code (processor, IO, sensor interfaces).

Our primary focus in phase 1 is to certify a single software component developed as new software within ArduPilot and which architecturally integrates with the rest of ArduCopter system and software in a seamless manner. We design and develop the Advanced Fail Safe (AFS) function in a principled manner i.e. with a view to ease certification by minimizing defect escapes and lowering verification costs through automation. The objective of AFS function is to take safe corrective action in abnormal situations – identifying the following \emph{six contingency situations} triggering  the appropriate recovery response actions:
\begin{enumerate}
\item Cylindrical Geofence breach: Max Altitude check violation
\item	Cylindrical Geofence breach: Range check violation
\item	Polygon Geofence breach: polygon boundary check violation
\item	GPS Lock Loss
\item	Ground Station Communication Loss
\item	Insufficient Battery
\end{enumerate}

A shown in the bottom right of  Figure \ref{fig:sys_sw_arch}, the AFS function is built as a separate software component in a companion computer i.e. separate computer hardware that also communicates with the ArduCopter flight software using MAVLink over another independent channel.  The channel itself can be a wired communication medium like a serial interface or ethernet interface or another potentially another wireless interface. Alternatively the AFS function can also be loaded on the ArduCopter flight hardware and co-hosted as separate software software partition along with flight software partition and the inter-partition communication still leveraging MAVLink transport interface. This flexibility in architectural separation of AFS functionality allows us to develop the AFS software component independently and leveraging the time-space partitioning strategy (e.g. ARINC 653 partition \cite{Krodel2007RealTimeOS}) for demonstrating verification of the AFS software component in isolation. Subsequently, we would then focus on the dependencies of the AFS function with the rest of the system i.e. verification of the integrated system and it's multiple components in a systematic manner.  To reiterate, in phase 1 and in this report, we limit our focus still to generating evidence for a single AFS software component. 

AFS software component is built on top of Robot Operating System, version 1 (ROS1~\cite{book:ROS}) for a Linux operating system on top of which all ArduPilot software code is built. We leverage MAVROS, ROS-based extendable communication node, which enables MAVLink extendable communication between computers running ROS (1) for any MAVLink enabled autopilot, ground station, or peripheral. MAVROS is the "official" supported bridge between ROS (1) and the MAVLink protocol. AFS Function is built using  Robot Architecture Definition Language (RADL) specification and associated Radler~\footnote[4]{Radler 1.0 documentation: \url{https://sri-csl.github.io/radler/}}~\footnote[5]{Radler examples: \url{https://github.com/SRI-CSL/radler}} code generation tool and build environment to generate executables that ensures the software code executing in a ROS environment adheres to the specification. We discuss in Section~\ref{sec:RadlerMain} the details of the RADL architectural specifications, the nice properties inherited due to the architectural paradigm when the system and associated software components strictly adhere to the specification and the Radler build tool that assembles the requisite software for execution on the ROS platform. The AFS specific RADL specification i.e \emph{afs.radl} is discussed in Section~\ref{sec:radlEvidence}. We design the AFS system architecture using Radler as a collection of nodes (with periods, publish/subscribe topics) and topics.  The system can be tested using the SITL simulator of ArduPilot.   

Radler architecture specification consists of the logical and physical parts. The logical part is specified in terms of node and topic similar to ROS. The nodes execute independently and periodically, and publish and/or subscribe topics. AFS Function node (afs\_function) executes its step function with period of 100 milliseconds and publishes 4 recovery actions subscriber ROS nodes: (i) AFS Geofence Breach Recovery for 3 different kinds (altitude, range, polygon) (ii) AFS GCS Comm Loss Recovery (iii) AFS GPS Lock Loss Recovery (iv) AFS Battery Insufficiency Recovery. The AFS Gateway node (afs\_gateway) acts as bridge between AFS Function and through the MAVROS/MAVLink connects to the on-board flight controls on the ArduCopter. 

More specifically, AFS Gateway node forwards back-and-forth messages between AFS Function and ROS/MAVROS/MAVLink interface on the companion computer. It remotely collects events from flight controls e.g. GPS Lock Loss, Remaining battery energy, Comm Loss, Geofence Breach events and forwards that to the AFS Function. The AFS Function implements the Recovery action logic and sends control commands back (e.g. change of flight mode ) to AFS Gateway which then send the command to the flight controls on the ArduCOpter via the MAVLink. 

Physical part of the AFS RADL specification map nodes to the process on specific machines, in this case companion computer. Radler build process, using source codes (afs\_gateway.h, afs\_gateway.cpp, afs\_function.h, afs\_function.cpp), does some explicit checks and also generates the glue code for scheduling, communication, and failure detection such as timeout or staleness of data. Note that in Phase 1, we exclusively focus on the AFS Function and not on the AFS Gateway.

\section{High Level Behavioral Description of AFS Function}
\label{sec:HLDR}

The primary objective of intent specification and formally capturing requirements is to be able to demonstrate, with confidence, evidence of implementation deployed on actual system meets the intent. We will be specifying the intent of the AFS Function and the associated requirements for triggering recovery actions during contingency events in the Constrained Language Enhanced Approach to Requirements (CLEAR) notation.  The formalized requirements are analyzed for consistency, completeness and other generic properties to ensure latent defects do not enter during requirements writing phase that can subsequently manifest in implementation where it can be trickier to diagnose. From requirements, we use Text2Test to automate generation of test cases and test oracles that are tightly traceable to these requirements.  The high-level requirements are also  model-checked using the Sally translation of the CLEAR requirements together with the node-level code contracts (specific properties) and architecture properties are also satisfied due to Radler specification and associated build.The node-level contracts are analyzed using the dynamic and static code-analysis tools.  The tools associated with such evidence evidence generation is the subject matter for Section~\ref{chap:tools} and the AFS specific evidence that is generated is also discussed in Section~\ref{chap:evidence}. In this section below we discuss the high level behavioral description of AFS Function that informs all the requirements discussed in subsequent sections.

Broadly there are  \emph{six AFS Monitor Events} due Off-nominal situations triggering the appropriate recovery/response actions below. Note that AFS Prioritization when multiple AFS events occur from high to low is: $(6), (4), [ (1), (2), (3) ], (5)$

\begin{enumerate}
\item  Cylindrical Geofence breach due to Max Altitude check violation
\begin{itemize}
\item \emph{Potential contingency scenario}: Weather/Wind causing trajectory to either overshoot the fence or too close to it with respect to tolerance margins to handle  navigation error.
\item \emph{AFS recovery/response action}: Try to drop to “Target Altitude” i.e. (Max\_Altitude - margin) within 5 sec. If Target Altitude is achieved, continue with mission. If not achieved, then land and communicate landing location to GCS and terminate mission.
\end{itemize}

\item	Cylindrical Geofence breach due to Range check violation
\begin{itemize}
\item \emph{Potential contingency scenario}: Weather/Wind causing trajectory to either overshoot the fence or too close to it with respect to tolerance margins to handle  navigation error.
\item \emph{AFS recovery/response action}: Try to drop to “Target Position” i.e. (Max\_Range - margin) within 5 sec. If Target Position is achieved, continue with mission. If not achieved, then land and communicate landing location to GCS and terminate mission.
\end{itemize}

\item	Polygon Geofence breach due to Polygon Boundary check violation
\begin{itemize}
\item \emph{Potential contingency scenario}: Weather/Wind causing trajectory to either overshoot the fence or too close to it with respect to tolerance margins to handle  navigation error.
\item \emph{AFS recovery/response action}: Try to drop to “Target Position” i.e. (Polygon\_Boundary - margin) within 5 sec. If Target Position is achieved, continue with mission. If not achieved, then land and communicate landing location to GCS and terminate mission.
\end{itemize}

\item	GPS Lock Loss: Loss of GPS signal for 3 seconds
\begin{itemize}
\item \emph{Potential contingency scenario}: GPS Denied/Degraded Environment, navigation drifts when coasting on pure IMU/Inertial without GPS fixes.
\item \emph{AFS recovery/response action}: First loiter or hover at the current location i..e waypoint or in transit between way points for 5 seconds. If GPS recovered within 5 seconds then resume mission and proceed to next waypoint. If GPS not recovered within 5 seconds, then go to last completed “Waypoint” with last known GPS Fix and  using ONLY IMUs without GPS fixes (coasting drifts) in transition and then at that position land and communicate landing location to GCS and terminate mission. Note that last waypoint has better potential than rallypoint in this scenario for GPS fix possibility while GPS signal availability at rallypoint is unknown. 
\end{itemize}

\item	Ground Station Communication Loss: does not receive a heartbeat message for a period of 3 seconds
\begin{itemize}
\item \emph{Potential contingency scenario}: MAVLink loss of Heartbeat messages from/to GCS, outside radio communication range, no cell towers etc.
\item \emph{AFS recovery/response action}: Go to a pre-configured Rally Point (Emergency Hovering point) and “loiter” and Try to re-establish communication connection for 5 seconds duration without loss. If still no communication reestablishment, then return to launch and terminate mission. If communication established, then complete the remaining mission; Increment strike counter to track communication disruptions history. If strike counter $\geq 3$ then return to launch as comms deemed unreliable and terminate mission.
\end{itemize}

\item	Insufficient Battery
\begin{itemize}
\item \emph{Potential contingency scenario}: Over consumption of energy due to difficult vehicle maneuvers during the mission, battery drainage/leaks etc.
\item \emph{AFS recovery/response action}: If battery level $\leq T_{RTL}$ and $\geq T_{LAND}$ then return to launch and terminate mission.
If battery level $<  T_{LAND}$ then land immediately to the ground and communicate landing location to GCS after you have landed and terminate mission.

\end{itemize}
\end{enumerate}

\chapter{DesCert Evidence Ontology}\label{chap:ontology}  %Devesh
%Description of Evidence Ontology

% chapter-level into paragraph
In this chapter, we describe the DesCert ontology for the evidence that
can represent all aspects of assurance as illustrated in Figure~\ref{fig:assurance-driven-development}.
The ontology allows the specification of the following types of evidence:

\begin{itemize}
  \item Evidence of \textit{development} activities and artifacts such as requirements, architecture, design, code.
  \item Evidence of \textit{verification} activities and artifacts such as requirements analysis, test generations, architecture analysis, code analysis, etc.
  \item Semantic relationships of these evidence items to each other and to assurance claims and arguments.
  \item Use of a tool to perform (automate) an activity, using input evidence artifacts and producing output artifacts.
\end{itemize}

Our approach enables property-based assurance by providing classes and rich semantics to specify \emph{properties} supporting an assurance claim. Properties typically hold over a development artifact such as requirement set, architecture, design, code --- i.e., the artifact satisfies the property.

Many additional concepts are needed (e.g., use of tools) for various aspects of assurance that are not supported by specific classes in the base RACK ontology. This didn't allow us to capture the proper semantics needed to connect different parts of the evidence. Thus, in the SRI overlay, we added several classes (and relationships with specific semantics) to represent the evidence. The following sections describe each type of evidence and its relationships.

\section{DesCert Enables Property-Based Assurance}
\label{sec:property-desc}

Figure~\ref{fig:assurance-driven-development} provides our approach to assurance where tools are used to do verification on artifacts produced by the development process. Both testing based methods (shown on left) and formal methods (shown on right) are used. To enable property-based assurance, the formal methods are based on establishing properties of a particular development artifact (or a set of artifacts) and the proving them using a tool.

\begin{figure}[!htb]
\centering
\includegraphics[scale=.7]{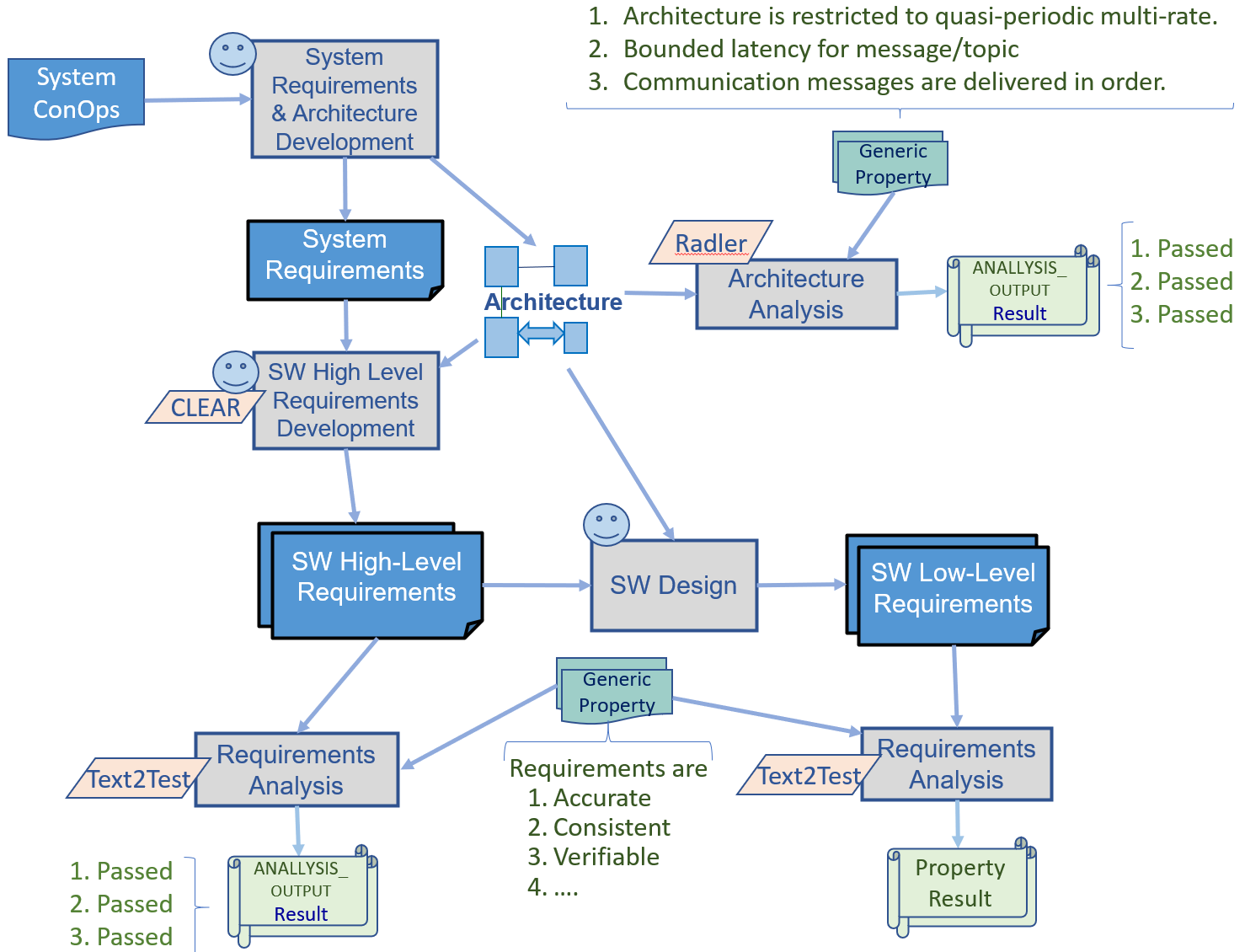}
\caption{Property Specification and Analysis for Architecture and Requirements}
\label{fig:arch_req_property_flow}
\end{figure}

Figure~\ref{fig:arch_req_property_flow} shows the flow of evidence to capture properties of architecture and requirements and using tools to perform analysis, resulting in proofs that the property holds. In this figure, development or verification \emph{activities} are indicated by grey boxes, \emph{development} artifacts are denoted by blue boxes whereas \emph{verification} artifacts are denoted by green shapes. A beige rhombus denotes the \emph{tool} used to perform an activity. (note: other aspects of system development and verification, e.g. code development and testing are not shown in this figure)

The system development starts with the system ConOps which are successively developed into a sequence of artifacts including system requirements, system architecture, software high-level requirements, software low-level requirements, and source code. At each stage of this successive development, it is essential to assure that a development artifact satisfies certain \emph{properties} to meet the following types of assurance objectives:

\begin{itemize}
  \item The artifact complies with a parent artifact from which it is derived/refined. E.g.: software high-level requirements comply with system requirements and system architecture.
  \item The artifact mitigates certain specific hazards. E.g., speed never exceeds more than 120 miles/hr, copter doesn't run out of battery power in air (it would fall to ground and cause hazard otherwise).
  \item The artifact's specification doesn't exhibit any abnormal behavior that can cause a hazard. E.g.: source code doesn't contain any null-reference exception.
  \item The artifact's specification satisfies certain properties that support generic assurance claims/arguments. E.g.: communication messages are delivered in order, requirements are consistent with each other.
\end{itemize}

To this end, we have defined the ontology for specification and analysis of properties, giving a first-class status to properties and their relationships, as described below.

\paragraph{Generic and Specific Properties.}

\begin{sidewaysfigure}[!htb]
\centering
\includegraphics[scale=.65]{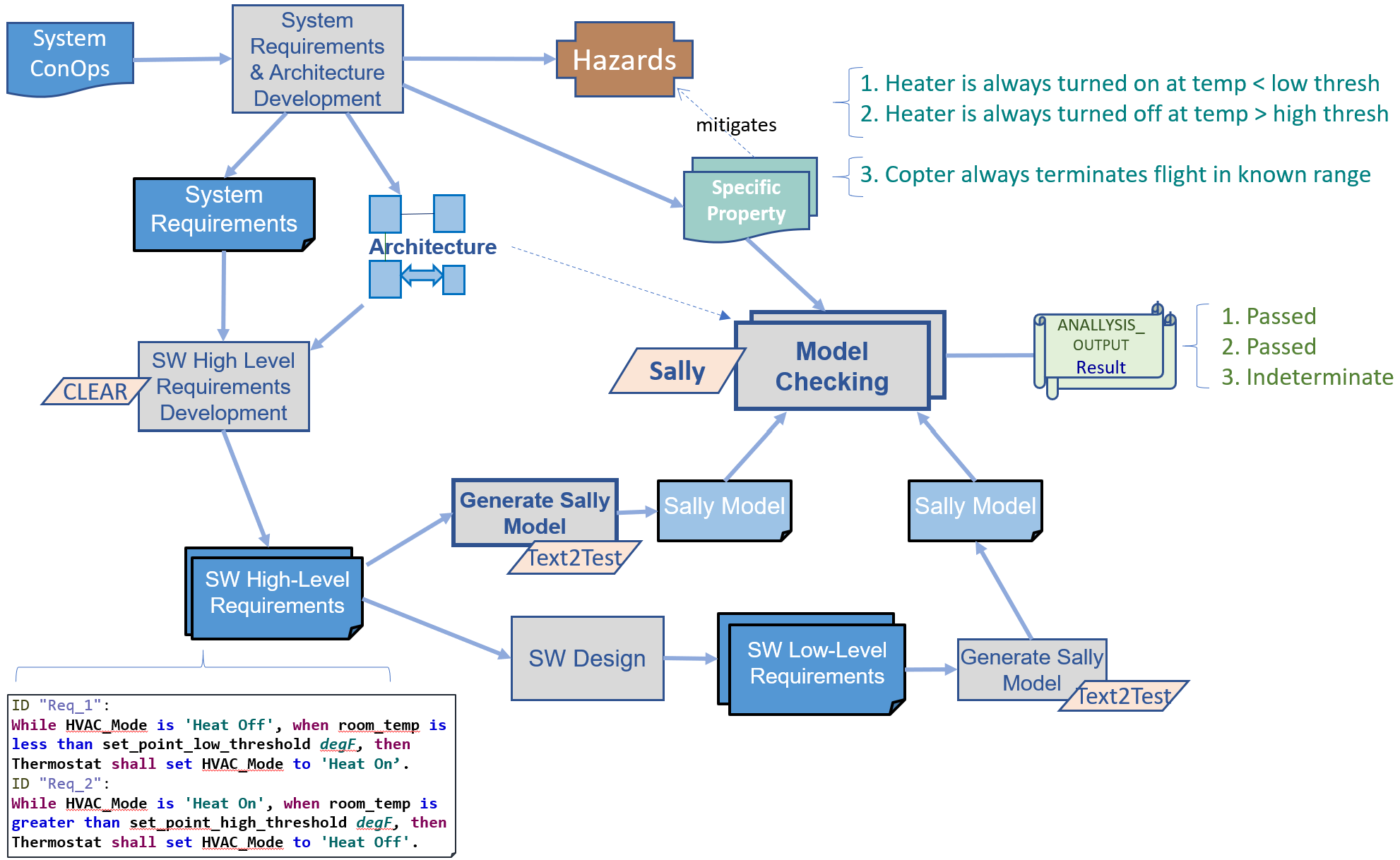}
\caption{Specification and Checking of Specific Properties of Requirements}
\label{fig:specific_prop_req}
\end{sidewaysfigure}

We classify properties into two broad categories: \textit{Generic Properties} and \textit{Specific Properties}:

\begin{itemize}
  \item Generic Property: A generic property is a declaration that an artifact (\textit{scope} of the property) must satisfy some general desired characteristics related to verification objectives, or must exhibit absence of certain generic defects that can cause hazards. Generic properties are established for the type of development artifact (e.g., requirements, architecture, design, code) and are automatically applied to that particular type of artifact regardless of the application. Figure~\ref{fig:arch_req_property_flow} provides examples of generic properties for architecture and requirements. Other examples: code doesn't exhibit numeric overflow, code doesn't exhibit null-reference-pointer exceptions, etc.
  \item Specific Property: A specific property is, by definition, application specific. It declares that an artifact (\textit{scope} of the property) must satisfy certain application-specific behaviors and must not violate certain application-specific constraints to prevent hazards. Examples: speed never exceeds more than 120 miles/hr, thermostat always turns heat on when temperature is below 50 degrees F. Figure~\ref{fig:specific_prop_req} shows the evidence flow for the specification and checking of specific properties of requirements.
\end{itemize}

As we have stated earlier, both generic and specific properties must be verified at each level of the development artifacts in order to meet different parts of the set of all assurance objectives. The type and technique of analysis used for each of these properties as well as the assurance objective they relate to are significantly different.

\begin{figure}[!htb]
\centering
\includegraphics[scale=.7]{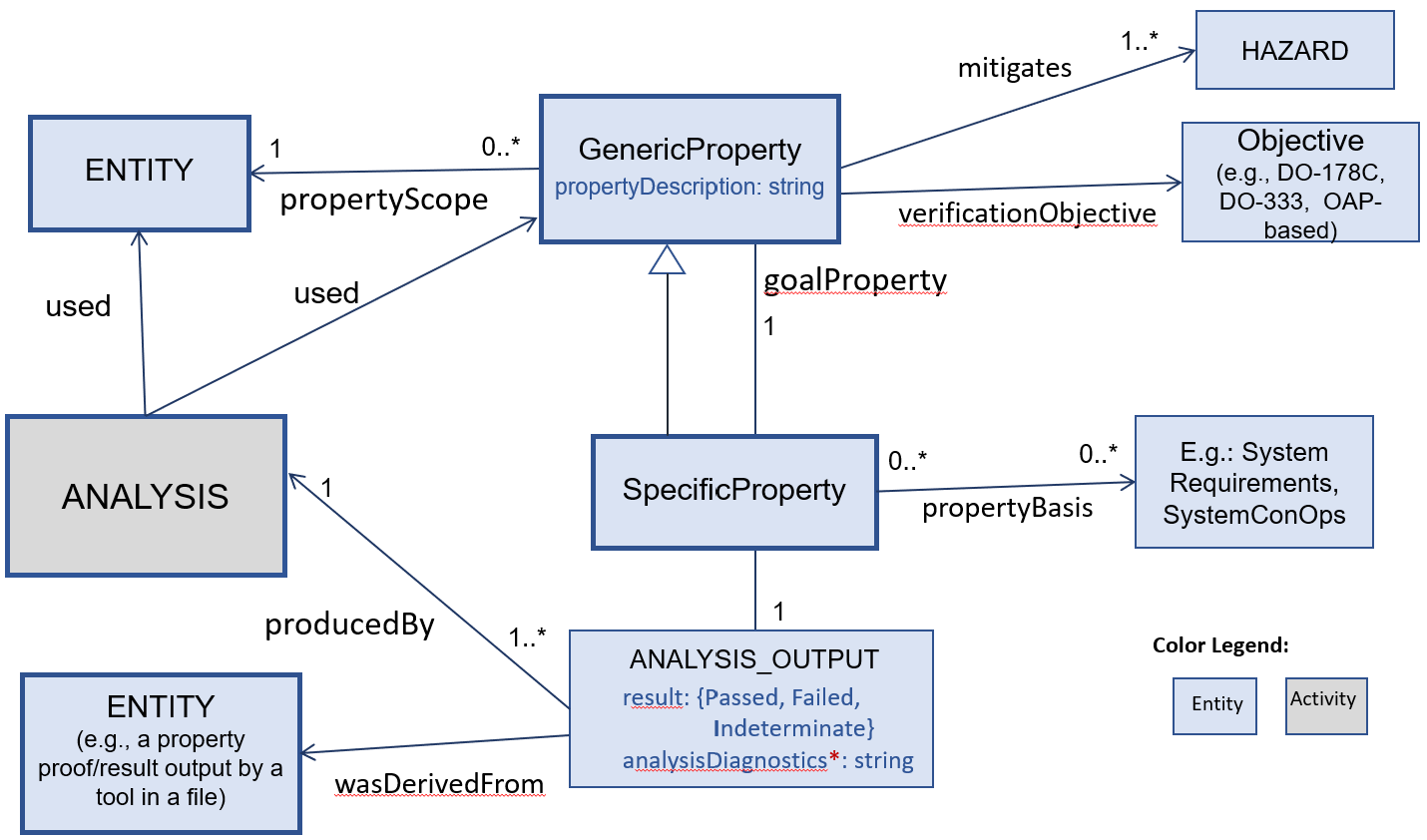}
\caption{Ontology Schema for Properties}
\label{fig:main_property_schema}
\end{figure}

Figure~\ref{fig:main_property_schema} shows the ontology schema for properties in UML notation, that derives from base W3C PROV schema described in section \ref{sec:ontology-desc}. The Generic Property class has a \emph{propertyScope} relationship to an ENTITY (e.g., a requirement set, architecture, code) over which the property must hold (note: Specific Property inherits all relationships and attributes of Generic Property). An ANALYSIS activity uses a property to try to prove the property(against the entity in its scope). A property can mitigate a potential hazard (via \emph{mitigates} relationship), satisfy a particular verification objective, and/or have a basis in a higher-level development artifacts from which the property is inferred. For example, a property to be verified on code could be inferred from software high-level requirements.

\section{DesCert Ontology Description}
\label{sec:ontology-desc}

\begin{figure}[!htb]
\centering
\includegraphics[scale=.8]{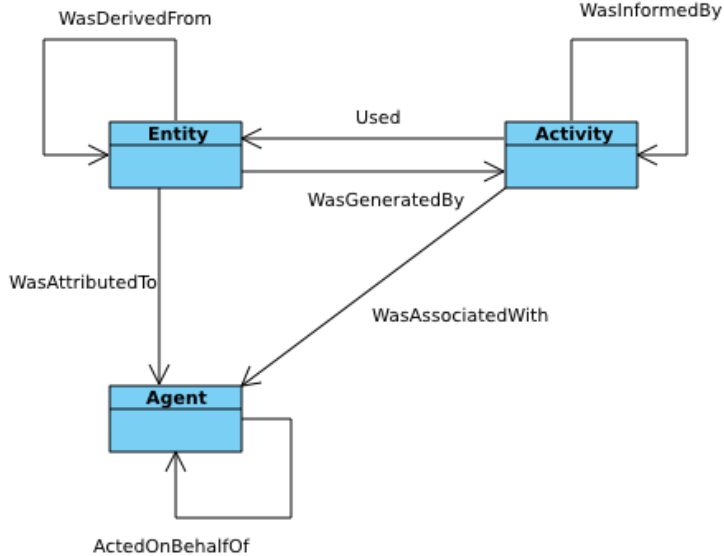}
\caption{The Core Structures in W3C PROV}
\label{fig:prov_core}
\end{figure}

This section provides a description of the salient aspects of DesCert Ontology. The DesCert ontology is based on the ontology classes in RACK (Rapid Assurance Curation Kit) database\footnote[3]{GE RACK: \url{https://github.com/ge-high-assurance/RACK}} being used in the ARCOS program. RACK in turn derives from the core \emph{provenance} model PROV defined by W3C.
Figure~\ref{fig:prov_core} depicts the core structures in PROV. An Entity captures a thing in the world (in a particular state) --- e.g., a particular version of a development or verification artifact. The entity was derived from some other entity, and was generated by an Activity that used other entities. An Agent (e.g. a person or tool) was associated with the activity, and the entity that was generated by the activity was attributed to that agent.

\subsection{Ontology for Checking Specific Properties against a Model of Requirements}
\begin{sidewaysfigure}[!htbp]
\centering
\includegraphics[scale=.65]{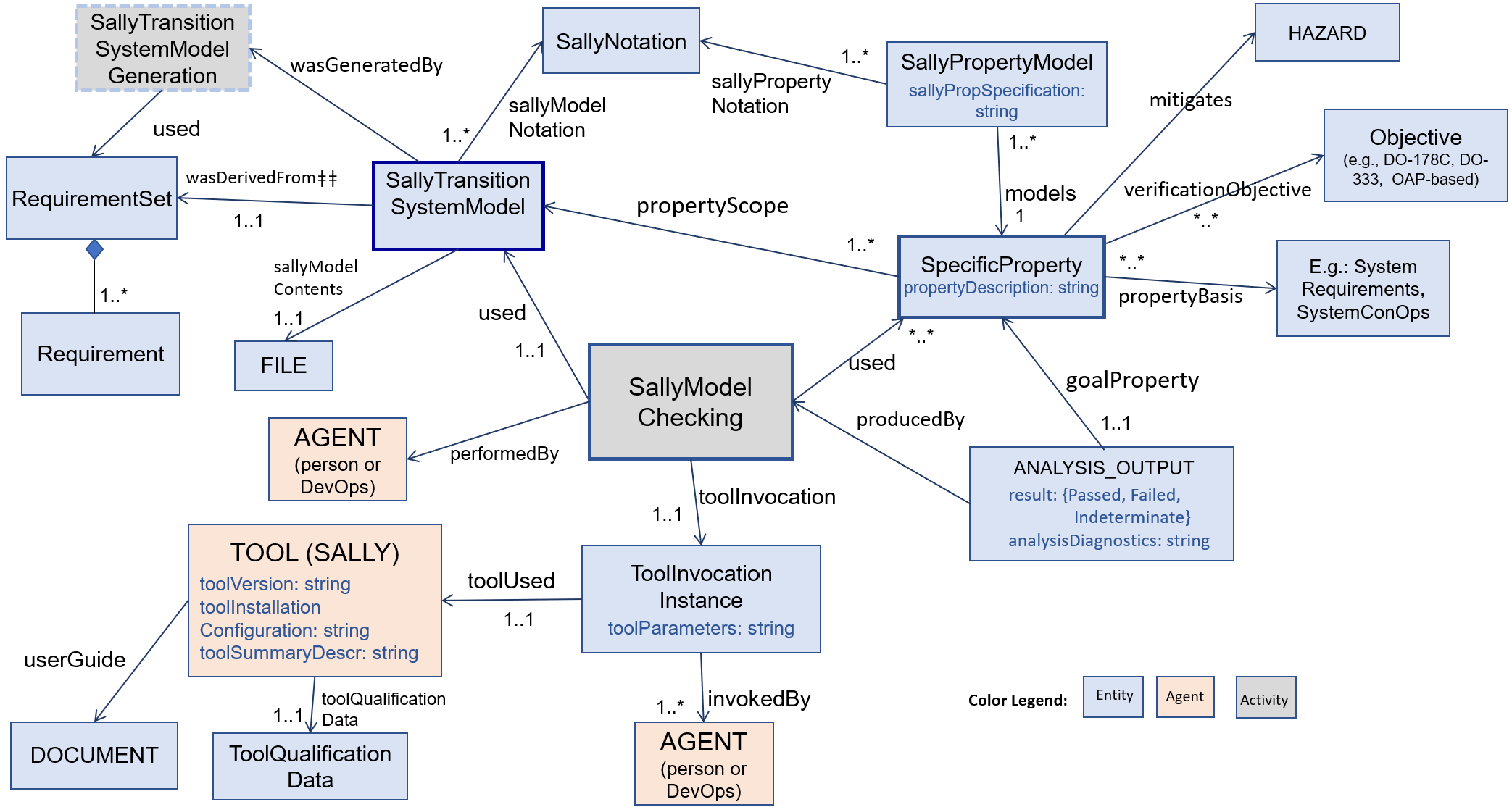}
\caption{Ontology for Checking Properties against the Sally Model of Requirements}
\label{fig:sally_model_checking_ontology}
\end{sidewaysfigure}

\begin{figure}[!htb]
\centering
\includegraphics[scale=.6]{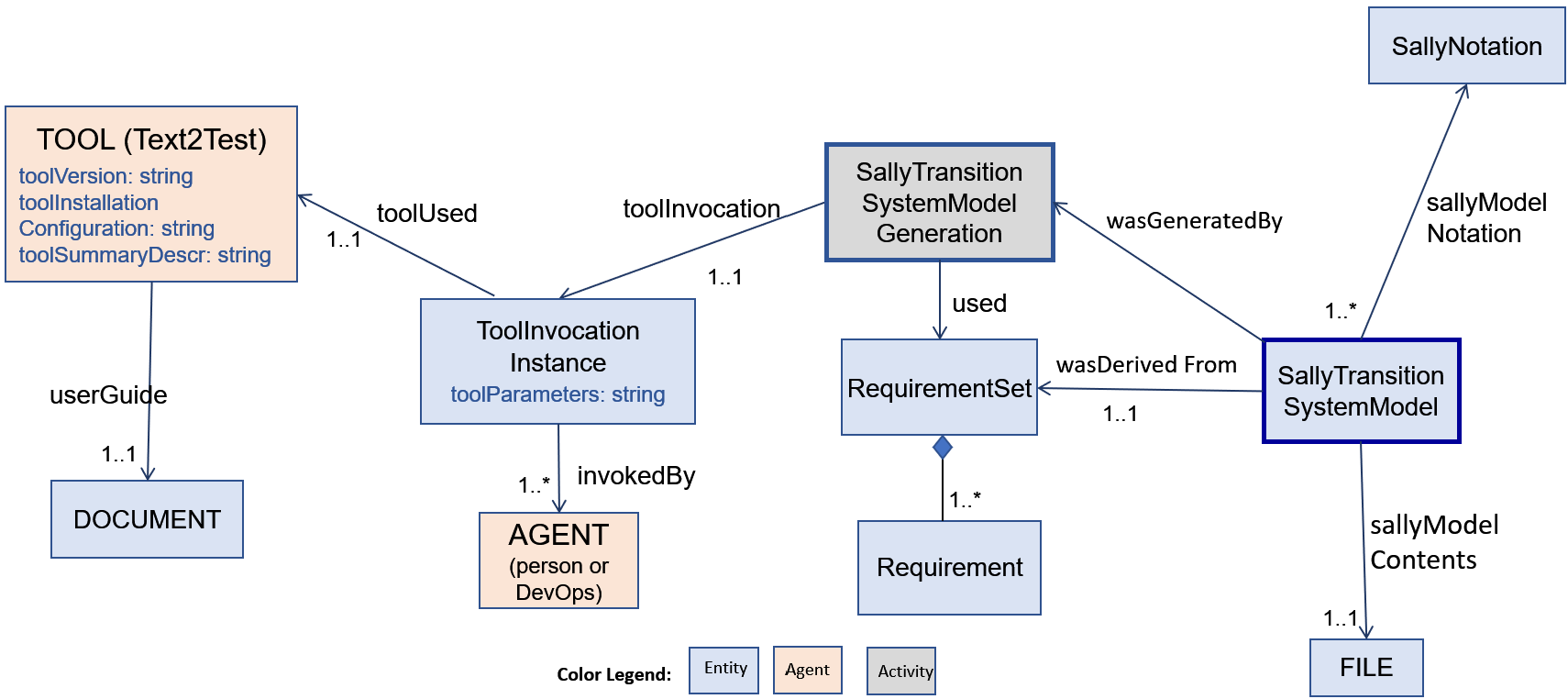}
\caption{Ontology for Generation of Sally Model from Requirements}
\label{fig:sally_model_gen_ontology}
\end{figure}

As mentioned in section~\ref{sec:property-desc}, \textit{properties} are central to the DesCert evidence flow; Figure~\ref{fig:main_property_schema} depicts the basic concept of Generic and Specific Properties. Using these basic concepts, Figure~\ref{fig:specific_prop_req} shows the evidence flow for the specification and checking of specific properties of requirements using the Sally tool. The ontology schema for this usage is shown in Figure~\ref{fig:sally_model_checking_ontology}. In the upper left quadrant of this figure is the \emph{SallyTransitionSystemModel}, which is generated by another activity (shown fully in Figure~\ref{fig:sally_model_gen_ontology}. The Sally model is completely based upon the \emph{RequirementSet}, as denoted by the \emph{wasDerivedFrom} relationship.Sally model uses the \emph{SallyNotation}. \emph{SallyModelChecking} activity uses the Sally model and a SpecificProperty to attempt to prove that the property holds on the Sally model, and by extension, on the RequirementSet. The results of the this are in ANALYSIS\_OUTPUT.

Figure~\ref{fig:sally_model_gen_ontology} shows the ontology for the auto generation of the \emph{SallyTransitionSystemModel} from a \emph{RequirementSet}. At the center of this figure is the activity that invokes the \emph{Text2Test} tool and pass it parameters that contain the name of the RequirementSet, execution period, and any requirements and dictionary files referenced by the RequirementSet. This produces that SallyTransitionSystemModel that traces to the RequirementSet via the \emph{wasDerivedFrom} relationship. The SallyTransitionSystemModel is then used for checking SpecificProperties against it as shown in Figures~\ref{fig:specific_prop_req} and \ref{fig:sally_model_checking_ontology}.

\subsection{Ontology for Checking Generic Properties of Requirements}

\begin{figure}[!htb]
\centering
\includegraphics[scale=.65]{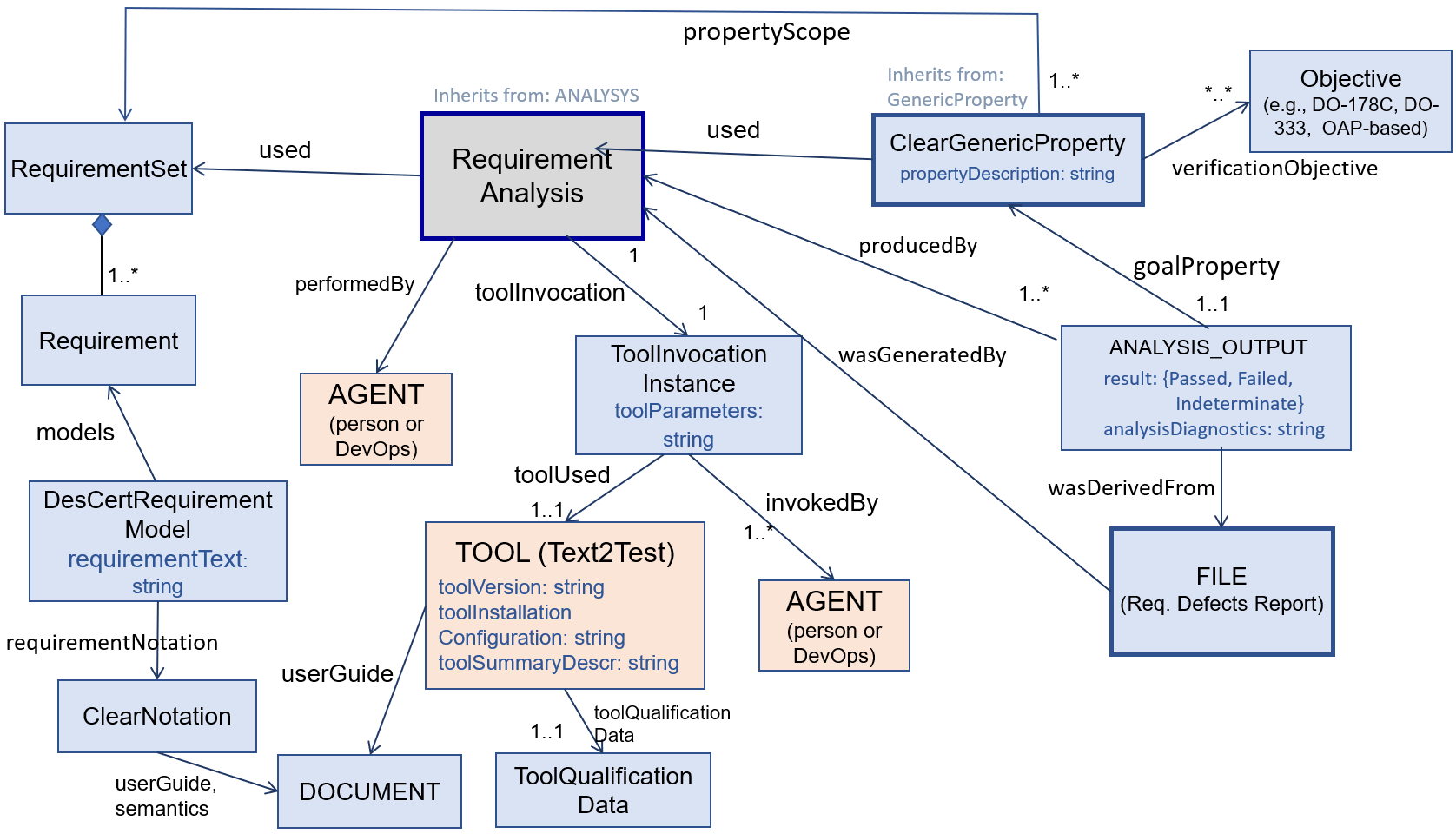}
\caption{Ontology for Checking Generic Properties of Requirements}
\label{fig:req_analysis_ontology}
\end{figure}

Figure~\ref{fig:arch_req_property_flow} depicts the checking of generic properties of requirements and the evidence flow. Figure~\ref{fig:req_analysis_ontology} shows the ontology for this evidence. The \emph{Text2Test} tool is invoked by the \emph{RequirementAnalysis} activity to analyze a \emph{Requirementset} against several predefined \emph{ClearGenericProperties}. The requirements contained in the Requirementset are expressed in the \emph{ClearNotation} which is supported by userGuide and semantics documents. The ClearGenericProperties trace to satisfy parts of DO-178C objectives. A high-level view of these properties is shown in Figure~\ref{fig:arch_req_property_flow}; details of the properties are described in Section~\ref{sec:clear-generic-properties}.

\subsection{Ontology for Test Generation from Requirements}

\begin{sidewaysfigure}[!htbp]
\centering
\includegraphics[scale=.7]{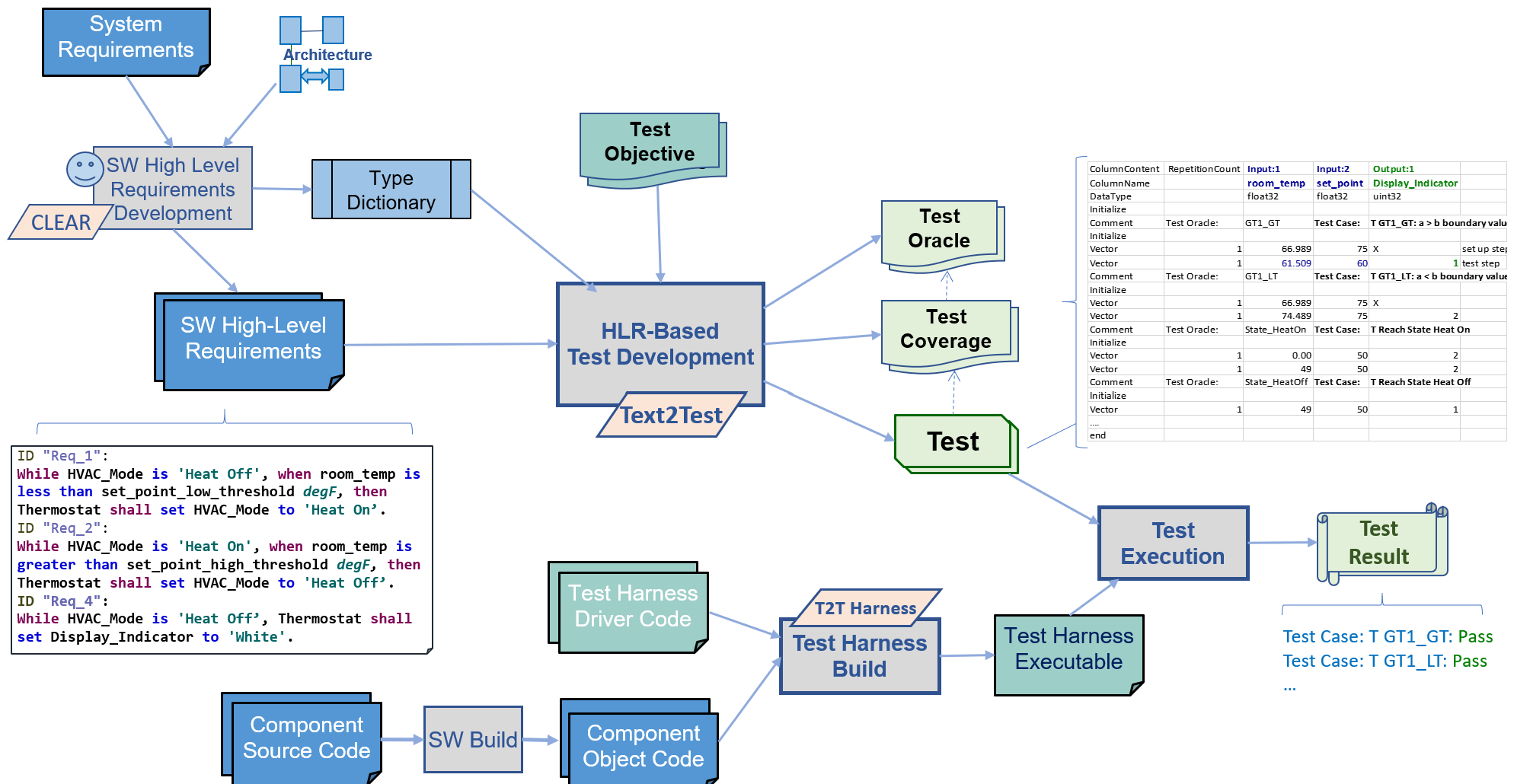}
\caption{Evidence Flow for Test Generation from Requirements}
\label{fig:testgen-evidence-flow}
\end{sidewaysfigure}

Figure~\ref{fig:testgen-evidence-flow} depicts the tool usage and evidence flow for generation of tests from software high-level requirements (HLR) and the execution of those tests. Software HLR for each software component are first developed using system requirements and architecture as inputs; the architecture provides an embedding context of the component within the system. The Text2Test tool is used for the generation of \emph{Test Oracles} and \emph{Tests} from the HLR.

\begin{figure}[!htb]
\centering
\includegraphics[scale=.45]{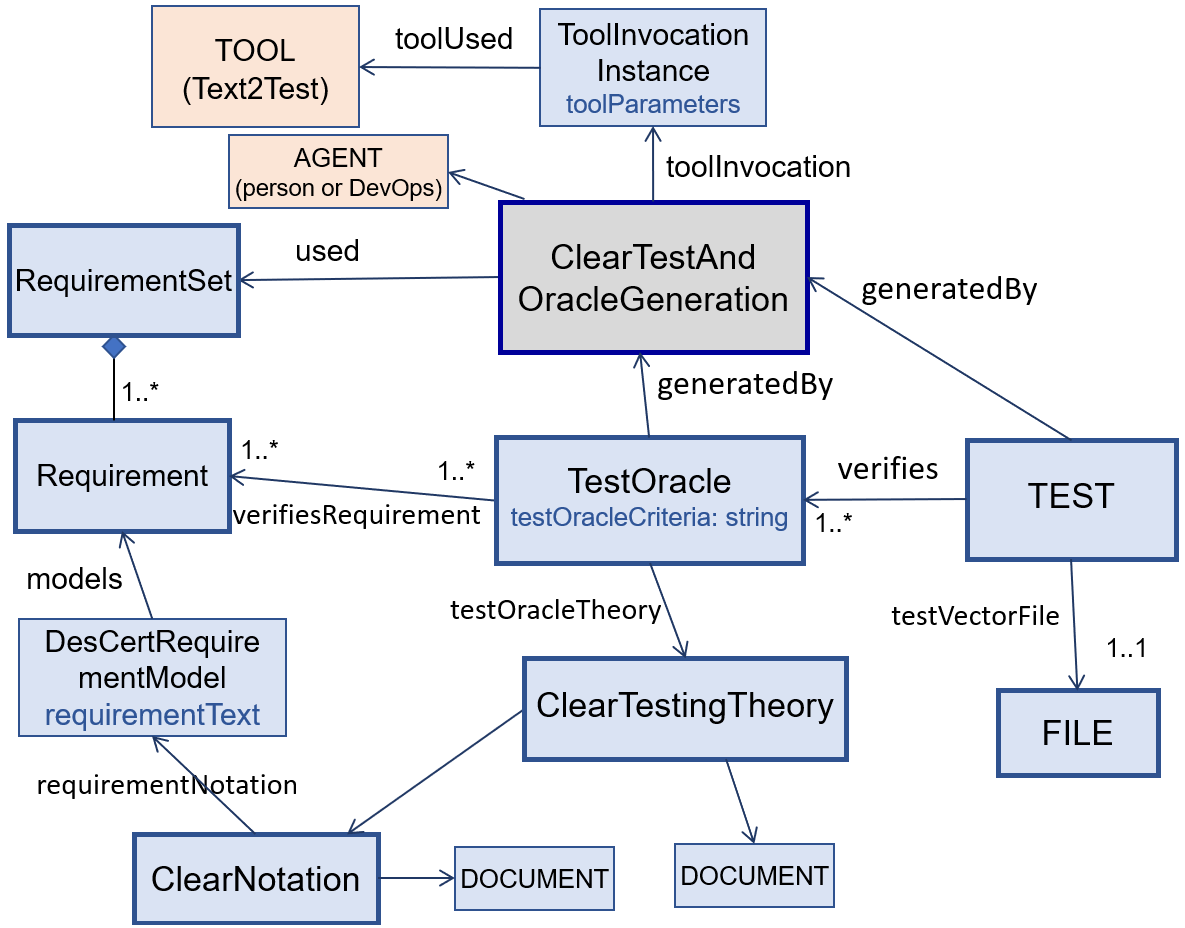}
\caption{Ontology for Test Generation from Requirements}
\label{fig:testgen-ontology}
\end{figure}

Figure~\ref{fig:testgen-ontology} depicts the ontology for generations of tests from software high-level requirements. Test Oracles and TESTs are generated by this activity. A Test Oracle traces to a requirement (and specific operator within that requirement) and is based upon the CLEAR Testing Theory described in more detail in Section~\ref{sec:clear-testing-theory}.

\chapter{Tools for Evidence Generation}\label{chap:tools} %Hao/Devesh/Jason/Anitha/Jorge/Mike
We describe the assurance-driven development tools used in the DesCert workflow.
These include tools for requirements capture and analysis, model checking, test generation,
software architecture, and code analysis.  
\section{Notation and Tools for Requirements Specification, Analysis, and Test Generation}

\begin{figure}[!htb]
\centering
\includegraphics[width=\columnwidth]{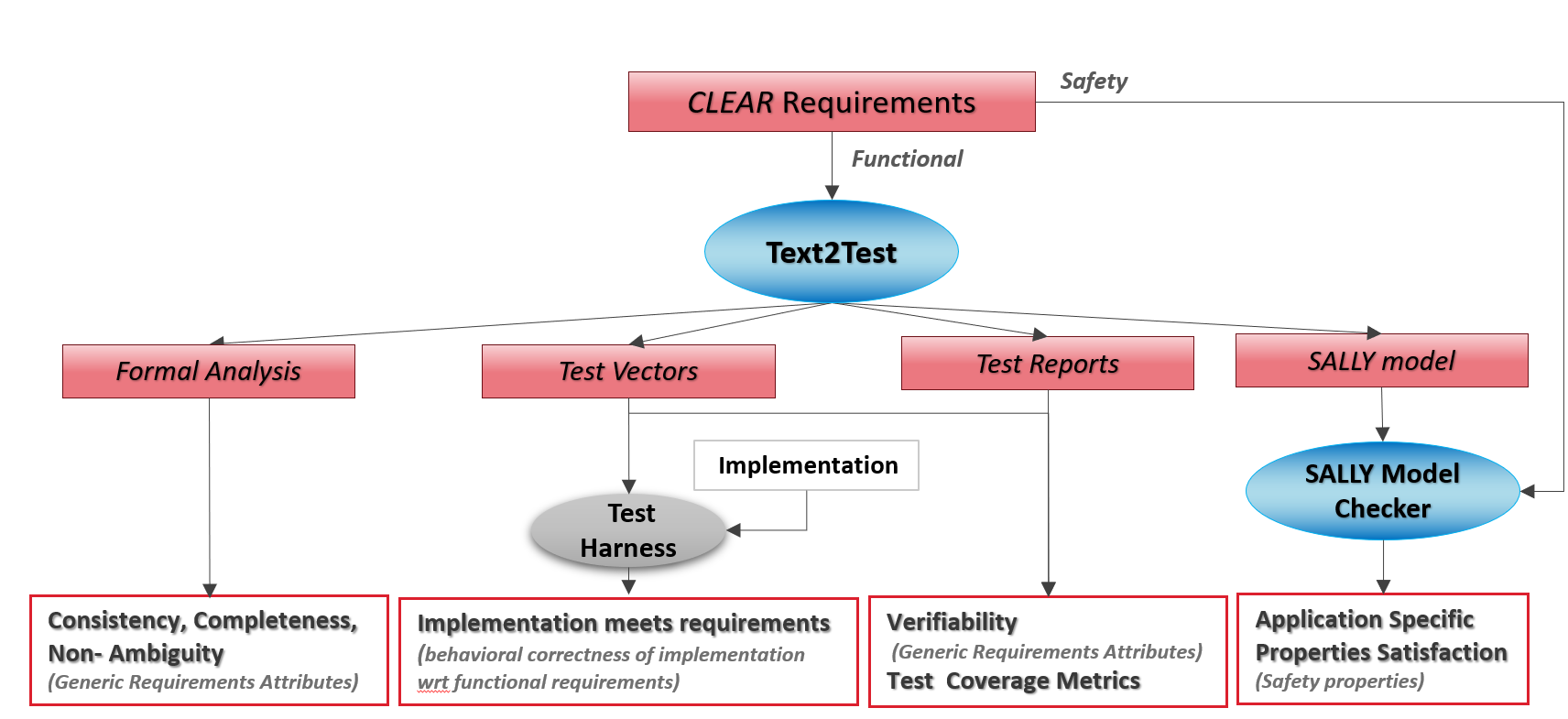}
\caption{Requirements Analysis Evidence Generation Overview}
\label{fig:ReqEvidenceGenOverview}
\end{figure}

Figure~\ref{fig:ReqEvidenceGenOverview} shows the high-level overview of requirement specification, the types of requirements analysis that and the respective evidence generated.

In essence, specifying requirements in the CLEAR notation allows us to use the Text2Test tool to automatically perform the following analysis and generate related evidences for assurance:

\begin{description}

\item [Formal Analysis]: The requirements are semantically analysed for consistency, completeness and non-ambiguity using formal solvers such as SMT at the backend. The results of the formal analysis are automatically captured in a tabular, easy-to-understand web page format.
    
\item [Test Vectors]: The tool automatically generates requirements-based test vectors in a Comma-separated textual file format. With the help of a test harness, these  test cases can be used to verify and assure that the implementation indeed meets the requirements.
    
\item [Test Reports]: Along with the test cases, the Text2Test also automatically generates various reports that quantifies the verifiability of the requirements as well as the provides a detailed explanation of which part of the requirement each test case checks. This serves as the traceability between the requirement and the test cases.

\item [Sally Model]: From the functional requirement specifications the Text2Test tool automatically generates Sally model, that can be used by the Sally model checker to verify specific properties (such as safety properties).

\end{description}
% !TEX root = Phase1Report.tex

\subsection{CLEAR: Constrained Language Enhanced Approach to Requirements}

\begin{figure}[!htb]
\centering
\includegraphics[width=\linewidth]{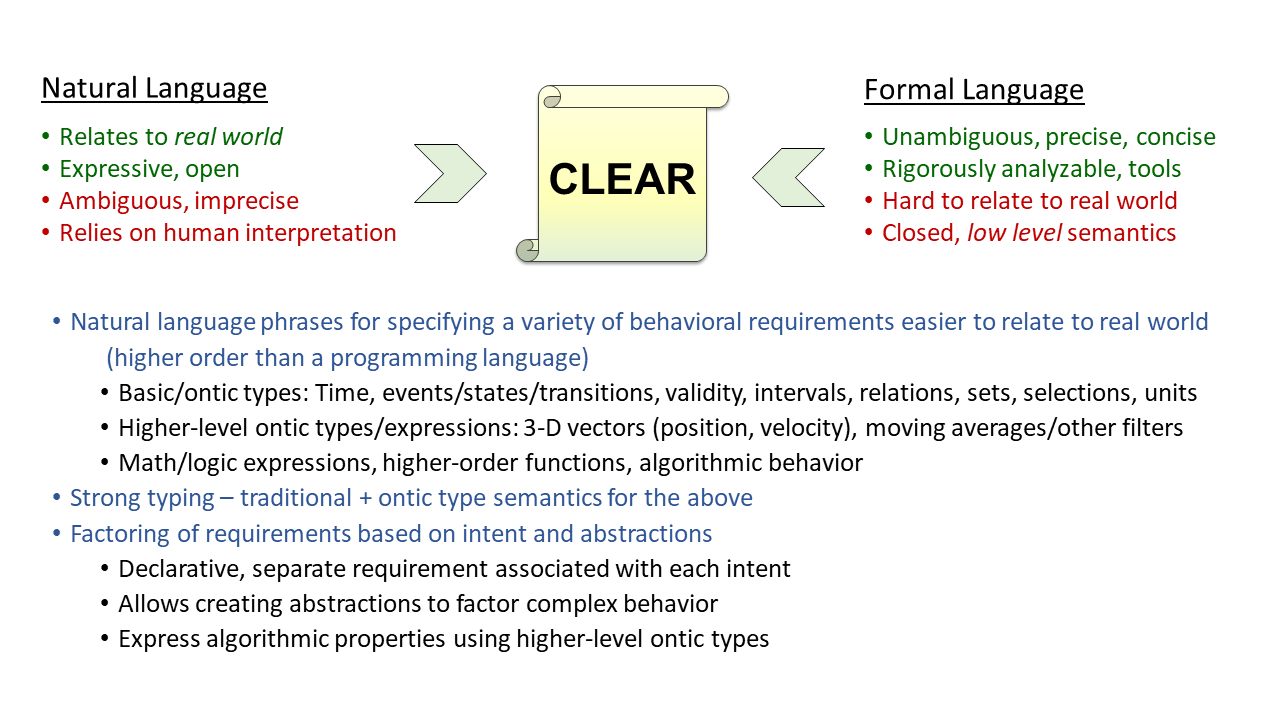}
\caption{An overview of the design principles of CLEAR.}
\label{fig:clear-principles}
\end{figure}

Typically, software requirements are written purely as natural language, which is expressive and, when written well, understandable without learning a specific notation. However, natural language is also subject to interpretation and ambiguity. Sometimes a formal notation is used to capture software requirements, which eliminates ambiguity and opens up possibilities for software analysis, but the syntax required is generally opaque to the untrained eye. In addition, formal software requirements can be hard to relate to the real world which is often much messier than the formal realm can easily capture.

CLEAR's approach attempts to combine the best of both worlds by defining a formal language that can generally be read as English, but in fact conforms to a traditional language grammar~\cite{arcosCLRTechReport}. A set of CLEAR requirements can generally be read and understood without any experience with the notation. The phrases and constructs available in CLEAR have been carefully considered to be as unambiguous as possible, which acts as a strictly-enforced style guide on requirements. Because it is a formal language, it can be parsed and analyzed without guessing at the intent or meaning of the requirements.

\pagebreak[1]

\paragraph{Single expressions of intent.}

For example, consider two requirements for a microwave:

\includegraphics[width=\linewidth]{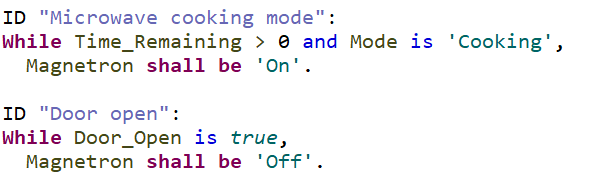}

Each requirement captures a single specific intent for the behavior of the microwave: when cooking, the microwave should be heating up the food. When the door is open, the magnetron must not be active. Both are reasonable expectations, but when considered together they are contradictory. If the door is opened while cooking, one requirement says to continue cooking while the other says to stop. In a large set of requirements, such conflicting requirements may be far apart in the document, making it difficult to spot the inconsistency. Text2Test performs this analysis and produces an error.

% Devesh: this is good way to distinguish requirements from designs or programs
The ability to capture individual intents and consider them both individually and in combination is a different approach than writing a program and allows for thinking about the software in a different way than programming. Much like a developer writing a test case may think of things that she did not consider when writing the implementation, defining the behavior of a program in individual requirement statements is a different thought process than writing code. By definition, a program operates on the complete input space; that is, for any possible input the code will respond to it in one way or another, even if that particular input was not considered by the author of the code. Requirements, on the other hand, just specify intended behavior. If there is a gap in the input space, it will become apparent that this behavior needs to be defined. The following Section~\ref{sec:clear-generic-properties} describes the analyses that are performed on CLEAR requirements for find inconsistencies (conflicts) among requirements and gaps.

Text2Test is the name of the tool which takes CLEAR files as input and parses, processes, and analyzes the requirements, producing reports and test cases. Each requirement is transformed into a data flow diagram. If multiple requirements define values for a particular output, these results are combined and checked to ensure that no contradictions exist (i.e., two requirements that prescribe different outputs under the same conditions). Text2Test can also analyze the conditions specified by the requirements for a particular output and produce a warning if there are inputs that have no applicable requirement. Text2Test is described in more detail in Section~\ref{sec:t2t-overview}.

\paragraph{CLEAR Notation Features.}

CLEAR provides requirements constructs to specify a variety of behaviors in a precise manner, while also providing a way to build abstractions and factory the requirements space for complex behaviors. The following are the salient features of the notation:
\begin{itemize}
\item	Each requirement is a declarative statement that can be independently reviewed and tested; requirements are additive --- adding requirements completes the behaviors required

\item  CLEAR allows specification of the following aspects of behaviors:
\begin{itemize}
   \item State-based, event-triggered, time-triggered behaviors and combinations thereof.
   \item Algorithmic aspects using combination of mathematical, relational, and Boolean expressions, sets/selections, interval arithmetic, interpolation tables.
   \item CLEAR includes a large library of math functions including numeric manipulation,  trigonometric, exponential, filters, etc.
\end{itemize}

\item Constructs for creating abstractions and for factoring of complex behaviors:
\begin{itemize}
  \item Tabular format for generalized truth tables, precedence tables (as alternative to while requirements and clauses) and interpolation tables
  \item Creating common definitions of specific behaviors or input conditions referenced from several requirements
  \item Creating complex types of objects and conditions/behaviors associated with them
  \item Creating intermediate abstractions (states, variables) to create a framework for the specification of complex behaviors
\end{itemize}

\item Support for ontic concepts and types:
\begin{itemize}
\item Basic ontic concepts of real-world time intervals, events, sensor input validity specification, and units.
\item Higher-level concepts such as XYZ Vectors of position, velocity, vector difference, concept of moving average of sensor values, etc.
\item Future extensions: Ontic type system around basic physical (and cyber) concepts such as position, velocity, time, temperature, pressure, angles and attributes of units and relationships applied to them. Users will create application-specific subtypes.
\end{itemize}

\end{itemize}

\paragraph{CLEAR Semantics.}

The CLEAR semantics provide a number of useful concepts and structures useful for writing software requirements:

\begin{itemize}
\item Objects have (discrete or continuous) time-varying values
\item Conditions are predicates over the values of objects at a given time
\item Events are changes in conditions
\item Actors are systems or components
\item Functions are computed on object values at a given time or over a time interval
\item Responses are functional updates to values of internal or output objects of an actor
\item Responses can be condition-based, event-triggered, time-triggered (or combination): CLEAR is agnostic about the Model of Computation (MoC)
\item Functions and structural aspects
\begin{itemize}
\item Algorithmic aspects using combination of mathematical, relational, and Boolean expressions, sets/selections, interval arithmetic, generalized truth tables, interpolation tables.
\begin{itemize}
\item CLEAR includes a large library of math functions (including transcendentals, filters, integrators)
\end{itemize}
\item Type system with support for ontic types (more future work)
\item Factoring of complex behaviors into multiple definitions/requirements
\end{itemize}
\end{itemize}

\paragraph{CLEAR: Structure of Requirements.}

\begin{figure}[!htb]
\centering
\includegraphics[width=\linewidth]{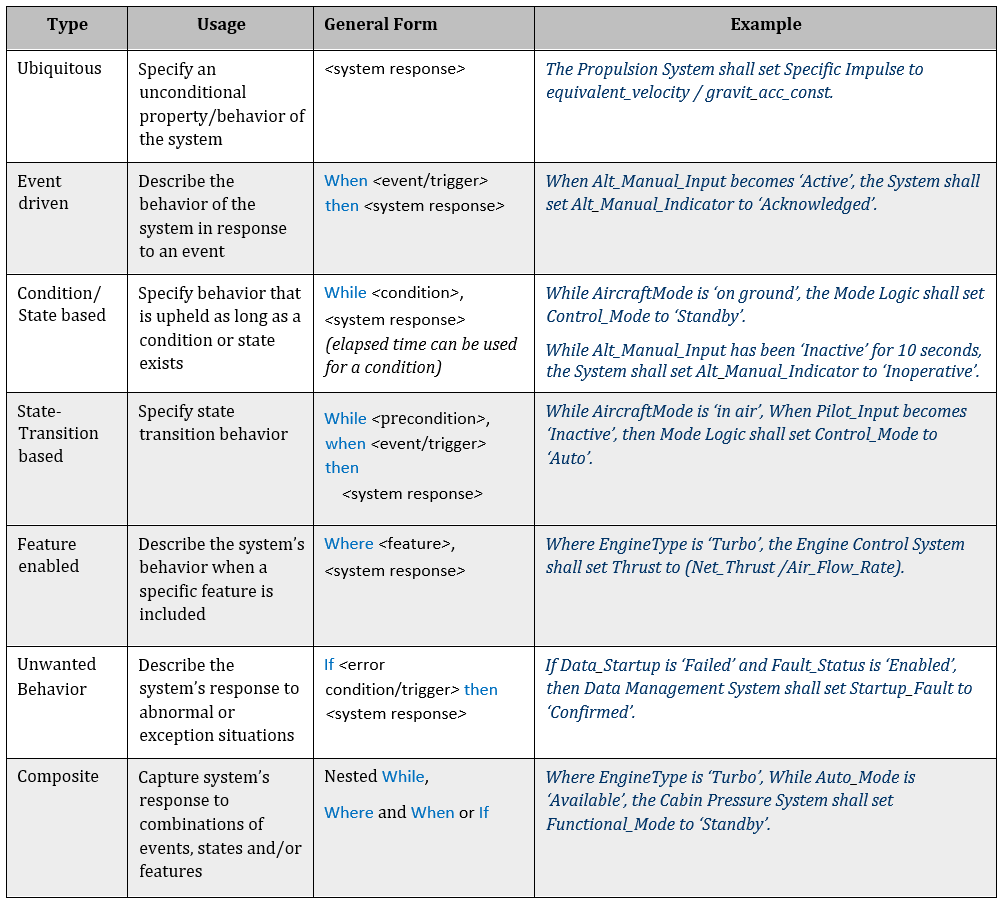}
\caption{Requirement Types/Structure in CLEAR}
\label{fig:clear-req-types}
\end{figure}

CLEAR requirements can take one of a few general forms which are fully described in Figure \ref{fig:clear-req-types}. If the system must do something unconditionally, a ``shall" statement can simply specify the output and its value. Responses to events, such as an input crossing a threshold or changing to a particular value, use the word ``when" followed by the condition and then the response to that event. Variables can be marked as state variables, and requirements that apply in a particular state can be specified using a ``while" statement. If the software can be configured with features enabled or disabled, the word ``where" is used to specify that a requirement only applies when a particular feature is enabled. Exceptional or abnormal conditions can be specified with the word ``if" instead of ``when", overriding other requirements even if they would normally apply. By giving ``if" statements priority, requirement authors can write the majority of requirements to deal with nominal cases without having to specify a lack of error in every single requirement, and only need to address the exceptional cases in ``if" statements. Finally, many of these features can be composed together in various configurations.

%We defined the formal semantics of several CLEAR constructs (for specifying requirements) and their mapping to the input model for the Sally model checker.

\subsection{Text2Test Tool Overview}
\label{sec:t2t-overview}

\begin{sidewaysfigure}[!htbp]
\centering
\includegraphics[width=\linewidth]{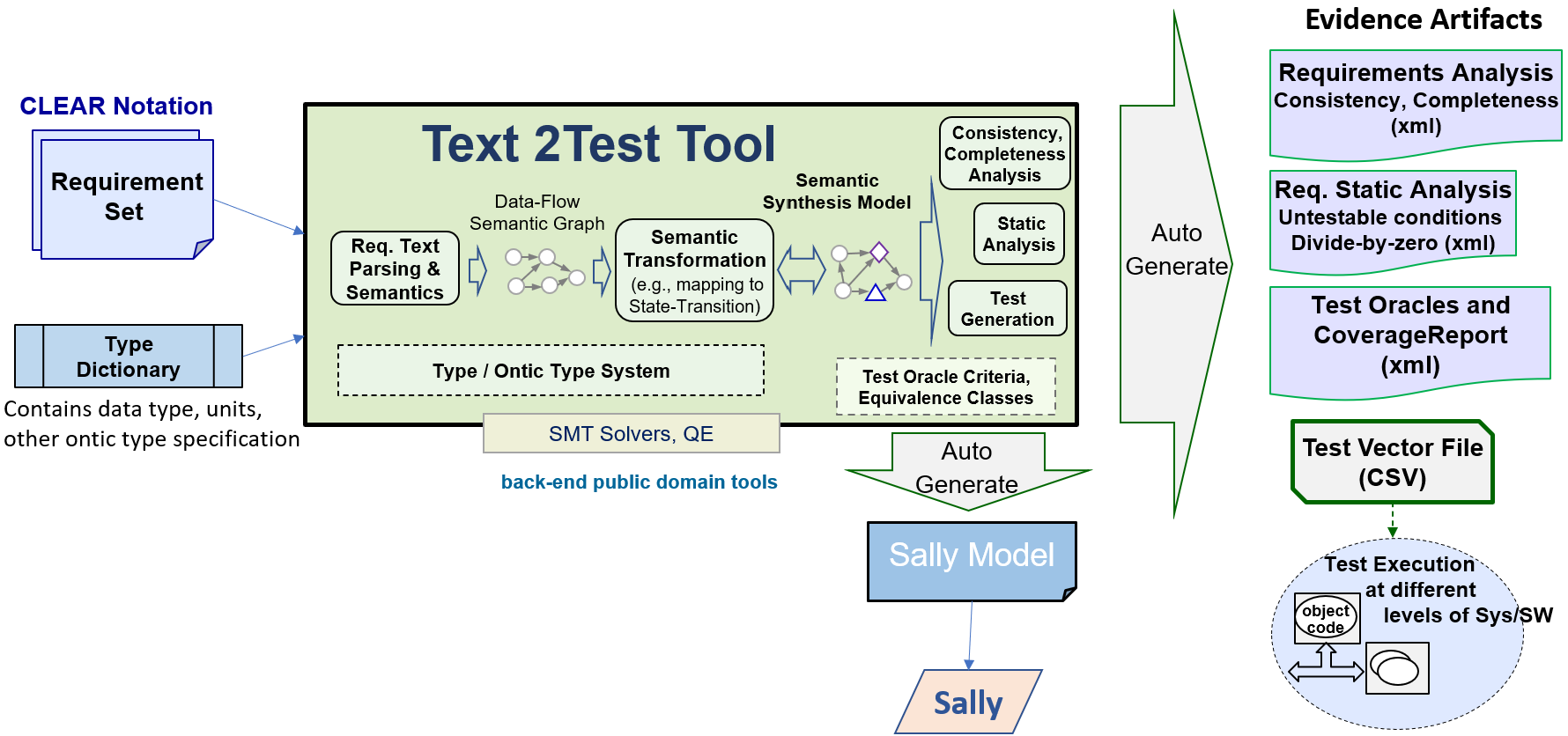}
\caption{Architecture of the Text2Test Tool}
\label{fig:t2t-tool-arch}
\end{sidewaysfigure}

Text2Test is a powerful tool capable of automatically generating evidence artifacts for CLEAR requirements in textual form.  Figure~\ref{fig:t2t-tool-arch} shows the architecture of Text2Test tool. It takes the CLEAR requirement set from front-end requirement editor, as well as metadata such as variable data type, units, and other ontic type specifications in separate dictionary files, as inputs. Then it creates a minimum-scale but semantic-equivalent internal block diagram model for the requirement set called \emph{Semantic Synthesis Model}. This internal model not only provides a visualization, but also is suitable for the model-based analysis including test generation, static analysis (e.g,  range propagation), and formal verification (against generic properties as listed in subsection~\ref{sec:clear-generic-properties}).  The internal model also serves as a medium which can be translated to models for other verification tools.

\subsubsection{From RequirementSet to Text2Test internal Semantic Synthesis Model}
\label{sec:t2t-internal-model}

\paragraph {Data-Flow Semantic Graph creation.}
A data-flow semantic graph (called ''raw model'') is first created from the requirement, capturing the data flow and semantics of the behavior operators used in requirement clauses and subexpressions. The raw model preserves the semantics of each and every individual requirement in the requirement set, but it is not fully functional as a whole in mainly two aspects. Firstly, a \textit{switch} block may have, at one of its data input ports, connection to the \textit{Invalid} block that cannot be executed or propagated through. This is due to the fact that the switch block is created from a selection requirement (for example, a ``when..., then...'' requirement) incomplete by itself.  Secondly, if a feedback path is formed, it often misses a unit delay implied by the keyword ``transition to'' without an explicit time shifting keyword ``previous'', which is needed for unit delay block creation in the raw model. Other than the system-level semantics imperfection, the raw model often contains redundant logical blocks, and lacks non-primitive blocks (state transition for example). These all make the raw model less user-friendly for reading and examining.

It is not hard to see that, the common root cause of these ``imperfection'' is that the incompleteness of individual requirement gets carried to the raw model whose creation process does not possess system perspective. To address these, a system-level aggregation will be performed on the raw model through a block merge process. Block merge is based on graph search and pattern recognition, aiming to fuse low level primitive blocks into blocks of richer semantics, filling semantic gaps as well as eliminating redundancy. The next subsection elaborates the merge of a set of blocks exhibiting state transition behavior, resulting in a fully functional state transition substructure of much more compact form.

\begin{figure}[!htb]
\centering
\includegraphics[width=\linewidth]{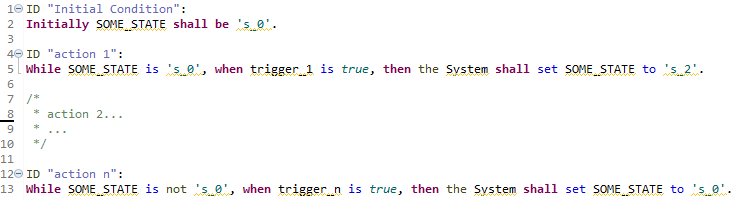}
\caption{Requirement set example for a state variable SOME\_STATE.}
\label{fig:st req set}
\end{figure}

\paragraph {State transition block merge to create Semantic Synthesis Model.}
A typical state transition behavior consists of state initialization, and a state transition function that determines the current state value based on the previous state value and/or external transition triggering signals (also called non-state trigger). In the CLEAR requirement set, the state transition behavior of one state variable is often distributed in multiple individual requirements, each of which contributes a partial statement of either state initialization or one transition action as the example shown in Figure~\ref{fig:st req set}. In a typical state transition action requirement, the ``while'' clause is to specify the previous (source) state value and the ``when'' clause is to specify the non-state triggering condition, followed by the clause that sets the current (destination) state value.

\begin{figure}[!htb]
\centering
\includegraphics[width=\linewidth]{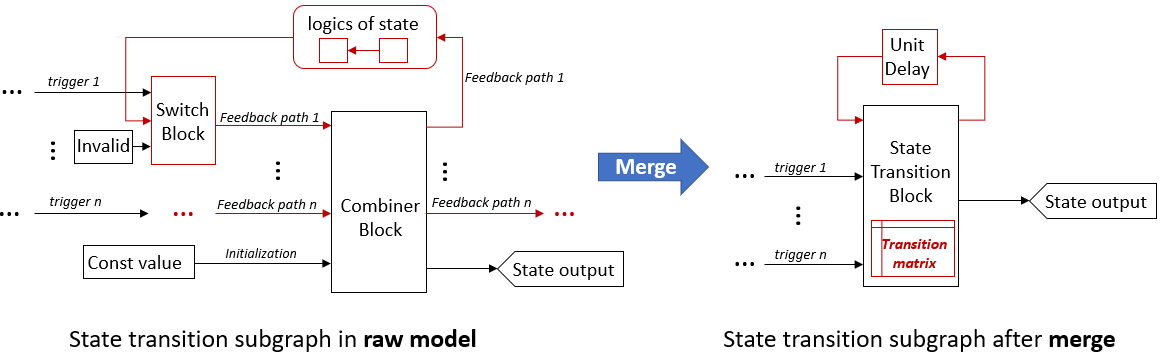}
\caption{State transition subgraph before and after merge.}
\label{fig:stb}
\end{figure}

The left hand side of Figure~\ref{fig:stb} shows the raw model subgraph representing the state transition behavior in Figure~\ref{fig:st req set}. In the raw model creation process, for each requirement, primitive blocks are created, explicitly mapping to CLEAR functional and logical keywords (e.g., \textit{switch} block maps to keyword``While/When'',  \textit{not} block maps to logical operator ``not''), then proper connection is added, forming a feedback loop path. A \textit{Combiner} block is created as a routing hub node aggregating all state value set and get for the common state variable. The initialization requirement is simply converted to a \textit{constant} block feeding to the \textit{Combiner}.

The state transition block merge starts from identifying the \textit{Combiner} block in the raw model and its associated feedback paths. Then the \textit{Combiner} block is replaced by a \textit{StateTransitionBlock} (\textit{STB}) block whose functionality is defined by an inherent transition matrix (initialized as an empty matrix) as shown in Table~\ref{st matrix}.  Next, each feedback path is analyzed to identify the associated non-state trigger and the source state value(s). Both of them are then recorded into the state transition matrix as row elements (for \textbf{semantic richness}) so that the entire feedback path is no longer explicitly needed thus removed from the subgraph (for \textbf{simplicity}), while the non-state trigger is reconnected directly to the \textit{STB} block input. Requirement ID is also recorded as row info (for \textbf{traceability}). The \textit{Invalid} block is removed (for \textbf{execution}), since the transition matrix is presumptive to be input-complete if the requirement set is input-complete. For assurance, the input completeness will be formally verified as a generic property in a later stage after merge. Lastly, a simple feedback path of \textit{unit delay} outside the \textit{STB} is added (for \textbf{temporal correctness}), providing the state value memory for the merged state transition subgraph. Note that, \textit{unit delay} is initialized by the initial state value, therefore, the initialization path in the raw model is also removed  (for \textbf{simplicity}), rendering the \textit{STB} itself as a memoryless non-timedependent block.

\begin{table}[h!]
\centering
\resizebox{\columnwidth}{!}{
\begin{tabular}{ |c|c|c|cV{3}c|c| }
 \hline
 \multicolumn{4}{|cV{3}}{\textbf{State transition matrix}} & \textbf{requirement ID}\\
 \hlineB{3}
\textbf{Initialization}&\multicolumn{3}{cV{3}}{$s_0$}& ``Initial Condition'' \\
\hline
&\textbf{non-state triggers} &\textbf{ source} &\textbf{destination}& \cellcolor{gray}\\
\cline{2-5}
\textbf{Transition} &trigger $1$   & \{$s_1$\}    &$s_2$& ``action 1''\\
\cline{2-5}
\textbf{actions}&... & ...& ... &...\\
\cline{2-5}
&trigger $n$&   \{$s_2, s_3, s_4$\}  & $s_0$   &``action n''\\
\hline
\end{tabular}
}
\caption{A complete state transition matrix for SOME\_STATE}
\label{st matrix}
\end{table}

The current CLEAR language supports a wide range of variations of state transition behavior specification other than the typical form in Figure~\ref{fig:st req set}, consequently corresponding to variations of subgraph structures in the raw model. For instances, 1) a state transition system can have a reset trigger that instantly overrides all other transition triggers once activated, 2) different triggers are associated with the same source state value(s), causing branching on a feedback path and/or intertwining among feedback paths, or 3) the current state value is determined by the non-trivial history of the state values instead of one-step memory of the previous state value, resulting in a time-dependent block such as \textit{timer} in the feedback path in the raw model. By analyzing the structural features of the raw model subgraph (e.g., feedback paths, branching nodes, locations of key functional blocks, etc.), Text2Test block merge is able to recognize those advanced state transition behaviors and achieve semantics-preserving and semantics-completing translation to the merged model. The merged model is the final stage semantically correct and complete model interpretation of the requirement set. It is the model that all the downstream formal analysis and test generation within Test2Test is based on. It also serves as the generic model from which semantically equivalent models for other tools (Sally for example) are obtained through proper translation. The sections below use the term ``internal model'' to denote the merged semantic synthesis model unless specified otherwise.

The following capabilities are implemented in Text2Test tool using the internal model; these are described in subsequent subsections:

\paragraph{Generic property checking and defect report generation} Text2Test utilizes the SMT-based property checking capabilities implemented on the internal model obtained above. Text2Test checks against a set of system generic properties as listed in subsection~\ref{sec:clear-generic-properties}, to detect possible fundamental defect(s) of the requirement set. Model/Requirement defects (with defect information such as defect type, root requirement IDs, variables and counterexample, etc.) are recorded in a .xml file for manual review purposes. These model checkings also serve the ``sanity check'' of the internal model,  only defect-free models proceed to be converted to other tool models (e.g., Sally model in the following subsection) for further model checking against specific properties.
\paragraph{Translation to Sally model} The Text2Test internal model is a correct and complete model for the requirement set. Its block diagram structure with well-defined block functionalities and clear connection relation make it easy to be further translated to Sally model as elaborated in subsection~\ref{sec:sally}.
\paragraph{Test generation} As a critical and legacy tool capability, test generation is elaborated in subsection~\ref{sub:testGen}.

\subsection{Checking Generic Properties of Requirements}
\label{sec:clear-generic-properties}

Section~\ref{sec:property-desc} introduces the notion of \emph{properties}, that we intend to analyze, into two broad categories: \textit{ Generic Properties} and \textit{Specific Properties}.

In the context of requirement specification, generic properties are those that are fundamental to any good requirements specification, irrespective of the system under consideration, such as consistency, completeness, verifiability, non-ambiguity, etc. We analyze the requirements for the following types of properties:

\begin{itemize}
\item  \textbf{Consistency}: that ensures that the requirements are free of
    \begin{itemize}
    \item Conflicts across multiple requirements such as two or more requirements specify different values for the same output variable under overlapping input conditions. For example, the following requirements of a thermostat system are inconsistent since they specify conflicting values for Display\_Indicator when the both their antecedents are true at the same time.
    \textit{\\REQ 1: While HVAC\_Mode is `Heat Off', the Thermostat shall set Display\_Indicator to `White'.
\\REQ 2: While HVAC\_SetUp is true, the Thermostat shall set Display\_Indicator to `Blue'.}

    \item Cycles in requirement data flow, without state transition
    \end{itemize}

\item  \textbf{Accuracy}: that assures that the requirements are accurately specified with specific logical and mathematical outcomes.

\item  \textbf{Non-Ambiguity}: that assures that the requirements do not use combinations of  natural-language phrases or mathematical expression that can interpreted ambiguously.

\item  \textbf{Completeness (``internal")} that identifies gaps in requirements wrt
    \begin{itemize}
    \item Input Gaps: Certain combinations of input conditions missing in requirement set. For example, the following requirement is considered input space incomplete if there are no other requirements that specify behaviours when HVAC\_Mode is `Heat ON'. \\\textit{REQ: While HVAC\_Mode is `Heat Off', the Thermostat shall set Display\_Indicator to `White'.}

    \item Output Gaps: Certain output values not produced by any requirement in the requirement set. For example, the above requirement is considered output space incomplete if the value of Display\_Indicator is defined to be an enumeration of several colors, but there are no requirements to set values of Display\_Indicator other than `White'.
    \end{itemize}

\item  \textbf{Requirements Verifiability/Testability} that ensures if there is any

    \begin{itemize}
    \item state in a requirement that is not reachable
    \item condition in a requirement that is not achievable
    \end{itemize}

\item  \textbf{Advanced properties} based upon Ontic type information helps ensures:
    \begin{itemize}
    \item Simple ontic-type violations – e.g., adding altitude to runway length

    \item Margins (values, time) used in decisions are not adequate based upon the real-world nature of inputs; e.g., improper time-debouncing or hysteresis of sensor input values (Margins are derived based on constraints associated with Ontic type for the sensor input)

    \item \textbf{Mode Thrashing} is an advanced  analysis based on margins, such as the system switching back-and-forth (metastable) between two states rapidly (general case: cycle through multiple states)

    \end{itemize}

\end{itemize}

Section~\ref{sec:clear-generic-prop-evidence} provide instances of generic properties that were checked on the AFS requirements.

% Text2Test/Sally integration goes here
\subsection{Sally Model Generation and Specific Property Verification}
\label{sec:sally}

%\subsubsection {Integration of Text2Test and Sally tools}
%\textcolor{red}{only talk about Sally model generation flow here}
%
%\begin{figure}[!htb]
%\centering
%\includegraphics[scale=.45]{figures/t2t_architecture_for_sally.PNG}
%\caption{The architecture of the Text2Test/Sally tool suite.}
%\end{figure}

Sally is a model checker for infinite state systems described as transition systems. Sally has both bounded model checking (BMC) engine and k-induction (kind) engine for the verification of transition systems. That is to say, integrating Sally with Text2Test provides advanced capability of specific property verification other than the generic ones for the requirement set, given that the requirement set can be modeled as a transition system. In our tool chain, the integration is achieved by translating Text2Test internal model to Sally model. This subsection introduces the model translation process, as well as the enhanced capability of specific property verification enabled by the tools integration and the extension.

\subsubsection{From Text2Test internal Semantic Synthesis Model to Sally model}
\paragraph{Model translation}
Sally model is a script model with inputs, states, and the state transitions. On the other hand, Text2Test internal model is a functional block diagram model which may or may not possess intended state transition behavior. But because there exists underlying system frequency for the internal model, it essentially can be remodeled as a transition system that responds to instantaneous changes. In this translation process, blocks/structures of the internal model are converted to the script segments (declaration, initialization, and state transition) in Sally model, following the summarized rules in Table~\ref{sally translation rules}. Because Sally script is based on the SMT lib 2.0 format, the translation order of the blocks/structures is irrelevant.

\begin{table}[h!]
\centering
\resizebox{\columnwidth}{!}{
\begin{tabular}{ |c|l|}
 \hlineB{3}
\multicolumn{2}{|l|}{\textbf{Case I:} A system \textit{input} block}\\
 \hline
 & \textbf{1) System input declaration:} A system input variable.\\
\cline{2-2}
\textbf{Sally} & \textbf{2) State declaration:}  An auxiliary input state variable.\\
\cline{2-2}
\textbf{counterparts:}  & \textbf{3) State initialization:} None.\\
\cline{2-2}
&\textbf{4) State transition:}  Next step value of the input state equals to input. \\
\hline
Note:& All states update simultaneously lagging by one step of system input(s).\\
 \hlineB{3}
\multicolumn{2}{|l|}{\textbf{Case II:} An \textit{STB} block and the associated \textit{unit delay} and \textit{output} block}\\
 \hline
\textbf{Sally}  & \textbf{1) State declaration:} A Sally state variable.\\
\cline{2-2}
\textbf{counterparts:}  & \textbf{2) State initialization:} Same value as the \textit{unit delay}'s initial value.\\
\cline{2-2}
& \textbf{3) State transitions:} Transition matrix in Sally language. \\
\hline
Note:&This structure as a whole corresponds to one single Sally state.\\
 \hlineB{3}
\multicolumn{2}{|l|}{\textbf{Case III:} A time-dependent block}\\
 \hline
\textbf{Sally} & \textbf{1) State declaration:} A Sally state variable.\\
\cline{2-2}
\textbf{counterparts:}  & \textbf{2) State initialization:} Same value as the block's initial value.\\
\cline{2-2}
& \textbf{3) State transitions:} Transfer function in Sally language. \\
\hline
Note:&\textbf{Case III} excludes the \textit{unit delay} block instances from \textbf{Case II} category.\\
 \hlineB{3}
\multicolumn{2}{|l|}{\textbf{Case IV:} A non-time-dependent block}\\
 \hline
\textbf{Sally} & 1) \textbf{State declaration:} A Sally state variable.\\
\cline{2-2}
\textbf{counterparts:}  & 2) \textbf{State  initialization:} None.\\
\cline{2-2}
& 3) \textbf{State transitions:} Block's math/logical function in Sally language. \\
\hline
Note:&A memoryless state has no initialization.\\
 \hlineB{3}
\end{tabular}
}
\caption{Text2Test internal model to Sally model translation.}
\label{sally translation rules}
\end{table}

Text2Test internal model responds to system input(s) instantaneous at the same time step, while in Sally model system input(s) are used to update system state(s) in the next time step. Therefore to eliminate the one-step response gap at the system inputs interface, a system input variable and an auxiliary input state variable are created in pair in Sally model , where the state is one step lag the input. Thus, the auxiliary input state(s) update simultaneously with all the other system states, all states as a whole exhibiting equivalent behavior of the corresponding Text2Test internal model.

\subsubsection{Checking specific properties against the Sally model}

\begin{figure}[!htb]
\centering
\includegraphics[scale=.27]{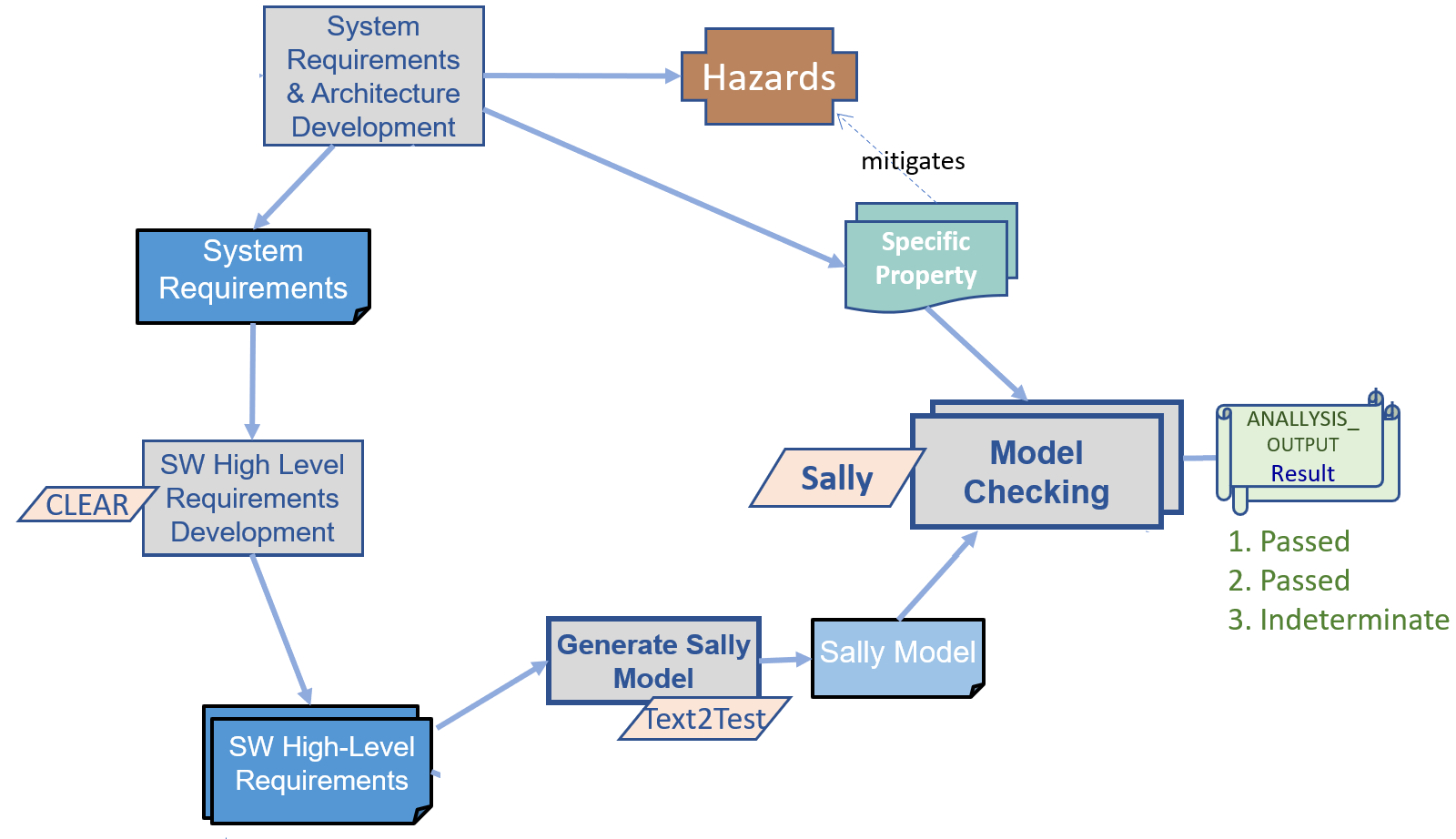}
\caption{Checking Specific Properties of Requirements using Sally}
\label{fig:specific_prop_req_sally}
\end{figure}

Figure~\ref{fig:specific_prop_req_sally} shows the evidence flow for checking a Specific Property against a Sally model (adapted from Figure~\ref{fig:specific_prop_req}). In a nutshell, the Sally model checker helps rigorously ensure that the specific properties (typically specified from system requirements to mitigate hazards) are satisfied by model (that is derived from the high-level functional requirements).

\paragraph{Specifying properties with system inputs}
The generic Sally query has one inconvenience on property specification with system inputs. In the query .mcmt file, an input condition cannot be directly encoded as part of the query formula, like the way that most other model checkers can.  Although Sally tool provides an assumption annex for the input conditions as a supplement to the query, so that the input conditions are treated as the extra system assumptions. But it is verbose and more importantly still does not address the issue that some property may have clause involving the relationship between input and state (something like ``State $s$ is always greater than input $x$.''). Nevertheless, in our Text2Test Sally integration, each system input has a corresponding auxiliary input state in the Sally model created during model translation (as shown in Table~\ref{sally translation rules} Case I). Therefore, it is strongly recommended to use the input state variable, rather than the input variable, in the property query formula whenever possible to be more flexible in expressiveness.

\subsubsection{Examples of Properties}

\paragraph{Mode-thrashing} Mode-thrashing is a hazardous phenomenon that often occurs in cases involving continuous value triggered mode switch. For example in the thermostat example in Section~\ref{sec:thermostat} Figure~\ref{fig:8var-thermo}, if the maximum sensing fluctuation (denoted by $\mathit{fluc}_{max}$) of the thermometer sensed temperature (denoted by $t_{sensed}$) is larger than the thermostat threshold margin (difference of ON/OFF temperature threshold values), then thermostat may send frequently oscillating ON/OFF control signals (denoted by $ctrl$) to the heater when the actual room temperature (denoted by $t_{room}$) is an unchanging value in-between or near the thermostat threshold values. In the extreme case where $\mathit{fluc}_{max}$ is as small as exact half of the threshold margin, mode-thrashing could still occur when $t_{room}$ happens to stay at the midpoint value of two threshold values.  Absence of potential mode-thrashing has an LTL formulation:
\begin{equation*}
(t_{room}-\mathit{fluc}_{max}\le t_{sensed}\le t_{room}+\mathit{fluc}_{max})\Rightarrow \neg (ctrl=\text{ON}~\textbf{X}~ctrl=\text{OFF}),
\end{equation*}
for all unchanging $t_{room}$, where \textbf{X} is the temporal operator ``next'' in LTL language. The encoding of this formula in Sally requires temporal extension to Sally generic query, which is elaborated in the following subsection~\ref{subsec:temporal extension}. The reason that here $\mathit{fluc}_{max}$ is \underline{not} naively compared with the thermostat threshold margin (or its half) is that, in a more general case, the continuous condition value and its fluctuation may go through some non-trivial transform before becoming triggering signal of the mode switch, so directly checking if undesired mode switch can occur is the only universal detection method. Detecting potential mode-thrashing is non-trivial but also not too complex to be encoded as a generic property of requirement set in Text2Test. Therefore, it is performed on both Text2Test internal model and the corresponding Sally model, the results are compared to showcase the semantic equivalence between two models.

%\textcolor{red}{Other examples. Need inputs from Anitha, also put forward reference to the evidence subsection for AFS}

\subsubsection{Temporal extension to Sally query}\label{subsec:temporal extension}
\paragraph{Motivation} Generic Sally query does not  allow basic temporal operators such as ``next.'' and ``prev.'', although ``next.'' is used in the Sally model script to denote the next time step. Queries are checked at all time step without an explicit temporal operator. This is to say, one cannot write a generic query formula about a state  $s$ in specific future or past time step(s). This is due to implementation limitation rather than the power of reasoning engines. A workaround solution is to create time-shifted auxiliary state variables (for instances, $prev\_s, next\_s$), and add  their proper declarations and state transitions into the Sally model, so that they can be used in the query. It is non-automatic, and it is tedious and error-prone in the case of specifying a property across multiple time steps that requires one auxiliary state variable for each related state at each time step. To enrich temporal logic semantics and take most advantage of the reasoning power, temporal extension to Sally query is developed as part of the tools integration.

\paragraph{Approach}Two basic temporal operators \textbf{X} and \textbf{F}, denoting ``next'' and ``eventually'' respectively, and time step syntax sugar are introduced to augment the query language. Let $t$ and $t'_{>t}$ denote the beginning and end time of the temporal domain, and $\mathit{generic}\_pred$ be an SMT Lib 2.0 format Boolean predicate of the state variables from the generic Sally model, we have the following general form of temporal extended predicates below:

\begin{itemize}
\item \textbf{X}($t,t'$][\textit{generic\_pred}]--- meaning that ``$\mathit{generic}\_pred$ holds for \underline{all} time steps in between $t$ (not included) to $t'$ (included).''\\
\item \textbf{F}($t,t'$][\textit{generic\_pred}]--- meaning that ``$\mathit{generic}\_pred$ holds for \underline{any} time step in between $t$ (not included) to $t'$ (included).''
\end{itemize}

Note that both $t$ and $t'$ are integer multiple of the system period. They can be negative, 0, or positive numbers, corresponding to past, current, or future time respectively.  A temporal extended predicate can be embedded in a larger query formula the same way a generic predicate does.  As a simple example, the property ``In system $sys$, when state $p$ is $true$, state $q$ shall be $true$ for the next 2 seconds.''  can be formulated as:

\begin{equation}\label{eq:extendedExample}
\mbox{(query }sys\mbox{ }(\Rightarrow\mbox{ }p\mbox{ }\textbf{X}[0,2][q])).
\end{equation}
\paragraph{Property translation and Sally model augmentation}A temporal extended property is firstly translated into an equivalent generic Sally query before Sally tool takes it as input. The translation is a straightforward process of temporal unfolding and (sometimes) shifting. Supposing system period is 1 second, Formula~\ref{eq:extendedExample} is unfolded to
\begin{equation}\label{eq:unfolded}
\mbox{(query }sys\mbox{ }(\Rightarrow\mbox{ }p\mbox{ (and }q\mbox{ }next\_q\mbox{ }next2\_q))),
\end{equation}
where $next\_q$ and $next2\_q$ are auxiliary state variables denoting 1 and 2 time steps forward shifts of $q$ respectively. Note the difference between the prefix ``\textit{next\_}'' in the auxiliary state variable name and the temporal operator ``next.''. In case that the temporal operator is \textbf{F} instead of \textbf{X}, Formula~\ref{eq:extendedExample} is unfolded to
\begin{equation*}
\mbox{(query }sys\mbox{ }(\Rightarrow\mbox{ }p\mbox{ (or }q\mbox{ }next\_q\mbox{ }next2\_q))).
\end{equation*}
Each of the newly created auxiliary state variables needs to be declared and given a state transition in the Sally model. The state transition is given by the form of assigning the next time step state value. While an auxiliary state variable of the \underline{past} time step can be easily assigned as ``(= next.$prev\_s$ state.$s$)'', it is not easy to assign a \underline{future} state variable without introducing more auxiliary variables than what are needed in the property. Naturally, entire Formula \ref{eq:unfolded} can be shifted 2 time steps towards past, resulting in the plain Sally query in the generic form:
 \begin{equation}\label{eq:shifted}
\mbox{(query }sys\mbox{ }(\Rightarrow\mbox{ }prev2\_p\mbox{ (and }prev2\_q\mbox{ }prev\_q\mbox{ }q))),
\end{equation}
where $prev\_q$ and $prev2\_q$ are auxiliary state variables denoting 1 and 2 time steps backward shifts of $q$ respectively.

\begin{figure}[!htb]
\centering
\includegraphics[width=\linewidth]{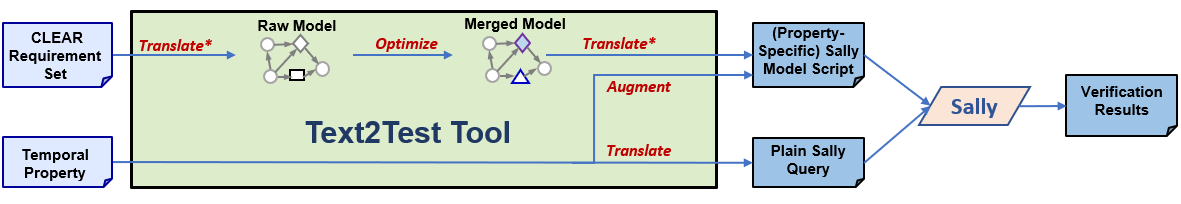}
\caption{Temporal property translation and Sally model augmentation.}
\label{fig:extended property check}
\end{figure}

Now, all state variables in Formula \ref{eq:shifted} are either on current or past time step. Their declarations and state transitions can be added to the original Sally model without introducing further more auxiliary state variables. The augmented Sally model is thereby a property-specific Sally model, because the choice of auxiliary state variables are property-specific. The entire process is done in an automatic fashion. Lastly, Sally tool verifies the property-specific Sally model against the plain Sally query. The complete data flow is summarized in Figure~\ref{fig:extended property check}.

\subsection{Test Generation from Requirements}\label{sub:testGen}

Figure~\ref{fig:t2t-tool-arch} provides an overview of the Text2Test tool. An important capability provided by Text2Test is the auto-generation of tests from a set of requirements \cite{ren-bhatt-2016-smt-hilite}.
Figure~\ref{fig:testgen-evidence-flow} shows the flow of evidence artifacts for test development from software high-level-requirements (HLR). The \emph{test development}, automated by Text2Test, generates two artifacts:
\begin{enumerate}
\item Test Oracles: The test obligations for specific behavior operator instances used in requirement clauses/subclauses across the requirement set.
\item Tests: Each test consists of a sequence test vectors containing input values and expected output values of the component under test. A test traces to the Test Oracles it satisfies.
\end{enumerate}

As described in Section~\ref{sec:t2t-internal-model}, the Text2Test tool creates an internal representation of the requirement set - the \emph{semantic synthesis model}. The nodes in this diagram are the \emph{behavioral operators} instances that are derived from the clauses/subclauses of all requirements in the set. Behavioral operators include Boolean logic operators, relational operators, mathematical operators and functions, time-based operators, selection operators, event-based operators, and state-changing operators.

This forms the basis for generation of test oracles and tests. As part of the Text2Test tool configuration, there is a formal definition of test oracle criteria and equivalence class definition of the required test obligation for each type of behavioral operator. The CLEAR testing theory, described in the next subsection.

\subsubsection{CLEAR Testing Theory}
\label{sec:clear-testing-theory}

\begin{quote}
  ``Testing can only reveal the presence of errors, not their absence.'' – Dijkstra (and DO-178C)
\end{quote}

\begin{figure}[!htb]
\centering
\includegraphics[scale=.9]{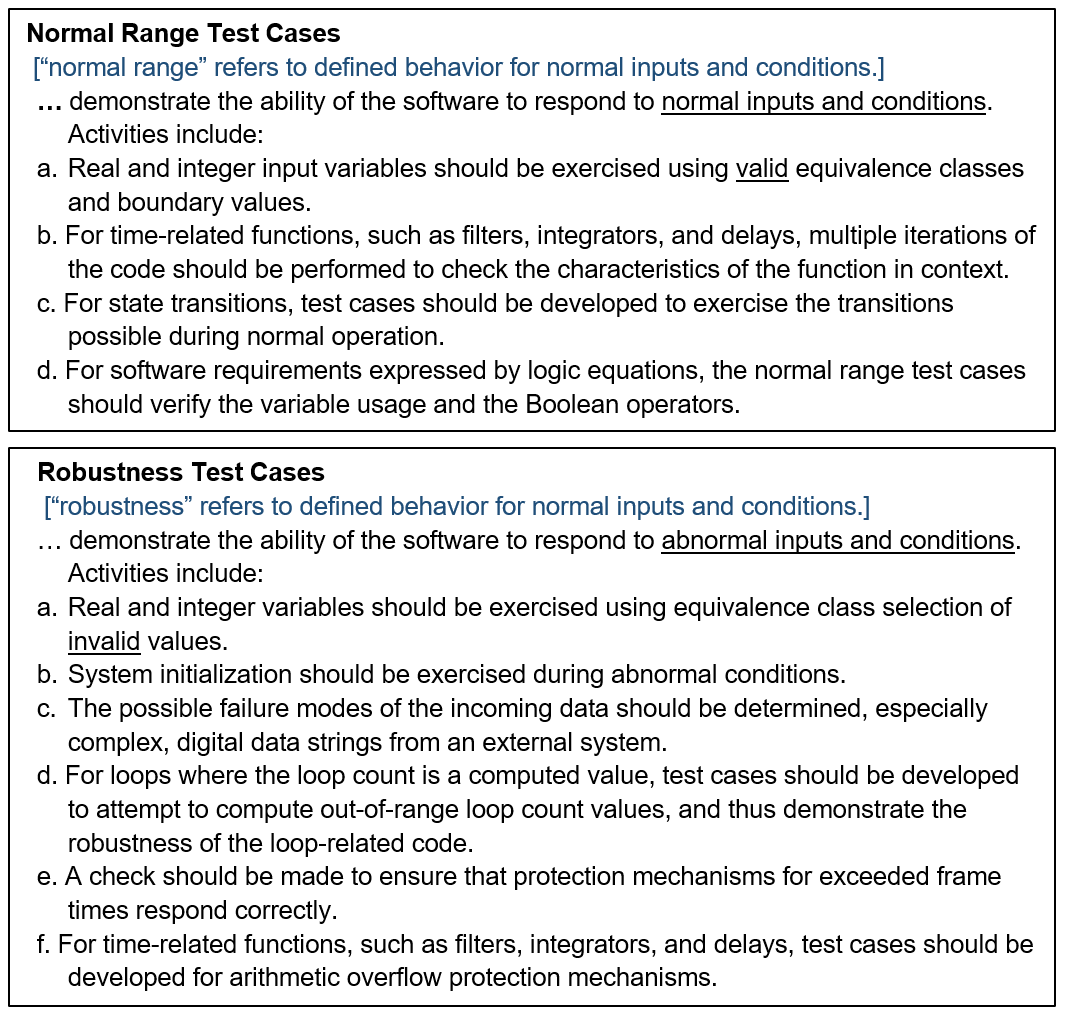}
\caption{Testing Criteria in DO-178C}
\label{fig:do178c-testing-criteria}
\end{figure}

Testing is inherently incomplete and cannot be formalized in logic. The challenge is to make testing more rigorous and bring some notion of “completeness.” One way to accomplish this is to base testing on some reasonable \emph{applicable criteria}. The CLEAR Testing Theory uses the well-established guidance in DO-178C~\cite{DO178C} to this end. DO-178C clearly establishes that tests must be based upon requirements and posits the criteria shown in Figure~\ref{fig:do178c-testing-criteria}, and brings notions of rigor and ``completeness'' within that framework.

The \emph{CLEAR Testing Theory} consists of a set of arguments and derivations to establish the claim that the \emph{Test Oracles} and \emph{Tests}, created for a given set of requirements in CLEAR notation, fully satisfy the ``applicable criteria'' for requirement-based testing for those requirements. A second claim is that a Test Harness correctly executes the Tests (and the implied test procedure) on the software component and produces pass/fail results of each test. The following is the summary of claims of CLEAR Testing Theory:

\begin{itemize}
\item Claim 1:
The Test Oracles and Tests satisfy all applicable criteria (e.g.: equivalence class / boundary value, time-related functions, state transitions) per DO-178C sections 6.4.2.1 and 6.4.2.2 for the set of requirements in CLEAR notation.

This claim is supported by the following subclaims:

Subclaim 1: A set of Test Oracle is created for each instance of an operator (behavioral sub-clause/subexpression) within a requirement in CLEAR notation.

Subclaim 2: The set of Test Oracles for an operator satisfies all applicable criteria (e.g.: equivalence class / boundary value, time-related functions, state transitions) per DO-178C sections 6.4.2.1 and 6.4.2.2.

Subclaim 3: A Test is created for each Test Oracle by 1) backward propagating the operator's inputs to component under test inputs, using principles of \emph{controllability}, and 2) by forward propagating the operator's output to an observable output of the component using principles of \emph{observability}.

\item Claim 2:
A Test Harness concretizes and correctly executes the Tests (and its implied procedure) on the target software and produces pass/fail results of each test case.
\end{itemize}

\begin{figure}[!htb]
\centering
\includegraphics[width=\textwidth]{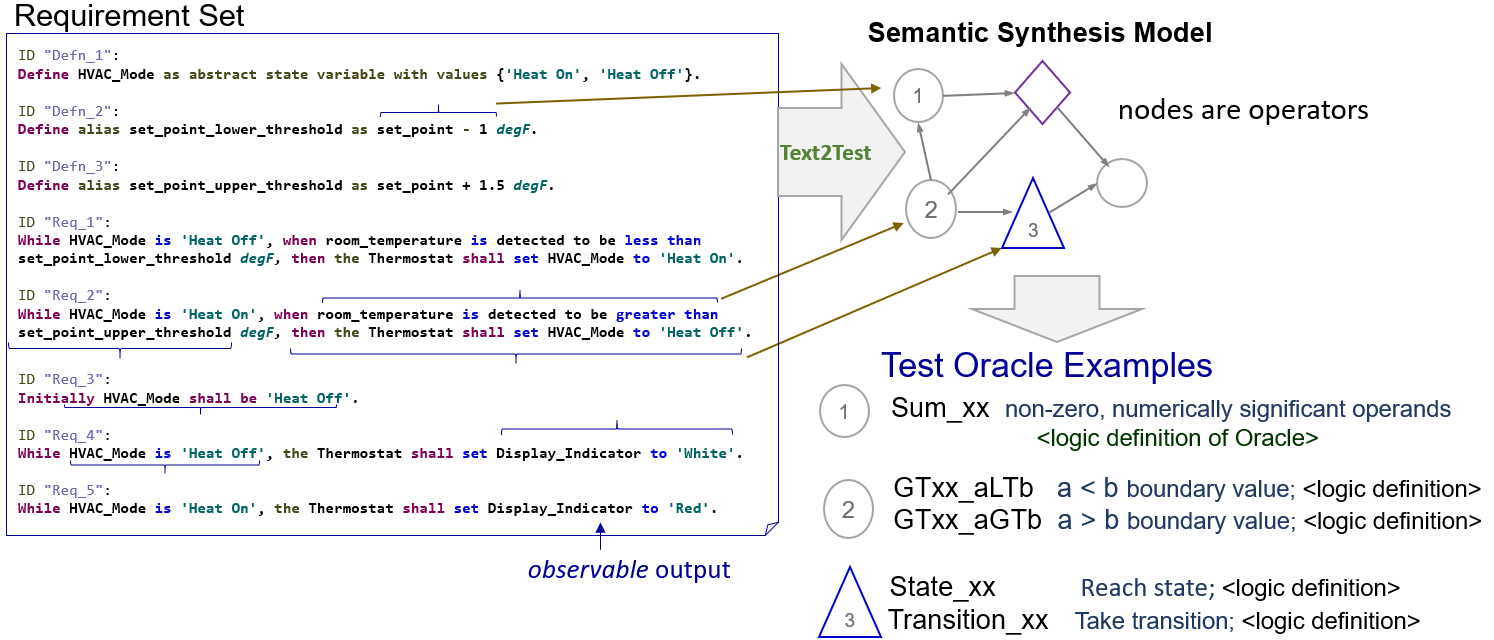}
\caption{Example of Derivation of Test Oracles from a Requirement Set}
\label{fig:test-oracle-examples}
\end{figure}

Figure~\ref{fig:test-oracle-examples} shows examples of derivation of test oracles from a requirement set using the Text2Test tool. As discussed previously in Section~\ref{sec:t2t-overview} and Figure~\ref{fig:t2t-tool-arch}, semantic transformations are applied to requirements to create a semantic synthesis model. The nodes in this model are the operators in the requirements (implied by natural language subclauses and mathmatical/logical subexpressions). For each operator, there is a set of test oracles, each oracle defines the equivalence class of input and output values for that a particular behavior of that operator.

\begin{figure}[!htb]
\centering
\includegraphics[width=\textwidth]{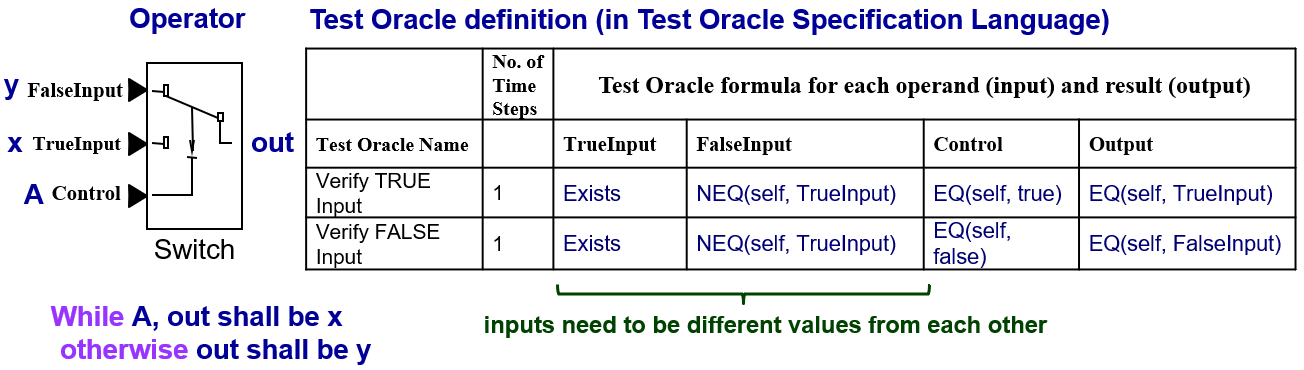}
\caption{Definition of Test Oracles for Switch Operator}
\label{fig:test-oracle-switch}
\end{figure}

\begin{figure}[!htb]
\centering
\includegraphics[width=0.5\textwidth]{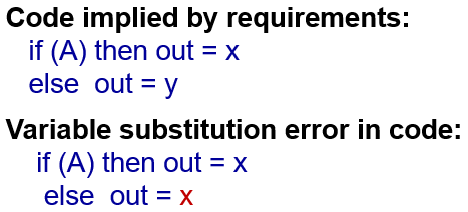}
\caption{Example of Code Error Caught by the Test Oracle of Switch Operator}
\label{fig:var-subst-error}
\end{figure}

Figure~\ref{fig:test-oracle-switch} shows the test oracle definition for a switch operator which represents the logic of a ``while ... otherwise ...'' requirement. The essential point to note here that the equivalence class definition specifies that the values at the two inputs of the switch (x and y in the requirement) need to be different so that one can ascertain that the proper branch in the code was chosen. Such test will detect the type of variable substitution error in the code shown in Figure~\ref{fig:var-subst-error}.

% Radler description goes here
\section{Architecture Specification and Analysis using Radler}\label{sec:RadlerMain}

The Radler Architecture Definition Language decomposes the design into a logical architecture
based on a pub/sub quasi-period model of computation.  Radler nodes execute periodically
and communicate with other nodes over bounded latency channels.  We are using the MAVROS
capability to support a Radler architecture for the ArduPilot.  In particular, the
ArduCopter Failsafe component will be implemented as an independent Radler node running
on a companion computer so that its failure modes are independent from the base platform.
The use of Radler also serves as a step toward the Phase 2 challenge problem where we have
to coordinate between multiple components while ensuring security properties.

We have the defined a Radler architecture for the ArduCopter with the AFS module running on an
independent \emph{companion} computer.  The AFS module uses MAVROS as the transport layer to subscribe to
messages from the ArduCopter module, the Ground Control Station (GCS), and the Remote Control (RC).
The AFS module receives status updates from the ArduCopter Base module and is able to change modes
and set certain flight parameters.  It is also able to annunciate warnings to the GCS and RC Pilot.

We have enhanced Radler to integrate nodes that are implemented in Java using the Java Native Interface.  We plan to use this capability
to define capabilities like logging and database access that leverage Java interfaces but are not
critical to the real-time responsiveness of the system.
We have developed a tutorial (both in the forms of a video and use case in Radler git repository) demonstrating the Radler code generation and its execution on SITL (software in the loop) simulator for the Arducopter AFS (advanced fail safe). In the tutorial, the Radler architecture consists of AFS gateway, battery, altitude, and log nodes communicating with Arducopter via MAVROS (ROS-based extendable communication node) on the companion computer.

We constructed a SITL/MAVROS/Radler deployment in a virtual machine environment using Vagrant and provided a sequence to create a pre-built image for the Boeing/TA4 Evaluation.  We are currently evaluating Java-based
runtime verification tools that we can use to monitor the behavior of the AFS module.

\subsection{The Radler Model of Computation}\label{sec:RadlerMoC}

A Model of Computation (MoC) specifies the execution of individual nodes in a distributed system
and their interaction through shared memory or message-passing channels.  
Radler implements distributed systems within a publish/subscribe architecture with a
quasi-periodic model of computation.  The nodes repeatedly execute a step function
with a minimum and maximum bound on the period between two successive execution.
The nodes communicate through topic channels with an associated message type.
Each topic has exactly one publisher node, but can have zero or more subscriber nodes.
Each node specifies a buffer size and a latency bound for the mailbox associated with
each of its subscription channels.   On each execution step, a node reads its input
mailboxes, applies its step function to these inputs, and then sends thes outputs to the
respective mailboxes for the topics on which it publishes.

\newcommand{\etime}[1]{\mathit{time}(#1)}
\newcommand{\evalue}[1]{\mathit{value}(#1)}
\newcommand{\estream}[1]{\mathit{stream}(#1)}
\newcommand{\rstream}[1]{\mathit{rstream}(#1)}
\newcommand{\estep}[1]{\mathit{step}(#1)}
\newcommand{\bstep}[1]{\mathit{bstep}(#1)}

There are several important theorems about RADL's  multi-rate, quasi-periodic model of computation that have been proved
in PVS\cite{conf/memocode/LarrieuS14}. These include: 
\begin{enumerate}
\item Bounded processing latency for message: A message sent by publisher node $P$ at time $t$ to subscriber 
node $S$ on topic $A$ with maximal latency $L(A,S)$ 
is processed by node $S$ within time $t + L(A, S) + \mathit{max}(S)$ unless it is superseded by a 
subsequent message from $P$\@.  The maximal delay occurs when a message sent by $P$ at time $t$ is received 
by $S$ at a time just after $\tau_S(i)$ and processed by $S$ at $\tau_S(i+1)$ where $\tau_S(i+1) - \tau_S(i) = \mathit{max}(S)$.  
\item No overtaking, with timing assumptions:  If $L(A, S) < \mathit{min}(P)$, then 
messages are received by $S$ in the order sent by $P$ since the $i$'th message from $P$
will be received by $S$ before the $i+1$'st message is sent.  
\item Bounded consecutive message loss: Assuming a buffer size of one, if $M$ is the small integer such that
$M. \mathit{min}(P) > L(A, S) + \mathit{max}(S)$, then 
at least one of $M$ consecutive messages is read by the subscriber (assuming no overtaking).  The fastest rate at which message 
can be sent is $1/\mathit{min}(P)$, and the 
maximum number of messages that can arrive in the
interval from $\tau_S(i)$ to $\tau_S(i+1)$ are those sent in the interval from $\tau_S(i) - L(A, S)$ to $\tau_S(i+1)$\@. 
\item Bounded queue length to eliminate message loss: Under the same assumptions as the previous bullet, 
with  a queue length of $Q$,  at most $M - Q$ consecutive messages are lost. 
\item \label{prop:bounded_message} Bounded age $\mathit{MA}(m)$ of a message input $m$ used by subscriber $S$ in a step function:
$\mathit{MA}(m) < L(A, S) + \mathit{max}(P)$ (without overtaking)\@.   The quantity $L(A, S) + \mathit{max}(P)$
is the biggest gap between $\rstream{P}(i)$ and $\rstream{P}(i+1)$\@.  
\end{enumerate}
Bounds can also be computed for the scenario where overtaking is possible~\cite{conf/memocode/LarrieuS14}\@.

\subsection{The Radler Architecture Definition Language}\label{sec:RADL}

A RADL description file is a text file, preferably encoded in utf8, with
extension ``.radl''. Such a file is called in short a radl file.
A radl file defines a module whose name is the file name. 
A file \texttt{MODULE}.radl defines an implicit scoping region for its
content whose name is \texttt{MODULE}.

\paragraph{Meta Syntax Rules}

A module is a set of value declarations. A declaration may be an alias, used to
declare a value equal to another one but with a new name. The name of a
value may be omitted, note that it'll prevent the user to reference this
value. The type of a value may also be omitted, note that if the type is
ambiguous, the inference will choose one. When defining a class field,
one may declare a new value or refer to another value by its identifier.

\paragraph{Grammar} 

A RADL specification is a set of typed
value definitions following this simple grammar:
\scriptsize
\begin{Verbatim}[frame=single,samepage=true,xleftmargin=5mm]
  module := decl*
  decl := alias | NAME? (':' TYPE)? TYPEVALUE | NAME? (':' CLASS)? class_value
  alias := NAME = identifier
  class_value := '{' field * '}'
  field := FIELDNAME value + 
  value := identifier | decl
  identifier := root_name | identifier '.' NAME
  root_name := NAME 
\end{Verbatim}
\normalsize

The two main classes used to describe the logical level are node and
topic. Their full definitions may be found in \url{https://github.com/SRI-CSL/radler/blob/master/radler/radlr/language.py}. 
The most important elements are:

\scriptsize
\begin{Verbatim}[frame=single,samepage=true,xleftmargin=5mm]
  class node
    PUBLISHES publication *
    SUBSCRIBES subscription *
    CXX cxx_class
    PERIOD duration
    WCET duration
  class topic
    FIELDS int8/uint8/int16/uint16/int32/uint32/int64/uint64/
           float32/float64/
           bool/struct/array/duration/time +
\end{Verbatim}
\normalsize

When creating a node, Radler will construct one instance of the provided
C++ class (the \texttt{CXX} field). The step method of this instance
will be called at a fixed frequency defined by the node's period (the
\texttt{PERIOD} field). This step function needs to be proven to have a
worst case execution time no greater than the one described here in the
\texttt{WCET} field. At each call, the step function is provided with
the messages received from its subscriptions and is required to write
the messages it has to publish.

A topic (defined uniquely by its name) is a purely logical way to define
point-to-point communications between one producer and multiple
consumers. There can be exactly one node publishing to a topic while
many nodes can subscribe to it.

Any user code needing libraries has to be declared as using a library
with the \texttt{LIB} field. This field allows two forms of libraries,
\texttt{cmake\_library} and \texttt{static\_library}:

\scriptsize
\begin{Verbatim}[frame=single,samepage=true,xleftmargin=5mm]
  class cxx_file
    PATH string ?
    FILENAME string *
    LIB cmake_library/static_library *
  class cmake_library
    PATH string ?
    CMAKE_MODULE string
    CMAKE_COMPONENTS string *
    CMAKE_VAR_LIBRARIES string ?      #defaults to {CMAKE_MODULE}_LIBRARIES    
    CMAKE_VAR_INCLUDE_DIRS string ?   #defaults to {CMAKE_MODULE}_INCLUDE_DIRS 
  class static_library
    PATH string ?
    HEADER_PATHS string *
    CXX cxx_file *
\end{Verbatim}
\normalsize

Static libraries are of the simplest form, gathering a set of source
files in their \texttt{CXX} field while the library header files are
found in the \texttt{HEADER\_PATHS} paths.
The cmake library enables the use of arbitrarily complex libraries since
it is a user defined cmake script. The \texttt{CMAKE\_MODULE} field
provides the cmake module name used to find it in the working directory.
\texttt{COMPONENTS} are the required components from this module, used
when calling the cmake \texttt{find\_module} command. 

When a value is type annotated, it is checked to be of this type. If no
type annotation is provided, the kind of the value is extracted from the
ones possible in the value declaration context. 
Subtyping allows for example to use a 16bits integer (\texttt{int16})
where a 32bits integer is required (\texttt{int32}). For now, subtyping
is only done on \texttt{sized\_types} by allowing a type to be used in a
bigger version. \texttt{sized\_types} and possible sizes are explicit in
the language. User values are checked to fit the type's size with the
function \texttt{check\_type\_size}.

\paragraph{User code}

Each node is a Mealy machine. The user provides a class which will be
instantiated with the default constructor to generate an instance
representing the state of the machine. Then, the \texttt{step} method of
this instance will be called to execute one step of the machine. The
signature of the \texttt{step} method is required to be:

\scriptsize
\begin{Verbatim}[frame=single,samepage=true,xleftmargin=5mm]
  void step(const radl_in_t* in, const radl_in_flags_t* inflags,
            radl_out_t* out, radl_out_flags_t* outflags)
\end{Verbatim}
\normalsize

The four argument types are structures defined in the generated header
file.
The input structure \texttt{radl\_in\_t} 
has one field per subscription of the node. The field
name is the radl name of the subscription. Each field is in turn a
structure reflecting the topic of the subscription, whose fields are the
name of the radl topic \texttt{FIELDS}.
The output structure \texttt{radl\_out\_t}
is similar to the input structure except that is used by the step
function to publish its publications. To this effect, the step function
has to fill the output structure.

\paragraph{Freshness and timeout}

Similar to the input and output structures, 
the flag structures \texttt{radl\_in\_flags\_t} and \texttt{radl\_out\_flags\_t} 
have a field for
each subscription and publication, respectively. 
The main idea of flags is to have some Boolean metadata attached to
messages, which by default propagate through nodes. To this effect, the
default value of the output flags of each publication of a node are set
to the logical OR of all the flags of its subscriptions.
Subsequently, the input flags are given to the step function as read-only while
the preset output flags are provided to the step function to give it a
chance to turn on or off the desired flags. Therefore, if the step function
has not change the output flags, they will propagate the input flags.

The input flag \texttt{radl\_STALE}
has the broad meaning that its associated value
is not ``fresh''. More precisely, it either means that the publisher of
the value flagged it as stale (by automatic propagation or by choice) or
that no new message arrived since the last call to the step function. In
the latter case, the step function gets the same input value (the
mailbox has not changed) but it is flagged as stale. To check if a
subscription \texttt{s} is stale, one simply calls
\texttt{radl\_is\_stale(in\_flags->s)} which returns a
Boolean.

The input flag \texttt{radl\_TIMEOUT}
has the broad meaning that its associated
value has violated the timing constraints. More precisely, it either
means that the publisher of the value flagged it as timeout (by
automatic propagation or by choice) or that we have not received a
message since period of the publisher plus the \texttt{MAXLATENCY}. In the latter
case, timing constraints are exceeded and something unexpected is
happening. To check if a subscription \texttt{s} is timeout, one simply
calls \texttt{radl\_is\_timeout(in\_flags->s)} which
returns a Boolean.

The function \texttt{radl\_turn\_on} is used to turn on a flag. For
example to turn on the stale flag of the publication \texttt{p} the following should be used:

\scriptsize
\begin{Verbatim}[frame=single,samepage=true,xleftmargin=5mm]
  radl_turn_on(radl_STALE, &out_flags->p);
\end{Verbatim}
\normalsize

To turn off flags, the similar function \texttt{radl\_turn\_off} should
be used.

\paragraph{Plant} 

The physical specification describes
the machines used in the realization of the overall system. A
physical description is provided by a value of type \texttt{plant}.
The most important field is the \texttt{MACHINES} field listing the
machines of the system. A machine is typically defined by the
operating system that it runs (\texttt{OS} field). 
%Other properties like
%processor kind and device physical port are not described here.
Their use is mostly geared towards system code generation
and configuration, done by the Radler tool.

\scriptsize
\begin{Verbatim}[frame=single,samepage=true,xleftmargin=5mm]
  class plant
    MACHINES machine *
  class machine
    OS linux/lynxsecure/certikos
  class linux
    NODES_UID uint16
    IP ip ?
    IMG string ?
    NODES node *
\end{Verbatim}
\normalsize

\paragraph{Software Build System}

The Radler build tool takes the architecture definition and individual
local functions as inputs and generate executables for the overall
system as output.
Radler synthesizes glue code for the communication
layer based on the logical description and binds it to the source
code for node functions to generate a code executable. In
addition, Radler utilizes the physical architecture part of
RADL to realize the final system. During the build process, it
can instrument the system so that platform assumptions such as
node periods and channel latencies can be checked at runtime.

On the implementation side, Radler uses ROS as a backend. Each node and communication channels are dynamically setup by ROS. 
Radler generates files from a radl file into a ROS catkin structure, then a call to \texttt{catkin\_make} will generate the executables as usual. If a module defines a plant, Radler generates a ROS launch file to run the requested nodes.  

\paragraph{The Example of Room Temperature Regulation}

\begin{figure*} [h]
\centering
\includegraphics[width=0.7\linewidth]{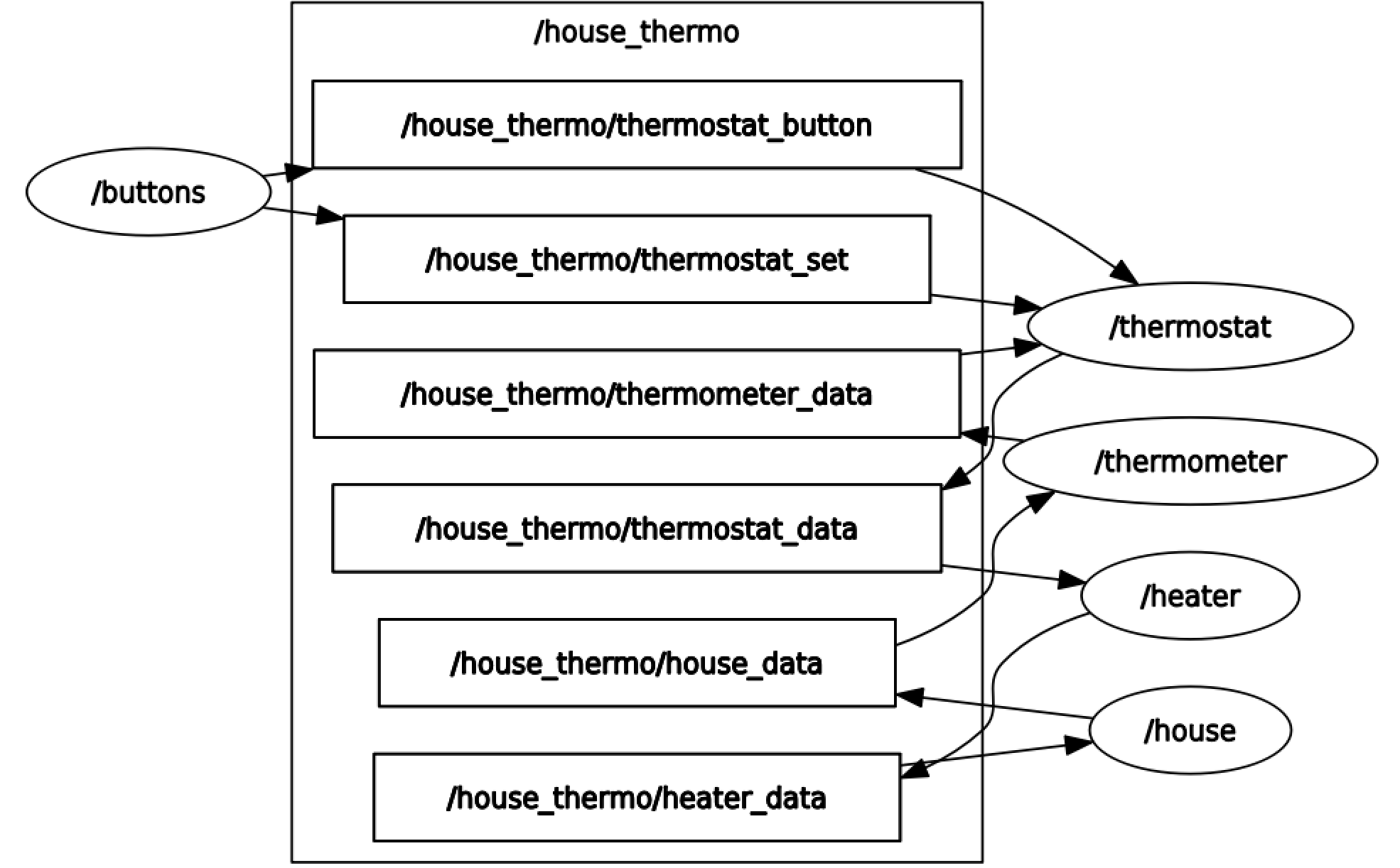}
\caption{Nodes (in ellipse) and topics (in rectangle) of the example of room temperature regulator.}
\label{fig:house_thermo_node_topic}
\end{figure*}

Figure \ref{fig:house_thermo_node_topic} shows the nodes and communications between them via topics for the example of room temperature regulator introduced in Section \ref{sec:thermostat}.

Five nodes (in ellipse) communicate through six topics (in rectangle). 
The \code{button} node publishes to \code{thermostat\_button} and \code{thermostat\_set} topics to generate signals that control the behavior of thermostat by means of switching on/off and setting the target temperature, respectively.
The \code{thermostat} node subscribes from \code{thermostat\_button}, \code{thermostat\_set}, \code{thermometer\_data} topics and publishes to \code{thermostat\_data} topic to regulate a room temperature within min/max. 
The \code{thermometer} node subscribes from \code{house\_data} topic and publishes to \code{heater\_data} topic to measure the room temperature with sensing noise. 
The \code{heater} node is the actuator that subscribes from \code{thermostat\_data} and publishes to \code{heater\_data}.
The \code{house} node provides the physical room temperature by subscribing from \code{heater\_data} and publishing to \code{house\_data} while the \code{thermometer} node provides the digital sampled sensed temperature.
A system consisting of these nodes was defined in a RADL description with user code in each step function. The Radler build process generates the glue code for scheduling, communications, and failure detections for executables. 

At the logical level, there are two main value types; node and topic. Excerpt from the example system's RADL description below:

\scriptsize
\begin{Verbatim}[frame=single,samepage=true,xleftmargin=5mm]
  basic_rate : duration 50msec

  thermometer_data : topic { FIELDS temp : float32 75 }
  thermostat_button : topic { FIELDS status : bool true }
  thermostat_set : topic { FIELDS temp : float32 75 }

  thermostat_data : topic { FIELDS switch_on : bool true }

  thermostat : node {
    SUBSCRIBES
      thermometer_temp { TOPIC thermometer_data MAXLATENCY 1msec }
      thermostat_switch { TOPIC thermostat_button MAXLATENCY 1msec }
      thermostat_set_temp { TOPIC thermostat_set MAXLATENCY 1msec }
    PUBLISHES
      heater_switch { TOPIC thermostat_data }
    PATH "src"
    CXX
      { HEADER "thermostat.h" FILENAME "thermostat.cpp" CLASS "Thermostat" }
    PERIOD basic_rate
  }
\end{Verbatim}
\normalsize

A node is described with fields such as \texttt{PERIOD}, \texttt{PUBLISHES}, and \texttt{SUBSCRIBES}. 
When the \code{thermostat} node is created, Radler constructs one instance of the provided C++ class specified in the \texttt{CXX} field. The step function of this instance will be called at a fixed frequency defined by the node's period (the \texttt{PERIOD} field). At each call, the step function is provided with the messages received from its subscriptions and is required to write the messages that it has to publish (the \texttt{SUBSCRIBES} and \texttt{PUBLISHES} fields).
A topic is uniquely defined by its name. 
For example, \code{thermometer\_data}, \code{thermostat\_button}, \code{thermostat\_set} topics are referenced as \code{thermometer\_temp}, \code{thermostat\_switch}, \code{thermostat\_set\_temp}, respectively.  
The \code{thermostat} node publishes to the \code{thermostat\_data} topic referenced as \code{heater\_switch} and \code{heater} node subscribes from it. 
A topic is a purely logical way of defining point-to-point communications between one producer and multiple consumers. That is, there can be exactly one node publishing to a topic (e.g., \code{house\_data}) while many nodes (e.g., \code{thermometer\_1} ... \code{thermometer\_n} in case of multiple sensors) can subscribe from it. 
The communication occurs via bounded latency channel (the \texttt{MAXLATENCY} field) for each topic.   
%Any user code can declare its dependency to libraries in the \texttt{LIB} field.
 
Below code segment shows the step function of the \code{thermostat} node, that is provided by the user (C++ class specified in the \texttt{CXX} field under the directory specified in the \texttt{PATH} field, as being exemplified in the RADL description above). 

\scriptsize
\begin{Verbatim}[frame=single,samepage=true,xleftmargin=5mm]
  void Thermostat::step(const radl_in_t* in, const radl_in_flags_t* inflags,
                        radl_out_t* out, radl_out_flags_t* outflags)
  {
    // change the set temperature 
    this->set_temp = in->thermostat_set_temp->temp;

    // set the status
    this->status = in->thermostat_switch->status;

    // decide whether to switch on the heater
    if (in->thermometer_temp->temp > (this->set_temp + this->tol)) {
      out->heater_switch->switch_on = false;
    } else if (this->status && (in->thermometer_temp->temp < this->set_temp)) {
      out->heater_switch->switch_on = true;
    } else {
      out->heater_switch->switch_on = this->status;
    } 
  }
\end{Verbatim}
\normalsize

A class will be instantiated with the default constructor to generate an instance representing the state of the Mealy machine. Subsequently, the step function of this instance will be called to execute one step of the machine. The signature of the step function should specify 
 the input (\texttt{radl\_in\_t*}) and output (\texttt{radl\_out\_t*}) structures define the node's subscription and publication, respectively. 
In the example, the step function of the \code{thermostat} node regulates the room temperature by switching on/off state of the heater.
The flag structures (\texttt{radl\_inflags\_t*, radl\_outflags\_t*}) can be used to check if a subscription, e.g., \code{thermometer\_temp}, is stale or timeout by calling \texttt{radl\_is\_stale(iflag->thermometer\_temp)} or \texttt{radl\\\_is\_timeout(iflag->thermometer\_temp)}, respectively.  These Boolean metadata attached to messages are by default propagate through nodes unless the explicitly being turned off (\texttt{radl\_turn\_off}).

A physical specification is provided by a value of type plant:

\scriptsize
\begin{Verbatim}[frame=single,samepage=true,xleftmargin=5mm]
  sys1 : linux {
    IP 192.168.1.201
    NODES heater
  }

  sys2 : linux {
    IP 192.168.1.202
    NODES thermostat
  }

  sys3 : linux {
    IP 192.168.1.203
    NODES thermometer house buttons
  }

  plant : plant {
    MACHINES
      house_heater { OS sys1 }
      house_computer { 
        OS l1 : lynxsecure {
          VMS
            vm1 { OS sys2 }
            vm2 { OS sys3 }
        }
      }
  }
\end{Verbatim}
\normalsize

The \texttt{MACHINES} field lists the machines that will be used by the system. In the example, three machines are defined by the
operating system that it runs (the \texttt{OS} field). The \texttt{IP} and \texttt{NODES} fields specify the IP address assigned for the OS and the nodes that run on the OS.  
In the example, three nodes (i.e., \code{thermometer}, \code{house}, \code{buttons}) run on the Linux with IP address of \code{192.168.1.203} while \code{heater} and \code{thermostat} nodes  run on a separate Linux machine. 
SRI's open-source release \url{http://radler.csl.sri.com} under permissive license together with documentations also provides use-cases for the heterogeneous platforms including AR.Drone, ArduPilot, RaspberryPi, and Android device. 
Some use-cases work with both core ROS and secure\_ros (SRI's fork of core ROS packages to enable secure communication among ROS nodes) \url{http://secure-ros.csl.sri.com}.

\paragraph{Inception of Java}

Radler supports embedding for the existing Java components using the Java Native Interface (JNI). 
An example of such existing Java components is BeepBeep stream processing engine \url{https://liflab.github.io/beepbeep-3} that allows processing log data and live event feeds. 
Specifically, BeepBeep can provide Radler with runtime monitoring capabilities by  
analyzing and transforming the event stream through a chain of basic event processors. 
Values subscribed from the topics can be piped as input to the stream processing engine.
Similarly, the realtime output event streams can be piped to the topics to be published. 

A simple use-case below shows a node monitoring point distance between two (x,y) positions.   
A radl file should include \texttt{cmake\_library} information for JNI module, which is passed as \texttt{LIB} field in the node description: 

\scriptsize
\begin{Verbatim}[frame=single,samepage=true,xleftmargin=5mm]
  jni : cmake_library {
    CMAKE_MODULE "JNI"
    CMAKE_VAR_LIBRARIES "JNI_LIBRARIES"
    CMAKE_VAR_INCLUDE_DIRS "JNI_INCLUDE_DIRS"
  }
  monitor : node {
    ...
    CXX { HEADER "monitor.h" FILENAME "monitor.cpp" CLASS "Monitor" LIB jni }
  }
\end{Verbatim}
\normalsize

Java Virtual Machine (JVM) creation should be in the class constructor by calling \texttt{JNI\_CreateJavaVM} with \texttt{JavaVMOption} to load and initialize the JVM. 
JNI calls, 
e.g., \texttt{FindClass}, \texttt{GetStaticMethodID}, \texttt{CallStaticVoidMethod},
are made inside of the step function. 
In the step function, a class loader (\texttt{FindClass}) has to find the right class by searching for an appropriate .class file 
that was provided at JVM initialization (\texttt{-Djava.class.path}). 
The class is then passed to \texttt{GetStaticMethod} to find the method of the class with the given signature. 
The method can be called with \texttt{CallStaticVoidMethod} since a static method is independent of any object. 

\scriptsize
\begin{Verbatim}[frame=single,samepage=true,xleftmargin=5mm]
  Monitor::Monitor()
  {
    ...
    JavaVMOption* options = new JavaVMOption[2];
    options[0].optionString = "-Djava.class.path=/path/to/BeepBeepCore&Plugin"
    vm_args.options = options;
    jint res = JNI_CreateJavaVM(&vm, (void **)&this->env, &this->vm_args);
  }

  void Monitor::step(const radl_in_t* in, const radl_in_flags_t* inflags, 
                     radl_out_t* oout, radl_out_flags_t* outflags) 
  {
    ...
    double x1 = in->position->x;
    double y1 = in->position->y;
    jclass cls = env->FindClass("PointDistance");
    jmethodID mid = env->GetStaticMethodID(cls, "getPointDistance", "(FFFF)V");
    env->CallStaticVoidMethod(cls, mid, 
                              jfloat(x1), jfloat(y1), jfloat(x2), jfloat(y2));
    ...
  }
\end{Verbatim}
\normalsize

% Randoop, Daikon, and CheckerFramework are described here with respect to the DesCert context
\section{Lightweight Code Analysis Tools}

% \textcolor{red}{TODO: need more content in this section}

The University of Washington research team led by Professor Michael Ernst have developed a suite of
lightweight analysis tools for Java.  These tools are used extensively in industry for checking
large codebases for bugs and software quality assurance.  The three main tools that we integrate in
the DesCert project are
\begin{enumerate}
\item The Randoop tool for synthesizing unit tests
\item The Daikon tool for the dynamic detection of program assertions
  \item The Checker Framework for pluggable and extensible typechecking
  \end{enumerate}

  These tools can be used for finding bugs as well as establishing generic and specific properties of
  code components.

  \subsection{Randoop: Feedback-Directed Random Test Generation}

  Randoop synthesizes unit tests for Java classes in the JUnit format~\cite{pacheco2007randoop}\@.
  It uses feedback-directed random test generation to build and compose sequences of
  constructor operations and method calls to explore portions of the state space that
  na\"{\i}ve test generation might not cover.  The unit tests synthesized by Randoop
  can be used to uncover and fix bugs or as regression tests that can verified whenever
  the software changes.  Randoop has been applied to large codebases and has successfully
  uncovered bugs in widely-used libraries including Sun's and IBM's JDKs and a core .NET component.

  Figure~\ref{fig:randoop} shows an example of an error-revealing unit test generated by Randoop
  in the OpenJDK library.  This test demonstrates the construction of a set \texttt{s1} on which 
  the generated assertion for the reflexivity of the equality test fails.  The unit test requires
  the construction of a \texttt{list} object
  and a \texttt{TreeSet} object (containing the list object) to which we apply the reflexive equality test.
  \begin{figure}
    \centering
  \begin{smallsession}
    // This test shows that the JDK collection classes
  // can create an object that is not equal to itself.
  @Test
  public static void test1() {

    LinkedList list = new LinkedList();
    Object o1 = new Object();
    list.addFirst(o1);

    // A TreeSet is an ordered collection. According to the API
    // documentation, this constructor call should throw a
    // ClassCastException because the list element is not Comparable. But
    // the constructor silently (and problematically) accepts the list.
    TreeSet t1 = new TreeSet(list);

    Set s1 = Collections.synchronizedSet(t1);

    // At this point, we have successfully created a set (s1)
    // that violations reflexivity of equality: it is not equal
    // to itself! This assertion fails at run time on OpenJDK.
    org.junit.Assert.assertEquals(s1, s1);
  }
  \end{smallsession}
    \caption{A Randoop Unit Test Example}
    \label{fig:randoop}
  \end{figure}

  Tests can reveal errors by failing generated assertions,  violating explicit contracts, or throwing unexpected
  exceptions.  The classification of test results  as errors or expected behavior is under user control since
  in many cases, the thrown exceptions might be appropriate given the inputs.  The generated tests can also be minimized
  to avoid redundant or irrelevant steps.  

  As described in Chapter~\ref{chap:baseline}, we have developed a Baseline DesCert plug-in for Randoop to
  collect the error-revealing and regression tests for Java class libraries.  These test suites are
  maintained as evidence artifacts along with the test analysis results.

  \subsection{Daikon: Dynamically Detection of Program Invariants}

  Unit tests such as those generated by Daikon can be used to generate traces for Java programs.  These traces
  can be mined for assertions that hold at specific program points including loop invariants and  
  preconditions and post-conditions for method calls.  The Daikon tool dynamically analyzes program behaviors
  to automatically learn useful assertions~\cite{ErnstPGMPTX2007}\@.  It is important to remember that these assertions might not
  valid since the only hold on the test runs, but in many cases, the information generated by Daikon is
  quite helpful.  Consider, for example,  a Java class that implements a stack with \texttt{push},
  \texttt{pop}, \texttt{top}, \texttt{topAndPop}, \texttt{isEmpty},  \texttt{isFull}, and \texttt{makeEmpty} methods,
  using an array representation for the stack content and a \texttt{topOfStack} slot that is \texttt{-1} when the
  stack is empty.  Daikon can detect that the array component is never equal to null and that the
  \texttt{topOfStack} slot is always at least \texttt{-1}\@.  It can also detect that the size of the array
  is always one more than the value of the \texttt{topOfStack} field.  These are indeed valid invariants.
  Daikon operates by generating and testing a number of assertions.  The generation of assertions is
  smart and based on instrumentation to track the interaction between values in a program.
  For instance, it will check if an array is sorted when the elements in the array look comparable,
  and the sortedness property appears to be relevant to the behavior of the program.  Once a set of useful putative
  invariants have been identified, other tools, including the Checker Framework can be used to verify if 
  the invariant is in fact valid.
  
  \subsection{The Checker Framework: Pluggable Extensible TypeChecking}

  Java is a strongly typed object-oriented programming language, but a number of common errors
  are not caught by the Java compiler.  Java 8 adopted the Type Annotation Specification
  as a notation for inserting annotations into comments.   The Checker Framework uses the
  Type Annotation Specification to support pluggable typecheckers for additional program properties~\cite{papi2008practical,dietl2011building}\@.
  One such example is the \emph{nullness} checker where the \texttt{System.out.println} method is invoked
  on a possibly null object \texttt{myObject.toString}\@.  The nullness checker flags this inconsistency.
  Fields and variables can be annotated with \texttt{@NonNull} types, but these can also be derived by the
  type inference tool which saves the labor of hand-annotating the code with types. 
  \begin{figure}[htb]
    \begin{smallsession}
public class NullnessExample {
    public static void main(String[] args) {
        Object myObject = null;

        if (args.length > 2) {
            myObject = new Object();
        }
        System.out.println(myObject.toString());
    }
}
    \end{smallsession}
    \caption{Checker Framework Example: NullnessExample.java}
    \label{fig:nullness}
  \end{figure}

  The Checker Framework has been extended with a number of other checkers.  The Regex checker ensures
  that regular expressions strings are actually well-formed regular expressions.   The Taint checker
  ensures that tainted user input does not pollute the input to sensitive operations such as a
  database query on a malicious query string.  The Encryption checker uses the \texttt{@Encrypted}
  annotation tag to ensure that sensitive data is encrypted when placed on a publicly visible
  channel.  The Checker Framework ships with over twenty five custom type checkers for properties such as
  initialization, resource leaks, locks, bounds of index variables, purity (absence of side-effects),
  and units of measurement.  There are also around twenty checkers that have been developed by third parties
  spanning information flow, determinism, immutability, and typestate.  Many of these are examples of
  Ontic type systems that are relevant to the DesCert project.

  The Checker Framework was applied BeepBeep3 runtime monitoring library~\cite{halle2017event} which is
  used to monitor runtime safety within our ArduCopter AFS subsystem.  The analysis revealed a number of issues.
  For example, it revealed an inconsistency in annotation of the \texttt{getProvenanceTree} which is
  declared to return a non-null result in one file but is defined to return a \texttt{null} result in another.
  These flaws have been reported to the developers of the BeepBeep3 tool.

%SeaHorn is described here. 
\section{SeaHorn Tool for Static Code Analysis}

% I mainly need a section on code analysis and code contract checking
% focusing on how you generate evidence.  This project is a real opportunity to define evidence
% formats for independently checkable witnesses from SAT, SMT, Horn-solving, static analysis, etc. 
% You can also write something about the EBPF verifier from this perspective.

% \textcolor{red}{TODO: need content in this section}

SeaHorn~\cite{seahorn} is a verification framework for LLVM-based
languages. SeaHorn performs a whole-program points-to
analysis~\cite{seadsa} to translate LLVM bitcode into a set of
verification conditions (VCs) whose satisfiability implies that the
program is safe. Currently, SeaHorn focuses on reachability
properties. Specifically, SeaHorn checks for predefined errors such as
memory errors and user-defined assertions. The details of these VCs
and how SeaHorn checks for predefined errors and assertions vary
depending on the back-end solver. SeaHorn provides three main back-end
solvers:

\begin{itemize}

\item a model-checker, called Spacer~\cite{KomuravelliGC14}, based on
  Property-Directed Reachability (PDR)~\cite{bradley2011sat} that
  produces (a) proofs, consisting of loop invariants and procedure
  summaries, (b) counterexamples if the program cannot be proven safe,
  or otherwise, (c) it returns ``unknown''. For practical reasons,
  Spacer does not precisely model certain aspects of the semantics of
  the program. For instance, \texttt{malloc} is modeled as a function
  that returns non-deterministically a pointer, ignoring important
  details such as alignment and memory layout. Alternatively, SeaHorn
  uses Sally~\cite{jovanovic2016property} if all functions and
  procedures can be inlined since Sally cannot model them. The benefit
  of using Sally is in the use of PDR combined with k-induction,
  a powerful technique for inferring loop invariants that Spacer does
  not support.

\item a bounded-model checker (BMC) that models more precisely the
  semantics of programs (including memory allocation) at the expense
  of producing bounded proofs (i.e., proofs that only hold for a
  finite number of loop iterations) but it is very effective at
  finding bugs in programs. 

\item a static analysis based on Abstract
  Interpretation~\cite{Cousot_POPL77}, called Crab, that produces
  proofs, consisting also of loop invariants and procedure summaries
  but it cannot  produce counterexamples. Due to the nature of
  abstractions, Crab is often more efficient than Spacer/Sally and BMC
  but it can produce many more false positives. Crab provides a rich
  set of fixpoint solvers, abstract domains and analyses. For
  instance, Crab provides a Memory analysis that can prove absence of
  memory errors such as null dereferences and buffer overflows.
\end{itemize}  

%% SeaHorn implements several communication mechanisms between all the
%% back-end solvers. For instance, Crab is used to provide loop
%% invariants to Spacer, discharging safe checks so that Spacer, Sally,
%% and BMC do not need to prove them, and providing approximate sizes of
%% memory objects to BMC.

In this phase, we have focused on Crab, the SeaHorn abstract
interpreter. More specifically, we have focused on the following tasks:

\begin{itemize}
\item Improve Crab capabilities to perform analysis of relevant C
  projects.

\item Implement a new Tag analysis that models many interesting ontic
  properties.

\item Start developing evidence formats (i.e., certificates) for
  independent checkers.
\end{itemize}

\subsection{Progress on Analysis of C code}
We have developed a new memory model, called \emph{region-based memory
  model (RBMM)}, that enables more efficient analysis of C code with
few restrictions with respect to the standard C memory model. This
work has been published in~\cite{crabir}.

Standard C memory model partitions memory into memory objects. A
pointer points to an offset within a memory object. RBMM further
partitions memory into regions. A region can span multiple
objects. The key difference is that pointers pointing to different
regions cannot alias even if they point to the same memory
object. RBMM makes two key assumptions in order to allow efficient
analysis of C programs: (a) matched pairs of memory writes-reads must
access the same number of bytes, and (b) it assumes that programs do
not have undefined behavior (UB). Assumption (a) excludes non-portable
code. Although it may seem counter-intuitive, assumption (b) does not
limit our analysis from proving absence of UB.  As shown by Conway et
al.~\cite{ConwayDNB08}, conditionally sound analyses can prove absence
of errors (e.g., memory violations) or otherwise, produce one
counterexample although it cannot produce all possible
counterexamples. In summary, the approach described in~\cite{crabir}
works as follows:

\begin{enumerate}
\item apply SeaHorn whole-program pointer analysis on the LLVM program so
  that memory is statically partitioned into regions. The analysis
  also identifies which parts of the program might not satisfy the
  assumptions of RBMM;

\item translate the LLVM program into a novel
  intermediate-representation (IR), called CrabIR, where all LLVM
  memory instructions are translated to instructions over regions and
  references;

\item perform abstract interpretation on CrabIR using Crab fixpoint
  solvers and abstract domains.
\end{enumerate}

In~\cite{crabir}, we apply Crab using our new region-based aware IR
(CrabIR) on five popular C projects: \texttt{bftpd}, \texttt{brotli},
\texttt{curl}, \texttt{thttpd}, and \texttt{vsftpd}. Sizes vary from
5K to 50K lines of code. The results show that Crab can prove that
around $60\%$ of all non-trivial\footnote{A memory dereference is
  considered trivial if LLVM analyses can already prove that it is not
  null (e.g., global variables).}  pointer accesses cannot be
null. This number is promising considering that the environment for
these programs is conservatively ignored and no specialized abstract
domains are used.

\subsection{New Tag Analysis}
In this phase, we have also developed a new Tag analysis in Crab,
where memory locations can be tagged with a numerical identifier
(i.e., tag) and then the Crab forward analysis propagates those tags
through memory. Upon completion of the analysis, Crab clients can ask
whether a memory location is tagged with a particular set of
tags. Taint analysis is an example of tagging.

As a pilot study, we have applied the Tag analysis on \texttt{thttpd}
(8K lines of code) in order to perform taint analysis where sources
are systems calls such as \texttt{read} and \texttt{mmap} and sinks
are systems calls such as \texttt{write} and \texttt{writev}. Each
source and each sink is assigned a different tag. The analysis runs in
few seconds and infer 24 of the 72 relationships between the 7 sources
and 8 sinks in \texttt{thttpd}. In the next phase, we plan to focus on
more specific ontic properties (e.g., security related properties) and
apply the Tag analysis on a broader set of relevant programs.

\subsection{Progress on Evidence Formats for Independent Checkers}

We have developed an approach that uses program invariants produced by
an abstract interpreter as evidence that can be checked by independent
checkers. We plan to start implementing a prototype in the context of
eBPF programs.

Any abstract interpreter produces program invariants at each basic
block or program location. More precisely, an abstract interpreter
produces an \emph{invariant map}, $InvMap$, that maps a location $l$
to $Inv_{l}$, a formula defined on a subset of program variables,
expressed in the restricted form allowed by the abstract domain
($Inv_l: AbsState$). The formula $Inv_{l}$ expresses a set of facts
that holds at location $l$ expressible in the particular abstract
domain.

In our approach, the tuple $\langle P, \sqcup, \Rightarrow,
\mathsf{TrFn}, InvMap \rangle$ constitutes the \emph{certificate} that
needs to be checked, where $P$ is the program, $\Rightarrow: AbsState
\times AbsState \mapsto Bool$ is the abstract implication, $\sqcup:
AbsState \times AbsState \mapsto AbsState$ is the abstract join, and
$\mathsf{TrFn}: AbsState \mapsto AbsState$ represents the abstract
transfer functions for each transition from location $l'$ to $l$.

Then, the checker receives a certificate and it must verify for each
location $l$ in $P$ that:

\[ \bigsqcup_{l' \in \mathsf{pred}(l)} (\mathsf{TrFn}^{l'~\rightarrow~l}(InvMap(l'))) \Rightarrow InvMap(l) \]

%% FIXME: we are abusing \Rightarrow as abstract implication. Better
%% use a different symbol. Note that it's not sqsubseteq

\vspace{2mm}
\noindent where $\mathsf{pred}$ returns the predecessors of a given
location.

Note that any other abstract interpreter different from the one that
generated the certificate can be used as checker.  Another possibility
is to translate program invariants ($AbsState$) into some first-order
logic fragment expressed by some combination of SMT theories so that
the checker can be replaced by a SMT solver. This would need also to
replace $\bigsqcup$ with logical disjunction, abstract implication
with logical implication, and the abstract transfer function
$\mathsf{TrFn}^{l'~\rightarrow~l}$ with the logical encoding of each
transition $l' \rightarrow l$.

We plan to implement the ideas from this section in the context of
eBPF programs. An eBPF program is a bytecode program that can be
executed in the kernel without recompiling the kernel. The price to
pay for such as flexibility is that the kernel needs to verify that
the program is memory safe before executing it. In a previous
work~\cite{prevail}, we developed an abstract interpreter based on
Crab, called Prevail, that outperforms the existing verifier in the
Linux Kernel. The main limitation is that Prevail must be run in user
space. Recently, Windows OS has adopted the eBPF technology and chosen
Prevail as the verifier. Currently, Windows OS runs Prevail in a
secure environment which is not an ideal solution either. To solve
this problem, we plan to use Prevail to generate certificates in user
space and implement a lightweight checker more suitable to be run in a
secure environment.

% \chapter{Tool Qualification}\label{chap:toolqual}%move to subsection in conops.
% \input{toolqual}

\chapter{Continuous Assurance Workflow: Baseline DesCert} \label{chap:baseline}%Huascar
A growing trend in software engineering is continuous development --- a practice
where software updates are made continuously, piece-by-piece, rather than in one
large batch. Continuous development considers the entire software life-cycle,
within which other continuous activities can be positioned, e.g., continuous
planning\cite{knight2001casper}, continuous
integration\cite{staahl2014modeling}, continuous
reasoning\cite{o2018continuous}, and so forth. Software projects that can
successfully implement continuous development can anticipate strategic
benefits\cite{fitzgerald2017continuous}, including:

\begin{enumerate}
  \item Maximize engineering productivity,
  \item Run more experiments before a general release, and
  \item Fix errors faster.
\end{enumerate}

The way humans assess software artifacts throughout the entire software
life-cycle, from system requirements to source code, is interactive. It
comprises a sequence of steps aimed at making sure that properties of these
artifacts meet certain assurance objectives (e.g., speed never exceeds more than
120 miles/hr). These steps are inter-connected and thus their execution closely
resembles the concept of continuous development. We believe that \textit{if
property-based assurance of software artifacts is done in a continuous fashion,
where formal methods can automatically establish properties of a set of
artifacts that we can prove using a program analysis tool, then we can
anticipate the same quality and consistency benefits of continuous development}.
Continuous assurance is done with artifacts changes, and will feed back
information to developers in tune with their development workflows.

In the following sections, we describe the Baseline DesCert workflow, a tool for
the continuous assurance of software as part of a build process.

\section{Continuous Assurance with Baseline DesCert}
\label{sec:baselinebackground}

In this section, we describe the \textit{Baseline DesCert workflow}, a tool for
monitoring and maintaining the status of the assurance artifacts during
continuous assurance. This tool also interacts the Rack system in order to
curate and ingest assurance artifacts in accordance with the DesCert ontology (see
Section~\ref{chap:ontology}).

% \textit{Baseline DesCert} is a family of Gradle plugins for the continuous
% assurance of software as part of a build process.

\begin{figure}[h!]\centering
  \includegraphics[width=13cm]{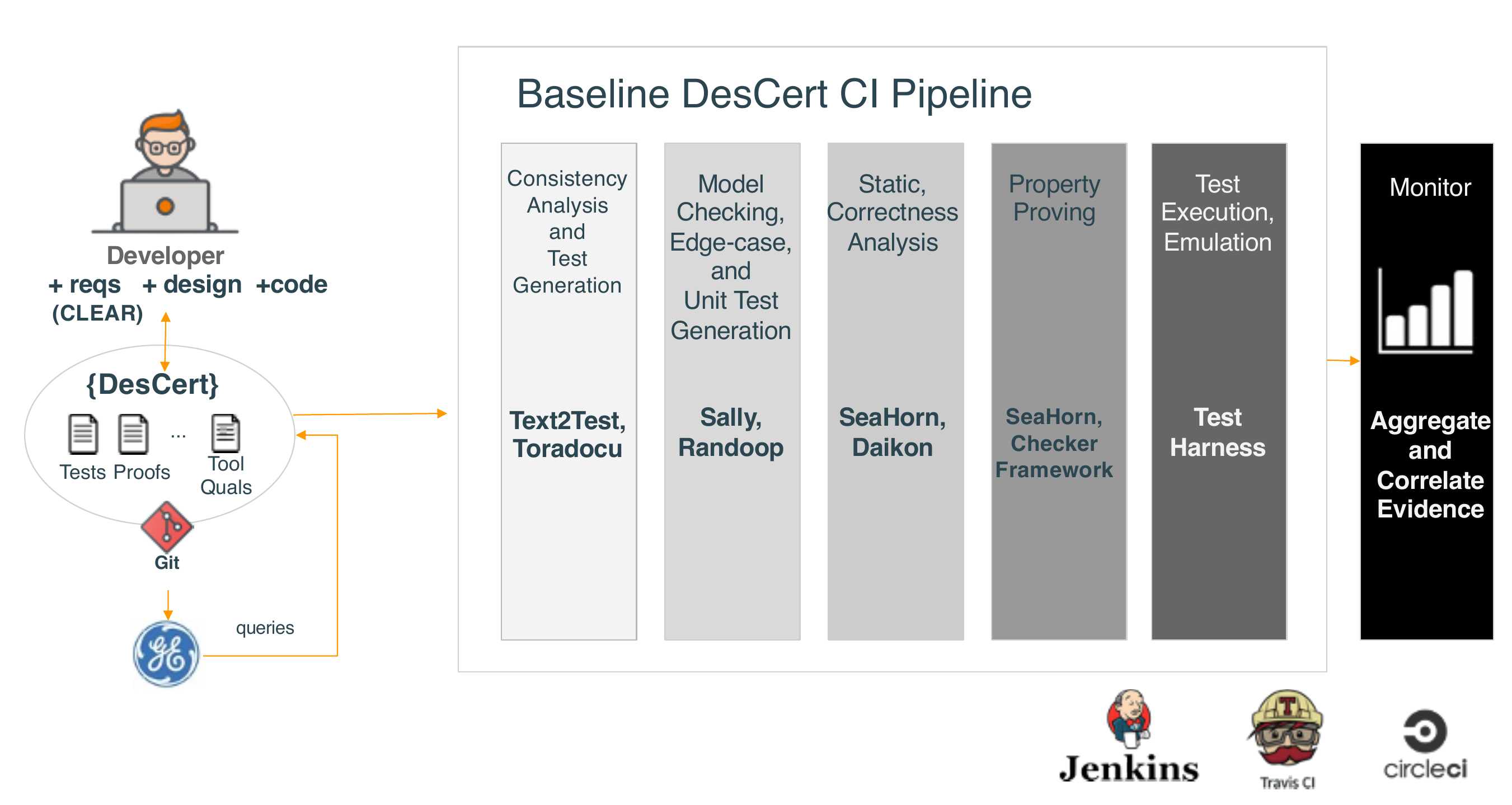}
  \caption{Baseline DesCert}
  \label{fig:baseline}
\end{figure}

In Baseline DesCert, continuous assurance starts with evidence generation.
Evidence generation occurs on a changing codebase in a fashion that mirrors the
continuous model of software development. Baseline DesCert views a codebase as a
changing artifact that evolves through modifications submitted by programmers.
\textit{Baseline DesCert}'s operating mechanism is centered around the use
Gradle and Gradle plugins to record evidence as part of the build process. It
integrates different tools into Gradle through plugins so that these tools can
be applied as software artifacts are created or updated. The recorded evidence
will be maintained within the GE TA2 Rack system. Figure~\ref{fig:baseline} shows
Baseline DesCert's main components.

We have chosen to develop the Baseline DesCert workflow on top of the Gradle
build system for a few reasons. Gradle is breaking new ground in open-source
software development. Its native support is constantly evolving. It can build
not only Java projects but also Android and C/C++ ones, giving us the
flexibility for project selection. Also, Gradle's plugins
portal\footnote{\url{https://gradle.plugins.org}} offers numerous plugins for a
wide variety of capabilities relevant to Baseline DesCert; e.g., C/C++ plugin.

% Built on top a build system like Gradle,
% it triggers evidence generation with the changes in source code. During a build,
% Baseline DesCert feeds back information to programmers without obstructing the
% programmers' workflow.

\sloppy{
As illustrated in Figure~\ref{fig:baseline}, \textit{Baseline DesCert} aims at chaining 
many evidence generation steps. Each of these steps is wrapped into a Gradle plugin.
Each plugin
\begin{inlinelist}
\item configures a set of evidence generation tools and
\item then executes them on codebase modifications
\end{inlinelist}. Particularly, Baseline DesCert aims at integrating
\begin{inlinelist}
\item Text2Test and Toradocu\cite{BlasiGKGEPC2018} for
consistency analysis and test oracle generation,
\item Sally\cite{dutertre2018verification} and Randoop\cite{pacheco2007randoop}
for model checking and unit test generation,
\item SeaHorn\cite{seahorn} and Daikon\cite{ErnstPGMPTX2007} for both static and
correctness analysis,
\item SeaHorn\cite{seahorn} and the Checker
Framework\cite{conf/icse/DietlDEMS11} for property proving,
\item Test hardness for test execution and emulation, and
\item a monitoring tool for assurance evidence management
\end{inlinelist}.
On each run, Baseline Descert is going to monitor not only the execution of its components
but also their output, aggregating and correlating recorded evidence in the process.
At this point, the developer who submitted software modifications can initiate the ingestion
of assurance artifacts into Rack in accordance with the DesCert ontology. } %% sloppy

The flow of evidence recorded by Baseline DesCert's executions is going to
capture properties of architecture and requirements of artifacts representing
the system ConOps. At each stage of this successive evidence generation process,
the objective is to work towards the construction of an assurance argument that
can be maintained along with the software's evolution.

\section{Baseline DesCert's Gradle Plugins}
\label{sec:baselinedescert}

Baseline DesCert is a work-in-progress tool. We have currently implemented
Gradle plugins for the University of Washington tools: Daikon and Randoop, and
are integrating the Checker Framework. Moreover, we have also implemented a Rack
data import plugin. This plugin is responsible for pushing the data generated by
the plugins into the Rack system. In addition to the Randoop, Daikon, and Rack
data import plugins, we are working on building more plugins for the other tools shown
in Figure~\ref{fig:baseline}. For example, we have started the definition of
plugins for CLEAR requirements checking, Text2Test, Sally, and SeaHorn. We have
documented their main API and how this API could be used and what type of
evidence it can generate. Plugins can be found at
\url{https://github.com/SRI-CSL/daikon-gradle-plugin.git } and
\url{https://github.com/SRI-CSL/randoop-gradle-plugin.git}.

%% Our continuous certification efforts are centered around the use of Gradle and Gradle plug-ins
%% to record evidence as part of the build process.  We are integrating the different tools into a
%% build system so that they can be applied as artifacts are created or updated.  The generated
%% evidence will be maintained within the GE TA2 Rack system.  We have implemented
%% Gradle plug-ins for the University of Washington tools: Daikon and Randoop, and
%% are integrating the  Checker Framework.
%% We are also starting to define plug-ins for CLEAR requirements checking, Text2Test, Sally, and SeaHorn.
%% We have documented their main API and how this API
%% could be used and what type of evidence it can generate. Plugins can be
%% found at \url{https://github.com/SRI-CSL/daikon-gradle-plugin.git  }
%% and \url{https://github.com/SRI-CSL/randoop-gradle-plugin.git}.  

The Randoop plugin, for example, integrates the
Randoop~\cite{pacheco2007randoop} tool to automatically create unit tests for a
set of classes, in JUnit format. The Randoop tool uses
feedback-directed random test generation to generate sequences of
method/constructor invocations for the classes under test. To use the plugin,
one must add {\small \texttt{apply plugin: ``com.sri.gradle.randoop''}} to the root
project's {\small \texttt{build.gradle}}. Individual Randoop settings should be
also specified in the {\small \texttt{build.gradle}}. At least, one should
specify, 
\begin{inlinelist}
  \item the path to the Randoop tool, 
  \item the JUnit package name (the location of the classes under tests), and 
  \item the Randoop output directory 
\end{inlinelist}. Figure~\ref{fig:randoopconfig} shows a complete configuration
of the Randoop plugin.

\begin{figure}[h!]\centering\scriptsize
  \begin{lstlisting}
    plugins {
      id 'java'
      id 'maven-publish'
      id 'com.sri.gradle.randoop' version '0.0.1-SNAPSHOT'
    }

    runRandoop {
      randoopJar = file("libs/randoop.jar")
      junitOutputDir = file("${projectDir}/src/test/java")
      // Maximum number of seconds to spend generating tests.
      // Zero means no limit. If nonzero, Randoop is nondeterministic: 
      // it may generate different test suites on different runs.
      timeoutSeconds = 30
      // Stop generation as soon as one error-revealing test
      // has been generated.
      stopOnErrorTest = false
      // What to do if Randoop generates a flaky test:
      // (1) halt, (2) discard, (3) output
      flakyTestBehavior = 'output'
      // A flag that determines whether to output 
      // error-revealing tests.
      noErrorRevealingTests = true
      // A flag that determines whether to use 
      // JUnit's standard reflective mechanisms
      // for invoking tests.
      junitReflectionAllowed = false
      usethreads = true
      outputLimit = 2000
      junitPackageName = 'com.foo'
    }
  \end{lstlisting}
  \caption{Randoop plugin configuration}
  \label{fig:randoopconfig}
\end{figure}

Once configured, the plugin can be run by invoking either the {\small
\texttt{runRandoop}} or the {\small \texttt{randoopEvidence}} tasks. The former
only generates the unit tests while the latter generates both the unit
tests and a set of evidence files summarizing the execution of the Randoop tool.
Figure~\ref{fig:randoopevidence} shows the output of the 
{\small \texttt{randoopEvidence}} task.

\begin{figure}[ht]\centering\scriptsize
  \begin{lstlisting}
    {
      "Evidence": {
        "RandoopJUnitTestGeneration": {
          "INVOKEDBY": "RandoopGradlePlugin",
          "AUTOMATEDBY": "RandoopGradlePlugin",
          "PARAMETERS": "[--time-limit:30, --flaky-test-behavior:output,
            --output-limit:2000, --usethread:true, 
            --no-error-revealing-tests:true, --stop-on-error-test:false, 
            --junit-reflection-allowed:false, --junit-package-name:com.foo, 
            --junit-output-dir:src/test/java]"
        },
        ...
        "RandoopTestsAndMetrics": {
          "BRANCH": "master",
          "EXPLORED_CLASSES": "2",
          "COMMIT": "6fb16d1",
          "PUBLIC_MEMBERS": "6",
          "NORMAL_EXECUTIONS": "314804",
          "REGRESSION_TEST_COUNT": "885",
          "ERROR_REVEALING_TEST_COUNT": "0",
          "AVG_EXCEPTIONAL_TERMINATION_TIME": "0.224",
          "MEMORY_USAGE": "4647MB",
          "EXCEPTIONAL_EXECUTIONS": "0",
          "GENERATED_TEST_FILES_COUNT": "3",
          "AVG_NORMAL_TERMINATION_TIME": "0.0572",
          "GENERATED_TEST_FILES": [
            "src/test/java/com/foo/RegressionTest0.java",
            "src/test/java/com/foo/RegressionTest1.java",
            "src/test/java/com/foo/RegressionTestDriver.java"
          ],
          "CHANGES": "local",
          "INVALID_TESTS_GENERATED": "0",
          "NUMBEROFTESTCASES": "885"
        }
      }
    }
  \end{lstlisting}
  \caption{Randoop plugin evidence}
  \label{fig:randoopevidence}
\end{figure}

The Daikon plugin integrates the Daikon\cite{ErnstPGMPTX2007} invariant detector
to report likely program invariants. Daikon runs a program (e.g., unit tests
generated by Randoop), observes the values that the program computes, and then
reports properties that were true over the observed executions. The plugin can
detect whether the unit tests to execute were generated by Randoop (executed
through the Randoop plugin) or not. If there were not generated by the Randoop tool, the
plugin would search for a test driver (i.e., a test class that contains a static
main method) it could run. If the plugin cannot find a test driver, then the Daikon
plugin assumes the project under test has its own unit tests and thus it generates
a test driver that can execute these tests.

Similar to the Randoop plugin, one must add {\small \texttt{apply plugin:
``com.sri.gradle.daikon''}} to the root project's {\small \texttt{build.gradle}}
and a few other settings to run the plugin. The Daikon plugin can be run by
invoking either the {\small \texttt{runDaikon}} or the {\small
\texttt{daikonEvidence}} tasks. Figs.~\ref{fig:daikonconfig} and
\ref{fig:daikonevidence} show a complete configuration of the Daikon plugin and
the generated evidence respectively.

\begin{figure}[t!]\centering\scriptsize
  \begin{lstlisting}
    plugins {
      id 'java'
      id 'maven-publish'
      id 'com.sri.gradle.daikon' version '0.0.2-SNAPSHOT'
    }

    runDaikon {
      outputDir = file("${projectDir}/build/daikon-output")
      // the project directory where daikon.jar, ChicoryPremain.jar,
      // and dcomp_*.jar files exist
      requires = file("libs")
      // *TestDriver package name. Daikon tool requires
      // a test driver. If you use Randoop,
      // then Randoop will generate one for you.
      // Otherwise, tell the plugin to generate a test driver,
      // simply by using the -Pdriver property when executing
      // the plugin.
      testDriverPackage = "com.foo"
    }
  \end{lstlisting}
  \caption{Daikon plugin configuration}
  \label{fig:daikonconfig}
\end{figure}

\begin{figure}[h!]\centering\scriptsize
  \begin{lstlisting}
    {
      "Evidence": {
        "DaikonPluginConfig": {
          "OUTPUT_DIR": "build/daikon-output",
        },
        ...
        "DaikonInvsAndMetrics": {
          "CORES": "16",
          "JVM_MEMORY_LIMIT_IN_BYTES": "477626368",
          "SUPPORT_FILES": [
            "build/daikon-output/RegressionTestDriver.dtrace.gz",
            "build/daikon-output/RegressionTestDriver.decls-DynComp",
            "build/daikon-output/RegressionTestDriver.inv.gz"
          ],
          "PP_COUNT": "5",
          "INVARIANTS_FILE": "build/daikon-output/RegressionTestDriver.inv.txt",
          "MEMORY_AVAILABLE_TO_JVM_IN_BYTES": "432013312",
          "CLASSES_COUNT": "1",
          "TEST_DRIVER": "src/test/java/com/foo/RegressionTestDriver.java",
          "TESTS_COUNT": "4",
          "INVARIANT_COUNT": "0"
        }
      }
    }
  \end{lstlisting}
  \caption{Daikon plugin evidence}
  \label{fig:daikonevidence}
\end{figure}

The third plugin is the Rack data import. This plugin is responsible for loading
the Randoop and Daikon evidence JSON files, transforming these files into a format that
the Rack system can ingest, and then (if the Rack system is running) pushing the
transformed files into the Rack system. The configuration is simple. This
plugin has only one dependency: the Gradle plugin {\small
\texttt{com.jetbrains.python.envs}}, version {\small \texttt{0.0.30}}. The Rack
data import uses this plugin to create an Anaconda environment that contains all
the dependencies needed for running the Rack CLI. The Rack
CLI\footnote{\url{https://github.com/ge-high-assurance/RACK/tree/master/cli}} is
a tool that exposes a set of APIs for setting up ontologies and also pushing
data into the Rack system. Once the Anaconda environment is created, the Rack data
import plugin uploads the DesCert ontology to the Rack system. If the Randoop
and Daikon evidence files are available, the plugin pushes the evidence data to
the Rack system. Otherwise, it immediately returns.
Figure~\ref{fig:rackingestconfig} shows the Rack data import plugin's
configuration.

\begin{figure}[h!]\centering\scriptsize
  \begin{lstlisting}
    plugins {
      ...
      id "com.jetbrains.python.envs" version "0.0.30"
    }

    envs {
      bootstrapDirectory = new File(buildDir, 'bootstrap')
      envsDirectory = new File(buildDir, 'envs')
    
      conda "Miniconda3", "Miniconda3-latest", "64"
      condaenv "descert", "3.8.5", "Miniconda3", ["numpy"]
    }

    task setupRackCli(type: Exec) {
      dependsOn 'build_envs'
      executable "./rack-descert.sh"
      args "cli", "--conda=Miniconda3", "--condaenv=descert"
    }

    task setupRackArcos(type: Exec) {
      dependsOn 'setupRackCli'
      executable "./rack-descert.sh"
      args "init", "--conda=Miniconda3", "--condaenv=descert"
    }


    task importData(type: Exec) {
      executable "./rack-descert.sh"
      args "import", "--conda=Miniconda3", "--condaenv=descert"
    }
  \end{lstlisting}
  \caption{Rack data import configuration}
  \label{fig:rackingestconfig}
\end{figure}

The {\small \texttt{rack-descert.sh}} script shown in
Figure~\ref{fig:rackingestconfig} contains functionality for
\begin{inlinelist}
  \item installing the Rack CLI,
  \item checking whether a Rack instance is running,
  \item setting up the DesCert ontology in Rack,
  \item curating evidence data, and
  \item ingesting the curated evidence data into the Rack system.
\end{inlinelist}

The curation step is a key operation of this plugin. It transforms recorded
evidence data to a new format that matches the DesCert ontology in the Rack
system. For example, it turns the Randoop's evidence data (See
Figure~\ref{fig:randoopevidence}) into a set of concepts that describe the
ontology of the Randoop evidence: the \textit{RandoopJunitTestGeneration}
activity is invoked by a user entity (e.g., developer). This activity is
automated by the \textit{Randoop tool} entity given some user-specified tool
configuration and a target \textit{source code} entity. The results of a
\textit{RandoopJunitTestGeneration} execution are captured in the
\textit{RandoopTestsAndMetrics} entity.

\begin{figure}[h!]\centering
  \includegraphics[width=12cm]{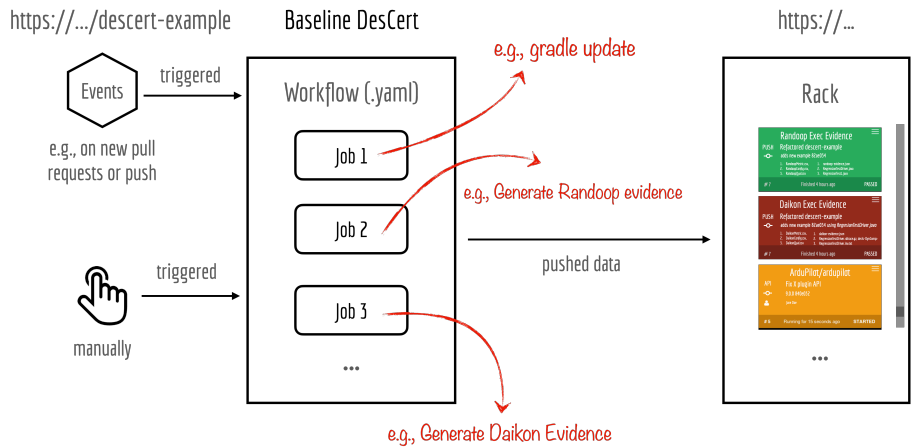}
  \caption{Baseline DesCert execution on descert-example repository}
  \label{fig:execbaseline}
\end{figure}

We have created an illustrative descert-example project (See
\url{https://github.com/SRI-CSL/descert-example.git}), its Docker image, its
configuration, and its documentation. This project generates evidence for a
basic Java project, as part of its building process, using the Randoop and
Daikon Gradle plugins. This project can algo transforms the generated evidence
into a format that the Rack system can handle during its data import task.
If a local version of Rack is running, a user can trigger the uploading of
the evidence data to the Rack system. Figure~\ref{fig:execbaseline} shows
Baseline DesCert's execution on the descert-example repository. 

\section{Track, Compare, and Visualize Your Evidence}
\label{sec:trackcompare}

\begin{figure}[h]\centering
  \includegraphics[width=13cm]{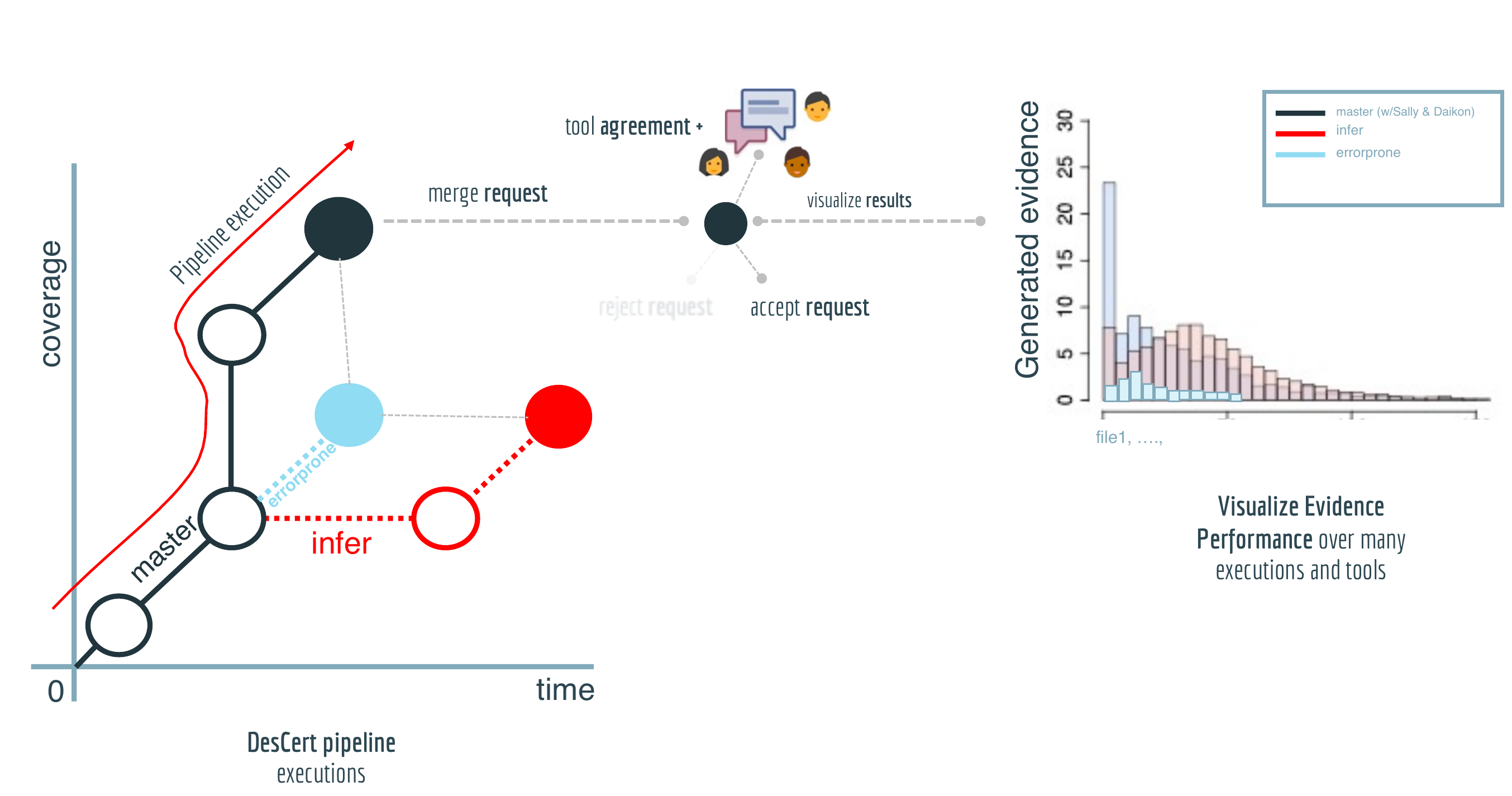}
  \caption{Experiment tracking Baseline DesCert}
  \label{fig:experiment}
\end{figure}

At a first glance, evidence generation with Baseline DesCert looks a lot like
continuous software development. But there are some key differences. Unlike
continuous software development, the essential unit of progress in Baseline
DesCert is an experiment, meaning programmers are going to be able to track what
they doing in the pipeline. With Baseline DesCert, we aim at automatically
linking programmers' experiments to latest git commits in the software
repository of a project under assessment. Our goal to facilitate to programmers
a mechanism for easily comparing any subset of experiments with visualizations,
trying to achieve tool agreement whenever possible. The agreement of tools on
specific generated evidence is an indicator of the quality of the generated
evidence. Figure~\ref{fig:experiment} illustrate our vision for experiment
tracking in Baseline DesCert.

With experiment tracking, programmers now have new capabilities. They can vary
evidence generation tools, change versions of a particular codebase, and even
adjust more general settings of each evidence generation tool. For example, on a
new experiment, one can change Sally and Daikon with SeaHorn or with Facebook's
Infer\cite{calcagno2015moving, calcagno2015open} and then track the quality of
newly generated evidence using the new tools.

As Figure~\ref{fig:experiment} suggests, comparing the results of each experiment
is going to facilitate collaboration between the users of Baseline DesCert.
Drawing inspiration from continuous integration systems like Travis-CI and
others, our goal is allow users, or different teams, to access the ``continuous
assurance'' history of a codebase at any time. Users can also watch other users'
experiments unfold in real-time and even provide feedback if requested. We
hypothesize that the ability to see evidence evolution over many executions and
many tool combinations (through experiment tracking and visualizations) is going
to provide a sense of progress with respect to the quality of the generated
evidence for a changing codebase.

\chapter{DesCert Evidence Generation}\label{chap:evidence}
%Evidence generated
\begin{sidewaysfigure}[!hbtp]
\centering
\includegraphics[scale=.55]{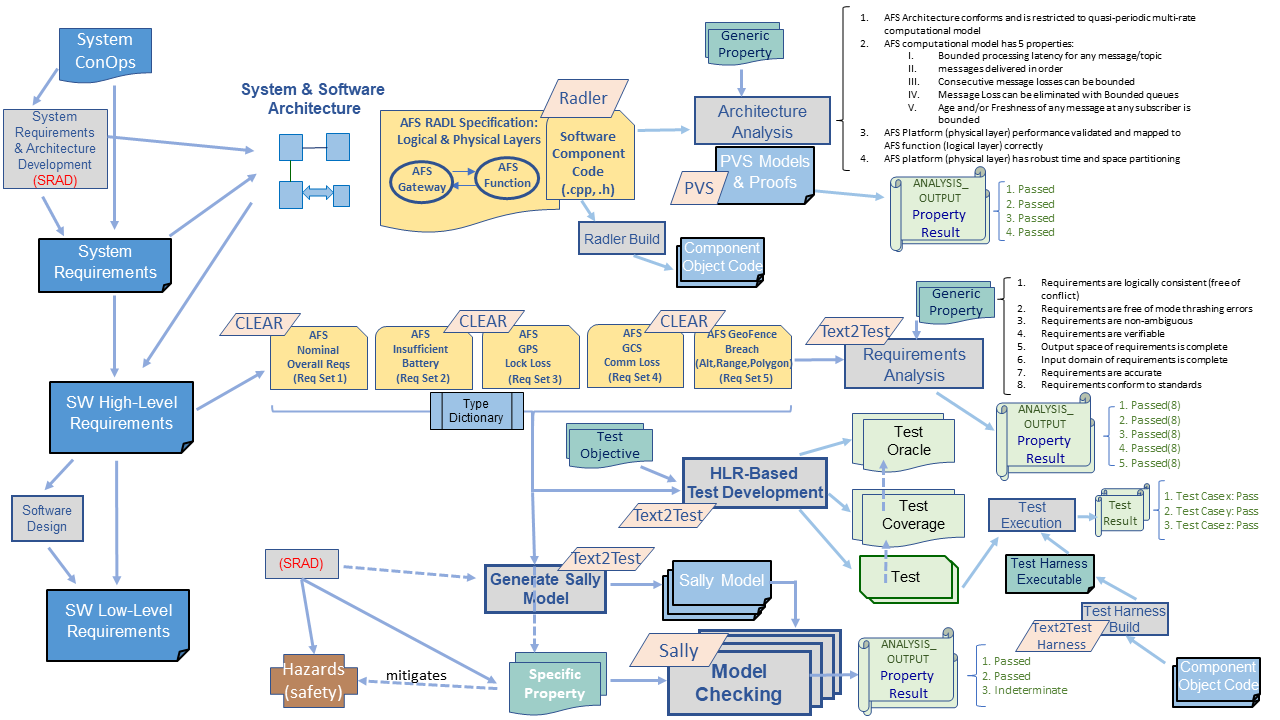}
\caption{AFS DesCert Evidence Generated and Populated in RACK}
\label{fig:descert_evidence_generated}
\end{sidewaysfigure}

In this section we show the different certification evidence for AFS, case study described in Section~\ref{chap:challenge}, that is generated using DescCert tools described in Section~\ref{chap:tools} and how that evidence is populated in RACK using the ontology described in Section~\ref{chap:ontology}. The evidence generation as well as the population of RACK with ingestion data is shown in Figure~\ref{fig:descert_evidence_generated}. 

At a high level, the left side of Figure~\ref{fig:descert_evidence_generated} shows typical design/development process starting from system ConOps definitions and from it developing system requirements and architectures (both system and software) and then specifying intent and software behavior in software high level requirements (HLRs) and after design of software capturing implementation details of software with low level requirements (LLRs) specifications. The middle parts of the Figure~\ref{fig:descert_evidence_generated} shows the different AFS development and evidence artifacts generated while the right side of the figure indicates the different ontological class items populated within the RACK.

The rest of this sections is organized as follows: the top layers of the Figure~\ref{fig:descert_evidence_generated}, left-to-right, shows DesCert evidence associated with generic properties of architectural RADL specifications and is described in detail in Section~\ref{sec:radlEvidence}; the middle layers within the figure that deals with AFS requirements and the associated requirements analysis evidence of satisfaction of certain generic properties are described in Section~\ref{sec:HLR} while the test evidence that are automatically generated from Software HLRs are detailed in Section~\ref{sec:TestEvidence}; finally the lower layers of the figure shows the automatic model generation from HLRs and verification of specific properties through model checking and is described in Section~\ref{sec:SallyEvidence}.

\section{AFS High-Level Software Requirements}\label{sec:HLR}

Without a doubt, the most fundamental step of any system development is capturing a good set of requirements in a systematic way such that it is accurate, complete, verifiable etc. However, this is not always an easy, straightforward task for most complex systems, including the AFS system. Though the ConOps described the contingencies and the recovery actions at a high level, specifying precise requirements for AFS such that it can be subject to a variety of analysis was not without challenges. We begin this section with a brief note about some of the challenges we encountered while specifying requirements, followed by details of the formalization and analysis of the AFS requirements.

Initially, we captured the requirements of each contingency independently, as specified in ConOps in Section~\ref{chap:challenge}. However, a deeper look at the failure conditions and the recovery actions reveled that (a) there was a lack of detail about desired behaviour when multiple failures occur at the same time; (b) some of the recovery actions had critical dependency on proper functioning of components whose contingencies and recovery actions have been defined within the same ConOps. For example, the copter is expected to return to launch, when the battery level of the AFS system is below a certain level; however, without proper functioning of GPS, the copter would not be able to return to launch. Despite having identified GPS failures as one of the contingencies in the ConOps, the respective recovery was assigned a low priority than battery. So, it was imperative that we not only capture the priority among the contingencies, but also take into account dependencies in the recovery actions. Also, on collectively analysing the recovery actions, we found that all of them describe response actions to the same actuator of the copter and broadcast messages to the same messaging channel. Hence, it was crucial for us to carefully specify a consistent and correct response when multiple failures occur. Further, the ConOps indicated that all the three geofence conditions have the level of priority. Thought this may be acceptable, it will lead to non-deterministic behaviours and, eventually implementation. Hence, we had to explore means to provide some sort of precedence among those equally prioritized contingencies.

Given all the above challenges, after several round of brainstorming and exploring multiple approaches, we finally decided to use the notion of \emph{states} to abstractly conceptualize the condition of the AFS during various contingencies and the \emph{transition conditions} defining when and how the system shifts between those states. However, to handle the priority and dependencies among contingencies, a flat state-transition was not adequate. We explored the notion of defining each contingency independently and specify an \emph{arbiter}~\cite{murugesan2013modes}, to finally decide a recovery action. However, the issue was that, to define the final recovery action, it became necessary to repeat the specification of each failure conditions again in the arbiter since there were dependencies. Hence, we took an hierarchical state machine approach~\cite{harel1987statecharts} where we encoded the notion of hierarchy, sequence and dependency of recover actions into the requirements.

It is worth to mention that, although state machines and hierarchies are often correlated with design models, we ave found that they also provide a comprehensible means to conceive and describe the requirements of complex systems. Hence, we have used it purely to capture the AFS contingency requirements in a concise and precise manner. This does not impose any restrictions on the way the system will be designed or implemented.

\subsection{CLEAR Features for AFS HLR}

We formally captured the AFS requirements using the CLEAR notation explained in Section~\ref{chap:challenge}. We defined the requirements of each contingency as separate \emph{requirement sets}.  A requirement set stands alone such that it sufficiently describes the necessary capabilities, and constraints of the contingency considered. All the variables and enumerations that are used across all the requirement sets are collectively defined in a common dictionary or definition files.

\paragraph{State Machines}

States are widely used as abstraction to capture mutually exclusive sets of discrete, dynamic system behaviors. The description of the states and the rules defining when and how the system transitions between those states, collectively known as \emph{state machines}, provided a coherent, concise means to express the AFS requirements.

%The discrete, dynamic behavior of complex systems are often expressed in terms of states -- that are mutually exclusive sets of system behaviors, and the rules defining when and how the system transitions between those states. The states and the transitions provide a coherent, concise way to express the requirements.

 \begin{figure}[!h]
  \centering
  \includegraphics[width=\columnwidth]{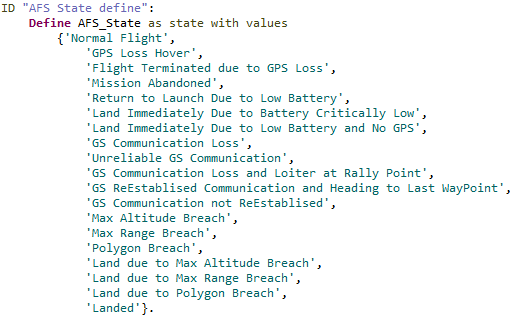}
  \caption{AFS State Definition}
  \label{fig:AFSState}
\end{figure}

To that end, using the state definition construct in CLEAR notation, we defined \texttt{AFS\_State} to represent the various contingency states of the AFS system, as shown in Figure~\ref{fig:AFSState}. This state definition has several benefits, such as succinctly specify the behaviours and helps define precedence among the behaviours in an easily understandable manner.

\paragraph{Precedence and Priority Among Requirements}
By definition, ensuring that all contingencies are addressed per the requirements is required or mandatory. However, at a given instant in time, when more than one contingency scenario occur, all the requirements are not all equal in terms of current importance or severity of the addressing the situation.  For example, handling contingency related to low/failed battery in a system has the highest precedence over failed ground communication. Further, to meet a specific contingency requirement (with higher precedence contingency management), correct functioning of another contingency (of lower precedence) may be necessary. For example, to safely land a flight at a emergency landing location in the event of low/failed battery situation requires the proper functioning of navigation aids; otherwise, there has to be an alternate course of contingency actions. Hence, at the time of specification, it is crucial to determine the relative necessity of the requirements given the possible combinations of contingency scenarios that may occur at operational time.

\begin{figure}[!htb]
\centering
\includegraphics[width=\columnwidth]{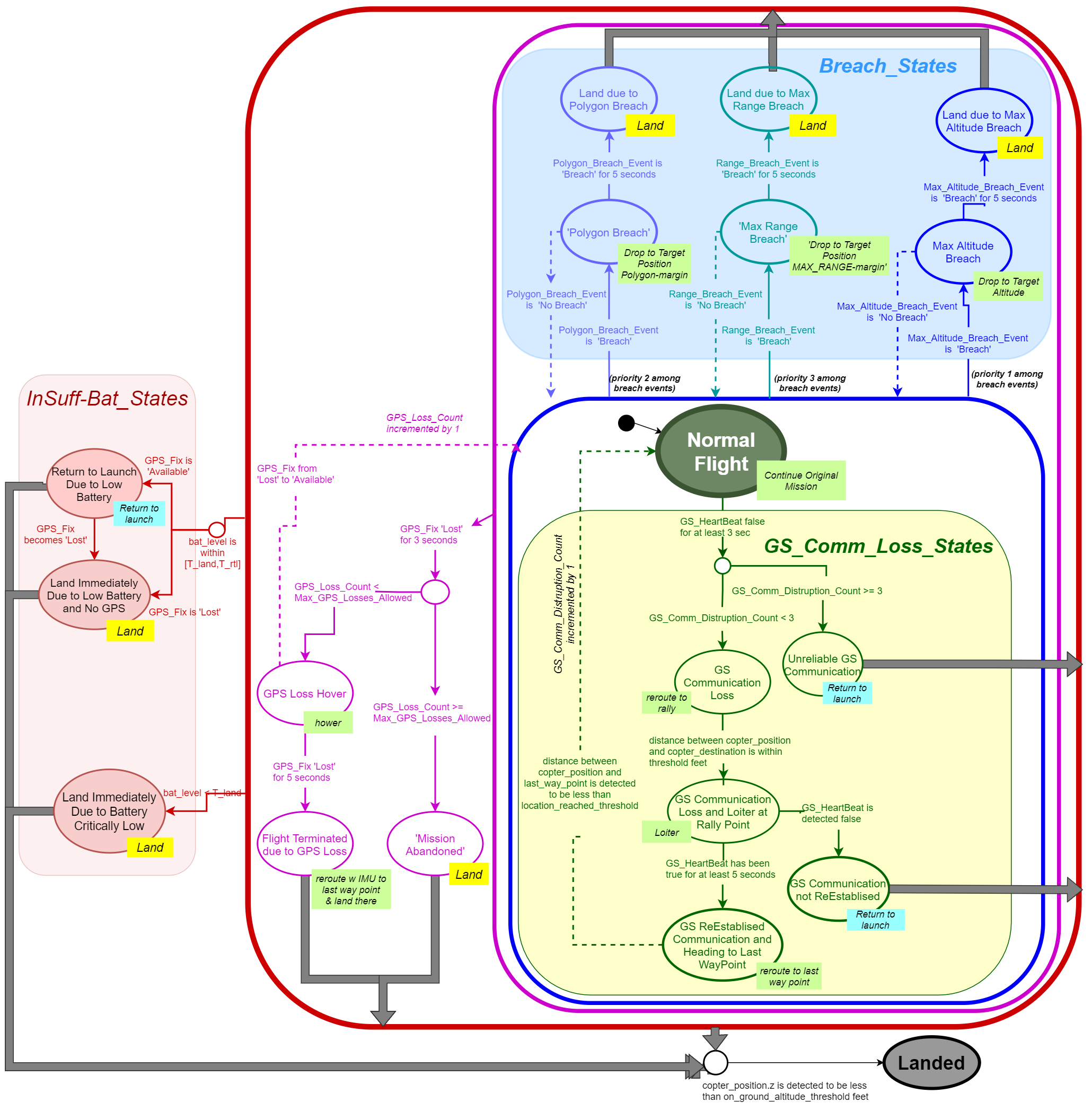}
\caption{Prioritized Contingency Management}
\label{fig:Priority}
\end{figure}

For AFS, as per the concept of operations discussed in Section~\ref{sec:conops}, we ranked the contingency scenarios in the order of precedence or priority. The Figure~\ref{fig:Priority}, visualizes the priority using grouping and hierarchical state machine representation. The insufficient battery contingency states (grouped and colored red are labeled InSuff-Bat\_States) were specified to have the highest priority, i.e., when the system is any non-low battery state and a battery contingency condition occurs, the system will transition to the low battery contingency states. The next in priority was the GPS Lock loss (states and transitions colored pink); i.e., when the system is in any of the Breach states (colored blue), or communication loss states (colored yellow) or normal flight state (colored green), if there is GPS Lock loss event when there are no  battery contingency, the system will transition to the GPS lock loss state. Similarly, we grouped and captured the other contingencies depending upon the priority identified in the ConOps. In the following sections, we delve into the details of specifying each of these contingency requirements in detail.

\newpage

\paragraph{Ontic Type Support for Navigation Entities}

\begin{figure}[!h]
  \centering
  \includegraphics[width=0.7\columnwidth]{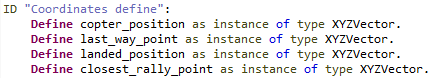}
  \caption{XYZVector Definition in CLEAR}
  \label{fig:XYZVecDefn}
\end{figure}

One of the new ontic types we defined in CLEAR for AFS is \emph{XYZVector} type, a coordinate system definition that allows uniquely defining the location of geographical and navigational entities. The coordinate system convention used for most vehicles, including aircraft, is based on an XYZ system, where $X$ and $Y$ represent the horizontal position on ground and $Z$ represents the altitude from ground. When specifying the requirements of aerospace systems, the notion of three dimensional coordinate systems is inevitable to express position of air vehicles in space, their distance to other entities in space, etc. In CLEAR notation users can define terms (or variables) of type XYZ coordinates as shown in Figure~\ref{fig:XYZVecDefn}.

To allow defining the notion of distance between two XYZ Coordinate terms, CLEAR notation provides the distance between construct. The distance between two XYZ points ($x_1, y_1, z_1 $) and ($x_2, y_2, z_2 $) are computed using the following formula,

$$ distance = \sqrt \{(x_1^2 - x_2^2) + (y_1^2 - y_2^2) + (z_1^2 - z_2^2)\} $$

\begin{figure}[!h]
  \centering
  \includegraphics[width=\columnwidth]{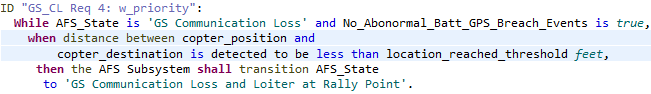}
  \caption{XYZVector terms and distance function usage in CLEAR}
  \label{fig:XYZVecReq}
\end{figure}

Figure~\ref{fig:XYZVecReq} illustrates the usage of this construct in one of the requirements of the AFS.

\subsection{AFS Requirements in CLEAR}

We now describe the CLEAR specification of the AFS contingencies in detail.

\paragraph{Insufficient Battery}

\begin{figure}[!htb]
  \centering
  \includegraphics[width=\columnwidth]{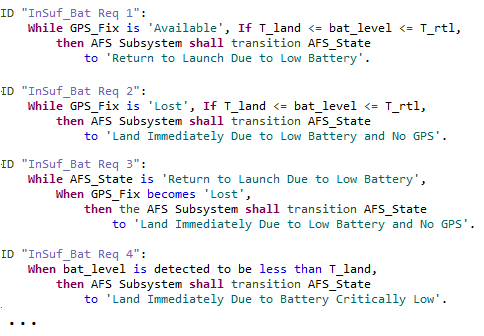}
  \caption{Low Battery Requirements}
  \label{fig:Insuff_Req}
\end{figure}

As mentioned earlier, the Insufficient Battery contingency has the highest priority. Per the ConOps, if the battery level is critically low (less than the \texttt{T\_land}), the system should land immediately; whereas, if the battery is between the \texttt{T\_rtl} (Return To Launch) and \texttt{T\_land}, the system shall return to launch.  While we originally defined two states for this contingency, as we were analyzing and formally capturing the requirements, we found that the system needs the GPS to function normally for returning to launch. Hence, to capture these scenarios, we has to define three different states for this contingency -- namely \emph{`Return to launch due to low battery', `Land immediately due to low battery and no GPS'} and \emph{`Land immediately due to Battery Critically Low'}. The conditions to each of these states as well as their behaviour (expected output) are different. A snippet of the formalization in CLEAR notation is shown in Figure~\ref{fig:Insuff_Req}.

\paragraph{GPS Lock Loss Requirements}

The next in priority was the GPS Lock loss (states and transitions colored pink in Figure~\ref{fig:Priority}); In other words, the GPS lock loss related contingency requirements need to come into play only when there are no battery contingency. To capture that condition within the requirement, we defined a boolean variable called \texttt{No\_Abonormal\_Batt\_Event} that would be true when the battery level is more than \texttt{T\_rtl}. The response for GPS loss requires that the system waits by hovering for a certain amount time to allow going back the mission when GPS recovers, or terminate and land otherwise. Further, the goal was also to abandon the mission if there are more than a certain number of GPS failures.

\begin{figure}[!htb]
  \centering
  \includegraphics[width=\columnwidth]{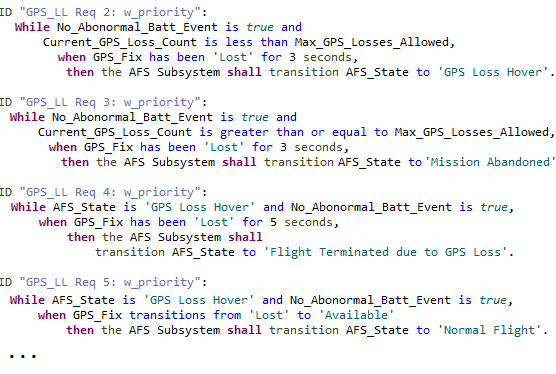}
  \caption{GPS Lock Loss Requirements}
  \label{fig:GPSLLReq}
\end{figure}

To capture these behaviours, we defined three states for the GPS lock loss -- namely \emph{`GPS Loss Hover', `Flight Terminated due to GPS Loss'} and \emph{`Mission Abandoned'} -- and the respective transition conditions. While we explicitly defined a counter (\texttt{GPS\_Loss\_Count}) to keep track of the number of GPS losses, the CLEAR notation construct \emph{`has been ... for ... seconds'} helps specify the notion of timers. Figure~\ref{fig:GPSLLReq} shows a small snippet of the requirements in CLEAR.

\paragraph{Geofence Breach Requirements}

The third in priority is the various breach contingency responses; In other words, when altitude, range or polygon breach occurred, if there are no battery or GPS related failures, then the system shall attempt to correct the breach situation within a certain amount of time, otherwise, the system shall land. While the ConOps indicated that the breach conditions have equal priority, in order to avoid non-deterministic implementation, it was decided that the altitude, polygon and range breach conditions will be respectively given precedence.

\begin{figure}[!htb]
  \centering
  \includegraphics[width=\columnwidth]{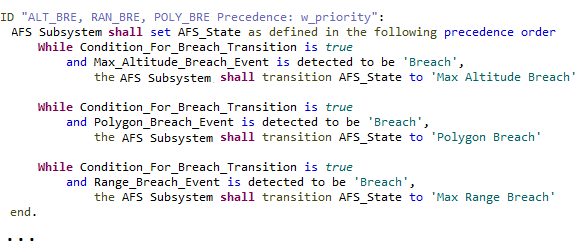}
  \caption{Breach Requirements}
  \label{fig:BreachReq}
\end{figure}

\begin{figure}[!htb]
  \centering
  \includegraphics[width=\columnwidth]{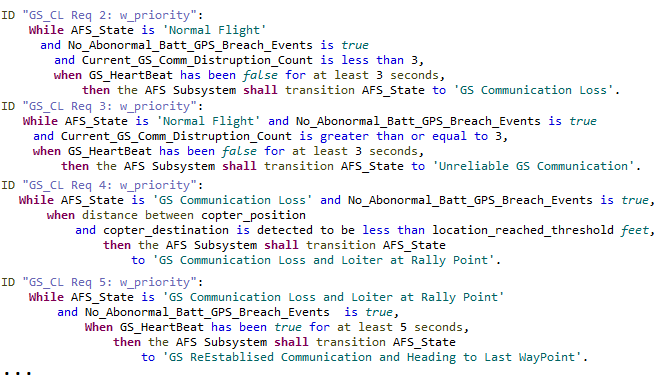}
  \caption{Communication Requirements}
  \label{fig:CommReq}
\end{figure}

To capture the notion of third in line priority while specifying requirements, we had to define a boolean condition whole value is true if there are no battery and GPS related failures at each instant in time. Further, to be able to specify that the system is not already in a state with higher priority, we used CLEAR notation's construct to define custom \emph{sets}, i.e, define a select sub-set of states among the \texttt{AFS\_State}. For instance, we defined sets such as \texttt{GPS\_Loss\_States and Breach\_States} indicating the group of all states related to GPS loss and breach contingencies respectively. This helped specify the requirements in reasonably modular and terse manner, as shown in the Figure~\ref{fig:CommReq}. Moreover, to capture the precedence among the breach conditions, we leveraged the \emph{`...as defined in the following precedence order...'} construct in CLEAR notation. This construct allowed us to specify the response in a precedence order, as shown the snippet~\ref{fig:BreachReq}. As mentioned earlier, instead of explicitly defining a timer to capture the temporal aspect of this contingency,  we used the in-built CLEAR construct \emph{`has been ... for ... seconds'} to concisely capture the requirements.

\paragraph{Communication Loss Requirements}

The communication loss related contingency has the least priority; this means that only when the system is in normal flight state when the communication failures, the related contingency behaviours are exhibited. Similar to the GPS loss requirements, the system shall attempt to reestablish the communication and get back to normal course within a certain amount of time, otherwise will return to launch. Moreover, when the total number of communication loss (denoted by \texttt{GS\_Comm\_Distruption\_Count} in requirement) incidents exceeds a certain threshold, the system shall return to launch. To capture this in CLEAR, we defined 5 different states. Since this had the least priority, we had to take into account the absence of other failures in every requirement relating the transition among the communication loss states. Hence, we defined a boolean variable for that purpose, called \texttt{No\_Abonormal\_Batt\_GPS\_Breach\_Events}, as the name suggests, indicates the absence of other failure conditions. A snippet of the formalization in CLEAR notation is shown in Figure~\ref{fig:CommReq}.

These requirements specified in CLEAR help automate a number of rigorous analysis and verification tasks. In the rest of this section, we describe in detail how we generate test oracles and test scenarios from these requirements.  We also translate these requirements to the Sally transition system language for checking against the architecture model and the low-level requirements.  We have been developing the formal semantics for the mapping between CLEAR and the input language of the Sally model checker using the Text2Test tool infrastructure.  We have also defined a precise ontology and representation format for the evidence artifacts at the requirements level within the TA2 provenance ontology. We have also developed a simple climate control example in order to explain the construction of the assurance argument.  This example is used to communicate how the different pieces of the assurance argument fit together.

% !TEX root = Phase1Report.tex
\section{AFS Evidence Generation from Sally Model Checking}\label{sec:SallyEvidence}

\begin{figure}[!htb]
\centering
\includegraphics[width=\linewidth]{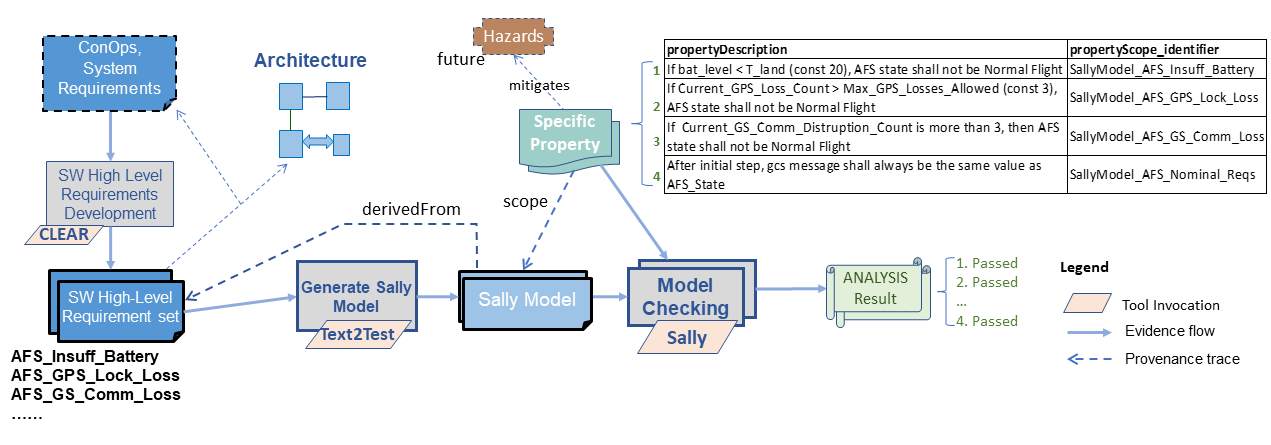}
\caption{An overview of the Sally model creation and checking process.}
\label{fig:sally_overview}
\end{figure}

Two main artifacts are required for checking properties against the software requirements: a Sally model, which is generated by Text2Test from a set of CLEAR high level requirements, and the properties themselves, which are first expressed in natural language and then formalized into a Sally query (Figure \ref{fig:sally_overview}). The Sally model checker takes the model and the query as inputs, using them to either verify that the property holds or produce a counterexample demonstrating that the property does not hold. (That is, assuming that a conclusion can be reached in a reasonable amount of time. The complexity of the model's state will grow exponentially as more steps are required to reach a conclusion.)

We developed a set of formal properties to check against the CLEAR requirements. This feature of Text2Test is under active development, and future versions will have more integration with the CLEAR document. For now, properties are formalized and specified as Sally queries.

As described in Chapter \ref{chap:ontology}, these properties fit into the DesCert ontology. For each, there is a set of software requirements, and a specific property that holds for those requirements. Two tools are used: Text2Test, which generates a Sally model based on the requirement set and the property itself, and Sally, which is run on the resulting model and property to verify the property's correctness.

For the Insufficient Battery requirement set, the model file is \texttt{AFS\_Insuff\_Battery\_Req\_sally.mcmt}. The property is found in the file \texttt{AFS\_Insuff\_Battery\_Property\_1.mcmt}, and states that if the battery level falls below a certain threshold, the AFS state must not be normal. The following command was used to verify this property with Sally:
\begin{lstlisting}
sally --engine kind -v 1 --show-trace \
        AFS_Insuff_Battery_Req_sally.mcmt \
        AFS_Insuff_Battery_Property_1.mcmt
\end{lstlisting}

The formal property for the Insufficient Battery set was as follows:

\begin{lstlisting}
(assume-input AFS_Insuff_Battery_Req_sys
  (= bat_level 19)
)

(query AFS_Insuff_Battery_Req_sys
      (=> (and not_initial_step true) (not(= AFS_State 0)))
)
\end{lstlisting}

To constrain an input variable, the \texttt{assume-input} function is required. A more flexible approach to constrain inputs is to use an internal variable which always mirrors the input, but that requires either tool support for generating such a variable with the model or modifying the model by hand. We must also exclude the initial step from checking because the initial values of all internal variables are undefined in the first step, which allows the model checker to choose anything and will always cause a counterexample to be generated. After the first step, the model's internal variables will have been initialized based on the first step's input variables and proper checking can begin. The \texttt{not\_initial\_step} variable is generated by Text2Test for this purpose, and simply starts out false for one step and then remains true forever.

For the GPS Lock Loss requirement set, the generated Sally model is found in the file \texttt{AFS\_GPS\_LockLoss.mcmt}. The property being tested is in \texttt{AFS\_GPS\_LockLoss\_Property\_1.mcmt}, and checks that if the GPS loss count exceeds the maximum, the AFS state must not be normal. At this stage in tool development, it was necessary to use an internal variable of the model in the property itself to access the previous value fo the GPS loss count. In the future, properties will be specified at the CLEAR level, and Text2Test will generate both the Sally model and the Sally query. This way, properties will be specified with expressions, independently of whatever internal variables are found in the requirement set. To verify this property, the following command was used:
\begin{lstlisting}
sally --engine kind -v 1 --show-trace \
        AFS_GPS_Lock_Loss_Req_sally.mcmt \
        AFS_GPS_LockLoss_Property_1.mcmt
\end{lstlisting}

For the GS Communication Loss requirement set, the generated Sally model is \texttt{AFS\_GS\_Communication\_Loss\_Property\_1.mcmt} and the property being tested is in \texttt{AFS\_GS\_Communication\_Loss\_sally.mcmt}. It states that if the communication disruption count has exceeded 3, the AFS state must not be normal. This property was verified with the following command:
\begin{lstlisting}
sally --engine kind -v 1 --show-trace \
        AFS_GS_Communication_Loss_sally.mcmt
\end{lstlisting}

The final requirement set defines nominal operation, with the Ground Control Station setting the state of the AFS. The Sally model is found in the file \texttt{AFS\_Nominal\_Requirements\_sally.mcmt}, and the property being checked is in the file \texttt{AFS\_Nominal\_Requirements\_Property\_1.mcmt}. This property states that under these requirements, the incoming GCS message should be set directly as the AFS state. The command used to verify this property was:
\begin{lstlisting}
sally --engine pdkind --solver yices2 -v 1 --show-trace \
        AFS_Nominal_Requirements_sally.mcmt \
        AFS_Nominal_Requirements_Property_1.mcmt
\end{lstlisting}
% Text2Test test evidence description goes here
\section{AFS Test Generation from Text2Test}\label{sec:TestEvidence}

One of the main drivers behind the CLEAR notation is to be able to automate formal-methods based analysis and test case generation.  Text2Test is the main tool that performs requirements analysis and test generation. Given a set of requirements, Text2Test performs a synthesis for this purposes from the data-flow model. It then uses Honeywell’s internal HiLiTE tool that provides comprehensive static analysis and data-flow test generation that has been used extensively for certification in many product lines.

\subsection{Analysis of Generic Properties of Requirements}
\label{sec:clear-generic-prop-evidence}

Text2Test  utilizes public domain SMT Solvers (e.g., Z3); analysis objective are formulated into an SMT problem and the solver is used to provide consistency analysis and
(limited) completeness analysis. A variety of arithmetic (including non-linear), logical, and time-based constructs are supported as part of these capabilities to
allow the tools to be used for large-scale industrial problems. Requirement analysis is enabled for each output variable. In
Text2Test data-flow synthesized model, all the requirements specifying the same output variable are combined and chained.

For each requirement set of the AFS system, we analyzed its generic properties, described in Section~\ref{sec:clear-generic-properties}, and shown below in Figure~\ref{fig:genPropertiesAFS}.

\begin{figure}[!htb]
\centering
\includegraphics[width=0.6\columnwidth]{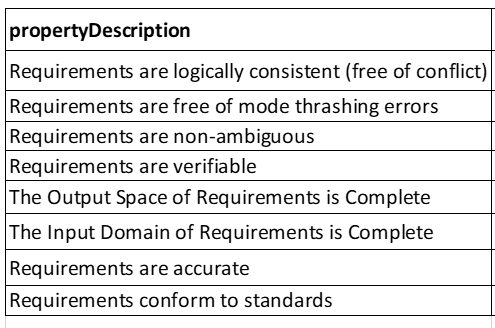}
\caption{AFS Generic Properties}
\label{fig:genPropertiesAFS}
\end{figure}

The errors found by the SMT solver (Z3) is reported in an XML file. The report is a consolidation of all the generic property requirements issues found in a given set of requirements. While there were no consistency and mode-thrashing errors, the tool reported the lack of input and output completeness along with the input and output domain values which made the set of requirements incomplete.

\begin{figure}[!htb]
\centering
\includegraphics[width=\columnwidth]{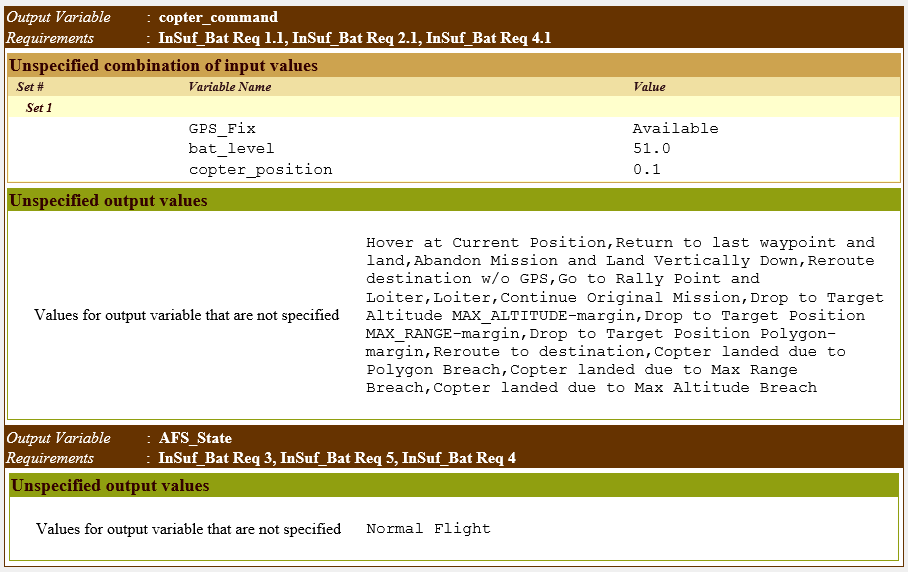}
\caption{AFS Generic Properties Analysis Output}
\label{fig:genPropertiesAnalysisOutput}
\end{figure}

The analysis is performed on groups of individual requirement statements that concern the same output attribute. The requirement defects analysis loops over each of the output variables and lists the generic issues found in each of them. Hence, the errors are displayed in a tabular format for every output attribute. The Figure~\ref{fig:genPropertiesAnalysisOutput} shows the report of the insufficient battery requirements analysis. In the figure, the \emph{Unspecified combination of input values} lists one combination of input values (aka. counterexample) for which the requirements set in consideration do not specify a value for a response. In other words, for the combination of inputs (\texttt{GPS\_Fix = Available; bat\_level= 51.0; copter\_position = 0.1}), there is no value specified for the output \texttt{copter\_command}. This does not indicate a problem with the system, rather this combination of input values occur when the system is in normal state; hence, this completeness issue reported shows that the requirement set under analysis does not include the set of requirement that specify normal system behaviours. Along the same lines, the \emph{Unspecified output values} lists all missing value assignments  to output \texttt{copter\_command} and \texttt{AFS\_State} (as per its definition) specified by the set of requirements. Again this is not an indication of problem with the AFS, rather the result of analysing individual set of contingency requirements, while the definitions include the values specified by other sets of requirements. In the next phases of the project, we plan to define a concrete approach that would compartmentalize the notion of requirement set, such that they can be verified in isolation (without these issues), as well as in a compositional manner with other requirement sets.

\subsection{Requirements-based Test Case Generation}

Using the HiLiTE tool at the back end, the Text2Test automatically generates requirements based test cases from the synthesized model of requirements set. HiLiTE generates specific tests at the model level  for each block embedded in the model, using either heuristic test case templates via backward propagation, or the formal specification of equivalence class via
SMT-solving. The theory behind the test case generation is explained in detail in Section~\ref{chap:tools}. %In the former approach, each equivalence class of a block’s behavior, which represents a test requirement, is represented by a set of test case templates heuristically selected as a specific combination of values for the block under test (BUT) input(s) and output that satisfy this equivalence class. Backward and forward propagation search through the computations of other blocks in the model generates a test vector in terms of model inputs and outputs to ensure controllability of the BUT inputs and observability of the expected BUT output. On the other hand, an equivalence class of a block behavior, is also expressed in HiLiTE with formalized rules on the block’s input and output ports. Each rule is a constraint must be satisfied by the block behavior. In the later approach, SMT-solving embodies formulating test case generation constraints from both equivalence class rules of the BUT and the blocks within its upstream sub-graph, as well as the connectivity, into an SMT problem. Therefore, constraints can be solved together to find a satisfying solution which excludes any conflicts. The formal specification of equivalence classes enables the search on a complete solution domain, empowered by the capability of SMT-solving on scenarios involving mixed data-types, linear/nonlinear computation, and complex model structure.
Once executed the Text2Test creates a number of report files and test vectors for each of AFS's requirement set.

\paragraph{Test Vector Report and Test Vectors}

\begin{figure}[!htb]
\centering
\includegraphics[width=\columnwidth]{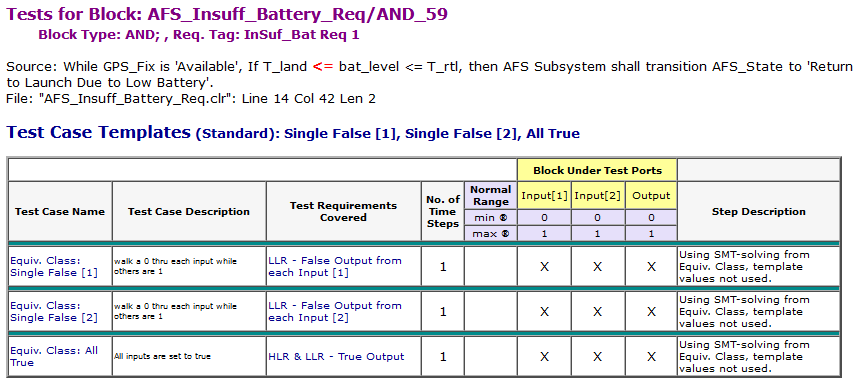}
\caption{AFS Test Report - Template}
\label{fig:AFS_TestReport}
\end{figure}

\begin{figure}[!htb]
\centering
\includegraphics[width=\columnwidth]{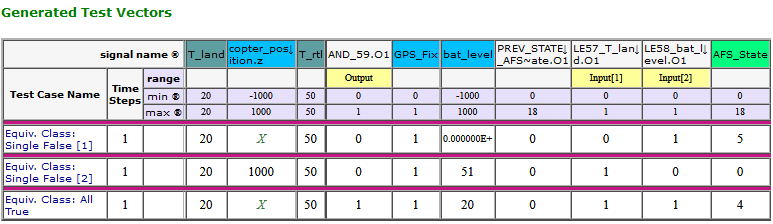}
\caption{AFS Test Report - Vector}
\label{fig:AFS_TestReport_Vector}

\centering
\includegraphics[width=\columnwidth]{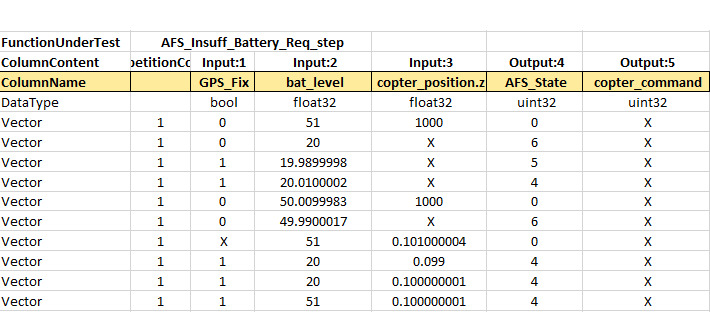}
\caption{AFS Test Vector}
\label{fig:AFS_TestVector}
\end{figure}

The test report contains tables of the test case templates and generated test vectors for each block. Figure~\ref{fig:AFS_TestReport} shows the templates part of the test vector report that was generated for AFS. Test case templates define the input and output requirements for each test for each block. In the portion of the Figure~\ref{fig:AFS_TestReport}, the template that specifies the input values that should produce the expected output values is shown. The column header in yellow shows the name of the variables for the block under test.

Figure~\ref{fig:AFS_TestReport_Vector} shows the vector part of the test vector report for AFS. The Generated Test Vectors show the actual values input to the block under test and the value at the block’s output port. The template requires that Input1 and Input2 have values of 0 and 1, but since these ports are not directly accessible, Text2Test must determine what values to give the model inputs to affect the correct values. In the generated test vector shown, notice that the block under test receives the system`s input (cells colored blue cells at the top of the Test Vectors table), constants (colored in teal color cells at the top of the Test Vectors table) and intermediate values (colored white). The system’s input values were determined from the requirements whereas the intermediate values are the result of intermediate computation performed before the values can be propagated to the output (cell colored bright green).

Further, Text2Test also generates test vectors in a machine-readable (csv) file contains all the actual vectors that can be used with a suitable harness to test the system in consideration. Figure~\ref{fig:AFS_TestVector} shows the comma separated vector file displayed in Excel.

\paragraph{Test Generation Status Reports}

\begin{figure}[!h]
\centering
\includegraphics[width=\columnwidth]{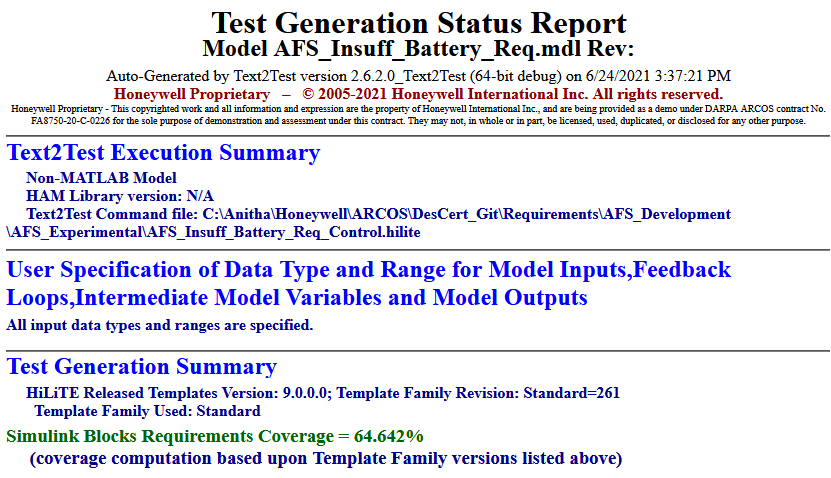}
\caption{AFS Status Report}
\label{fig:AFS_StatusReport}
\end{figure}

The Figure~\ref{fig:AFS_StatusReport} shows the status report for the AFS execution, that summarizes the execution and any issues encountered in analyzing the requirements and list which test cases were generated or could not be generated for each requirement. The detailed coverage metrics are displayed in this report.

% RADL evidence description goes here
\section{AFS RADL Architecture Evidence Generation}\label{sec:radlEvidence}

As mentioned in the beginning of Section~\ref{chap:evidence}, in this Section~\ref{sec:radlEvidence} we will focus on how DesCert AFS related evidence associated with generic properties of architectural RADL specifications are created and this is illustrated in the top layers, left-to-right, of the Figure~\ref{fig:descert_evidence_generated}.

We have described the RADL architectural language specifications, which consists of both the logical and physical parts/layers of the system,  the methodology to perform  software builds using Radler and associated architectural analysis in Section~\ref{sec:RadlerMain}. We have also introduced the ArduCopter system architecture and where the AFS subsystem (software component) resides in the larger system in Section~\ref{sec:SySoArch}. The specific RADL specification for the AFS subsystem is captured in \emph{afs.radl}
in the DesCert Github repository\@.\footnote{Contact the authors for permission to access the repository.} 

\begin{figure*} [hbtp!]
\centering
\includegraphics[width=0.7\linewidth]{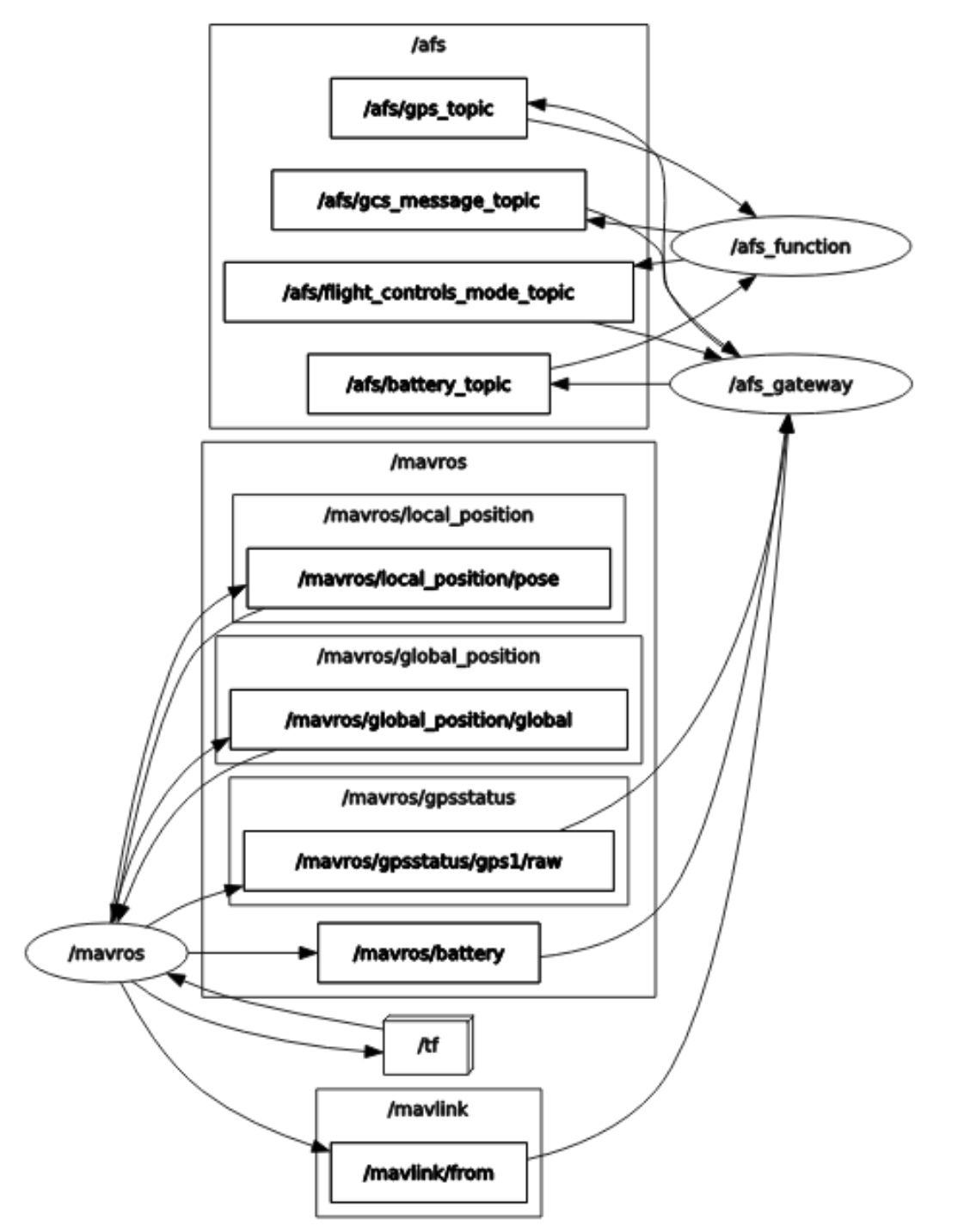}
  \caption{Nodes (in ellipse) and topics (in rectangle) of the AFS subsystem}
\label{fig:afs_node_topic}
\end{figure*}

Figure \ref{fig:afs_node_topic} shows the nodes and communications between them via topics for the AFS subsystem which consists of two nodes.  The \code{AFS Function \emph{(afs\_function)}} node both publishes to as well as subscribes from \code{AFS Gateway \emph{(afs\_gateway)}} node. \code{AFS Gateway} acts as a bridge between \code{AFS Function} and MAVROS which in turn interfaces with the ArduCopter flight code through the MAVLink. \code{AFS Gateway} captures all messages from ArduCopter and extracts battery information, raw GPS position information,Geo-fence breach events and other health of the ArduCopter health information, GCS heartbeat information etc and publishes these as topics to \code{AFS Function} subscriber. The \code{AFS Function} node then executes its step function with period of 100 milliseconds and publishes 4 recovery actions subscriber ROS nodes: (i) AFS Geofence Breach Recovery for 3 different kinds (altitude, range, polygon) (ii) AFS GCS Comm Loss Recovery (iii) AFS GPS Lock Loss Recovery (iv) AFS Battery Insufficiency Recovery. The \code{AFS Function} node then publishes as topics flight controls command (e.g. change in flight controls mode or switching sensor information etc) to \code{AFS Gateway} subscriber which translated the command appropriately and sends them via MAVROS to ArduCopter via MAVLink transport (potentially repeatedly messages may needed to sent to ArduCopter until ACKs on receipt of the message is confirmed).

We next describe below how the different \emph{generic properties associated with architecture} for AFS RADL specification, illustrated in the top layer of Figure~\ref{fig:descert_evidence_generated},  is proven using architectural analysis.  

\vspace{0.1in}
\shadowbox{
\begin{minipage}{0.9\textwidth}
\begin{center}
\emph{Architecture Generic Property 1:} 
\end{center}

Software architecture of ArduPilot Advanced Fail Safe (AFS) subsystem conforms and is restricted to quasi-periodic multi-rate computational model.
\end{minipage}
}
\vspace{0.1in}

\emph{Architecture Generic Property 1 met rationale}: ArduPilot Advanced Fail Safe (AFS) subsystem include afs\_gateway and afs\_function that are specified in afs.radl discussed above. Radler build process used the afs.radl specification which included multiple rates of different computational nodes (tasks running at fixed period) with communicating topics/messages between publishers and subscribers over channels with expectation of guaranteed delivery and latency/timing/delays.  Radler compiles the associated software source code related to afs\_gateway and afs\_function (RADL nodes of AFS) and then executes the code within a ROS environment with a fixed period that is may or may not vary but is bounded (i.e. period may vary between min and max clock).

\vspace{0.1in}
\shadowbox{
\begin{minipage}{0.9\textwidth}
\begin{center}
\emph{Architecture Generic Property 2:} 
\end{center}

Quasi-periodic computational model of any RADL architecture and in particular ArduPilot Advanced Fail Safe (AFS) subsystem has the following 5 properties:
\begin{enumerate}
\item Bounded processing latency for any message/topic between every pair of publisher and subscriber 
\item Communication messages are delivered in order i.e. no overtaking, with timing assumptions 
\item Given fixed buffer (including size 1), consecutive message losses can be bounded 
\item Message Loss can be eliminated with Bounded queue length which can be used to provision communication channel for transporting message/topic between publisher \& subscribers appropriately 
\item Age and/or Freshness of any message at any subscriber in terms of time steps is bounded.
\end{enumerate}
\end{minipage}
}
\vspace{0.1in}

\emph{Architecture Generic Property 2 met rationale}: 
Analysis results, based on PVS proofs, are presented in paper~\cite{conf/memocode/LarrieuS14} and discussed in Section~\ref{sec:RadlerMain}. 

\vspace{0.1in}
\shadowbox{
\begin{minipage}{0.9\textwidth}
\begin{center}
\emph{Architecture Generic Property 3:} 
\end{center}

Radler, validates platform/physical layer performance when mapped and bound to logical layer functions as well as any other assumptions on the architectural model used in verifying any other properties listed in \emph{Architecture Generic Property 2}. For any RADL Specification, and in particular ArduPilot Advanced Fail Safe (AFS) subsystem RADL specification,  the following 4 properties are always held at run-time due to Radler run-time checks/validation: 
\begin{enumerate}
\item No stale message received after it's latency and period 
\item Every message is never received more than it's timeout duration latency and period 
\item System healthy and no failures 
\item Max latencies, node periods, execution times adheres to the RADL specifications.
\end{enumerate}
\end{minipage}
}
\vspace{0.1in}

\emph{Architecture Generic Property 3 met rationale}: 
Radler enforces for each node, a state structure, an initialization function, a step function, and a finish function code structure as well as generates the communication layer (glue code), the scheduler, the overall compilation script and configuration files, and builds an executable for each machine. Radler also instruments the code for runtime monitoring and validation of platform performance and assumptions. Please see detailed rationale in  paper~\cite{conf/memocode/LiGS15} and discussions in Section~\ref{sec:RadlerMain}. 

\vspace{0.1in}
\shadowbox{
\begin{minipage}{0.9\textwidth}
\begin{center}
\emph{Architecture Generic Property 4:} 
\end{center}

Physical Layer of RADL Architectural Specification and in particular ArduPilot Advanced Fail Safe (AFS) subsystem physical layer has robust time and space partitioning.
\end{minipage}
}
\vspace{0.1in}

\emph{Architecture Generic Property 4 met rationale}: 
Robust space Partitioning based on the memory partitioning strategy in the specification depending on whether the two RADL nodes in  are in (1) same virtual machine within a single RTOS  or  (2) two different virtual machines with different individual RTOS but in the same hypervisor within a single machine or (3) two machines.
Radler will then ensure provisioning respectively for (1) a shared memory ring buffer (intra partition) or (2)  a ring buffer will be setup in a memory region shared
between those two virtual machine with message passing APIs (inter-partition) and (3) IP-based communication will be used with queuing and sampling ports (inter-partition) appropriately. 

Robust time partitioning with WCETs for every step function (computation duration) within each RADL node being specified in the RADL specification and also subsequently  Radler tool is also able to generate an instrumented version of the software code and monitor the system with runtime checks for violations of  timing properties such as maximum latency of channels, node periods, execution times, etc. The radler tool also generates local system log that radler can be analyzed and validated against the architecture specification. Such runtime checks are often the only way of validating the timing parameters assumed by the model. Please see detailed rationale in paper~\cite{conf/memocode/LiGS15} and discussions in Section~\ref{sec:RadlerMain}.

\chapter{Conclusions}\label{chap:conclusions}

In this report, we have spelled out the details of the DesCert approach to assurance-driven development
of safety-critical software.  The high-level goal of the project is to raise the level of rigor, composability,
efficiency, and automation in generating and composing evidence for software safety claims.
We adopt a continuous certification approach where the evidence is generated and curated during
the design lifecycle.  We organize the assurance argument using an Eight Variables Model for control
systems that interact with a physical plant in a physical environment through sensors and actuators
to execute commands from an operator.  The High-Level Requirement is that the control software must
carry out the planned mission while maintaining the safety of the pose of the Plant and responding to
the Operator commands.  It must do this in the  face of disturbances from the Environment and uncertainty
in the readings taken by the Sensor and the responsiveness of the Actuator.

Our goal is to support an assurance-driven development methodology
where the designs are optimized to deliver efficient arguments, where
any latent flaws can be easily unmasked by a skeptical evaluator.
In order to design for the efficiency of the argument, we have used
a precise formal language for capturing requirements, and a rigorous model
of distributed computation with multirate, quasi-periodic computation and
communication.  We have developed techniques for extracting models from the
requirements and analyzing them for generic and specific properties.
We have done some preliminary work on ontic type systems that capture the
intent underlying data representations in order to identify bugs due to data misuse
and deliver efficient arguments for the absence of such bugs.  We have also developed
support for evidence generation from code analysis tools including test generators,
dynamic analyzers, and static analyzers.  

We have undertaken a substantial case study of an Advanced FailSafe (AFS)
module for the ArduCopter rotorcraft that is required to execute a flight mission.
The AFS module implements recovery mechanisms
for handling failures due to GPS Lock Loss, Battery Insufficiency, Ground Communication
Loss, and  breaches of Altitude, Range, and Polygonal Geofences.  

The elements of our approach as captured in this case study include
\begin{enumerate}
\item Requirements definition using CLEAR
\item Generic and specific properties of artifacts
\item Ontic types to capture the interpretation of a data representations as a URL, IP address, (authenticated)
  User ID, SQL query, etc., applied in traceable manner across the design artifacts
\item Requirements analysis using model checking (with Text2Test and Sally)
\item Logical and Physical Architecture definition (using RADL)
\item Radler Architecture properties established using a proof assistant (PVS)
\item Radler glue code for scheduling and communication, and build process
\item Code contracts for the step functions in the Radler nodes
\item Static analysis for type correctness and other specific and generic properties
\item Dynamic property checking using Randoop and Daikon on Java source code, Text2Test
  on object code, and runtime safety monitoring on running systems. 
\end{enumerate}

We have instrumented the tools generated evidence data and metadata from this case study for
automatic ingestion into the ARCOS Rack Knowledge Base developed by GE Research.
This evidence has been employed in the construction of prototype assurance cases by the ARCOS TA4
teams.

Our future work within the ARCOS project focuses on
\begin{enumerate}
\item Broader ontic type analysis with traceability from requirements to code
\item Multi-component requirements and software architecture
\item Security and Fault-tolerance properties
\item Expressive static analysis of source code components
\item Proof-carrying code infrastructure for eBPF for certifying kernel modules
\item Automated continuous assurance workflow spanning the design lifecycle
\item Incremental assurance integrated with version control
\item Assurance dashboard for monitoring repairing evidence status
\item CLEAR customizations for domain-specific operations
\item Open-source evidential marketplace for software components/libraries
\end{enumerate}

To conclude, in Phase 1 of the DesCert project, we have developed the outline of a methodology
for continuous assurance using a rigorous workflow supported by tools that can generate and
compose semantically coherent evidence for constructing efficient assurance arguments.
Current and future work is directed at demonstrating the scalability of this approach
for multi-component systems with complex safety and security requirements.

\bibliographystyle{alpha}
\bibliography{biblio}

\end{document}